\documentclass[11pt]{article}

\usepackage{graphicx}
\usepackage{calc}
\usepackage{rotating}
\usepackage[english]{babel}
\usepackage{graphicx}
\usepackage{subfig}
\usepackage{float}
\usepackage{amsmath}
\usepackage{amssymb}
\usepackage{amsthm}
\usepackage{latexsym}
\usepackage{dcolumn}
\usepackage{geometry}
\usepackage{color}
\usepackage{hyperref}
\usepackage{mciteplus}
\usepackage{slashed}
\usepackage{multirow}

\def\gtwid{\mathrel{\raise.3ex\hbox{$>$\kern-.75em\lower1ex\hbox{$\sim
$}}}}
\def\vio{\mathrel{\hbox{$E$\kern-.60em\hbox{$/
$}}}}
\newcommand{\hobs}{\ensuremath{h_{\rm obs}}}

\begin{document}

\begin{flushright}
KIAS-P17020\\
\end{flushright}
\vspace*{2.0cm}
\begin{center}
{\Large \bf {Two Higgs bosons near 125\,GeV in the NMSSM: 
beyond the narrow width approximation} \\
\vspace*{0.8cm}
{\large Biswaranjan Das$^a$, Stefano Moretti$^b$, 
Shoaib Munir$^c$ and Poulose Poulose$^a$ } \\[0.25cm]
{\small \sl $^a$Department of Physics, IIT Guwahati, Guwahati, Assam 781039, India} \\[0.25cm]
{\small \sl $^b$School of Physics and Astronomy,
University of Southampton, \\ Highfield, Southampton 
SO17 1BJ, UK} \\[0.25cm]
{\small \sl $^c$School of Physics, Korea Institute for Advanced Study,
Seoul 130-722, Republic of Korea}\bigskip \\
{\small \url{biswaranjan@iitg.ernet.in}, \url{s.moretti@soton.ac.uk},
  \\ \url{smunir@kias.re.kr}, \url{poulose@iitg.ernet.in}}}
\end{center}
\vspace*{0.4cm}

\begin{abstract}
\noindent
In the next-to-minimal supersymmetric (NMS) Standard Model (SM), 
it is possible for either one of the additional singlet-like 
scalar and pseudoscalar Higgs bosons to be almost degenerate 
in mass with the $\sim 125$\,GeV SM-like Higgs state. 
In the real NMSSM (rNMSSM), when the mass 
difference between two scalar states 
is comparable to their individual total decay widths, the 
quantum mechanical interference, due to the relevant diagonal 
as well as off-diagonal terms in the propagator matrix, between 
them can become sizeable. This possibility invalidates usage of 
the narrow width approximation (NWA) to 
compute the cross section for the production of a di-photon 
pair with a given invariant mass via resonant Higgs boson(s) 
in the gluon fusion process at the Large Hadron Collider (LHC). 
When, motivated by the baryon asymmetry of the universe, 
CP-violating (CPV) phases are explicitly invoked in the Higgs 
sector 
of the NMSSM, all the interaction eigenstates mix to give five 
CP-indefinite physical Higgs bosons. In this scenario, the 
interference effects due to the off-diagonal terms in the Higgs 
mass matrix that mix the pseudoscalar-like state with the 
SM-like one can also become significant, when these two are 
sufficiently mass-degenerate. We perform a detailed analysis, 
in both the real and complex NMSSM, of these interference 
effects, when the full propagator matrix is taken into 
account, in the production of a  photon pair with an invariant 
mass near 125\,GeV through gluon fusion. We find that these 
effects can account for up to $\sim 40$\% of the 
total cross section for certain model parameter configurations. 
We also investigate how such mutually interfering states 
contributing to the $\sim 125$\,GeV signal observed at the LHC 
can be distinguished from  a single resonance.\end{abstract}

\newpage
\section{Introduction}
\label{sec:intro}

The discovery of a Higgs
boson~\cite{Aad:2012tfa,Chatrchyan:2012xdj} at the LHC provides
convincing evidence of  spontaneous electro-weak (EW) symmetry breaking (SB)
through the Higgs mechanism. It is also intriguing that the subsequent
measurements of its properties have shown their remarkable agreement
with the expectations from the SM of particle
physics. These measurements include the signal rates and coupling strengths 
in the various Higgs
boson production and decay channels that have so far been analysed, as well as its
spin and parity. However, the shortcomings of the standard Higgs
mechanism, including primarily the stability of the mass of the Higgs
boson against large quantum corrections, need to be addressed properly
in order to completely understand the
dynamics of EWSB. Putting this together with  other unresolved issues in the SM, such as its inability to explain the mass of neutrinos, the nature of dark matter (DM) and the large baryon asymmetry of the universe, compel us to believe that the elementary particle spectrum could be richer than the minimal one embedded in the SM. 

Supersymmetry (SUSY), proposed originally as a remedy for some of the
above problems faced by the SM, presents an appealing explanation for the
stability of the Higgs mass, while providing also a natural
candidate for DM. However, its minimal manifestation, known
as the minimal supersymmetric Standard Model (MSSM), is becoming 
increasingly constrained by the LHC measurements of the Higgs boson
properties, besides becoming more and more fine-tuned in explaining  
null searches for its own direct signatures
(i.e., of sparticle states). The MSSM is also troubled by issues arising
purely from naturalness considerations, like the presence of a
quadratic Lagrangian  term with a new mass parameter, $\mu$, which is
phenomenologically required to lie at the EW scale but has no
theoretical ground to do so~\cite{Kim:1983dt,Nir:1995bu}. The
NMSSM
was proposed to take care of this so-called $\mu$-problem
through the introduction of a new singlet scalar
superfield~\cite{Fayet:1974pd,*Ellis:1988er,*Durand:1988rg,*Drees:1988fc}. The
presence of this superfield leads to two additional neutral Higgs mass
eigenstates in the NMSSM (see,
e.g.,~\cite{Ellwanger:2009dp,Maniatis:2009re} for reviews) compared
to the MSSM. 
When all the parameters in the Higgs and sfermion
sectors are assumed to be real and hence CP-conserving (CPC), 
one of these two additional states is a scalar and the other a
pseudoscalar. With five neutral Higgs bosons in total, the 
real NMSSM (rNMSSM) provides some unique possibilities for collider 
phenomenology.

In multi-Higgs models like the MSSM and  NMSSM, the observed
signature near 125\,GeV can in principle be explained in terms of
a single Higgs resonance or, alternatively, two or more Higgs
resonances that cannot be individually resolved by the experiment as yet. 
In the MSSM, the possibility of the two scalars both lying near
125\,GeV is ruled out by the fact that such a mass for
one of the scalars generally necessitates the other scalar to be
essentially decoupled, and hence much heavier or lighter (see, e.g.,~\cite{Djouadi:2005gj}). In the rNMSSM, in contrast, it is still a plausible scenario,
as discussed in~\cite{Gunion:2012gc,*King:2012tr,*Gherghetta:2012gb,Wu:2015nba,*Domingo:2015eea}. The alternative possibility of a pair of $\sim 125$\,GeV scalar and pseudoscalar has also been studied in~\cite{Munir:2013wka}.

In the NMSSM, to address the baryonic asymmetry of the
universe, CP-violation can be invoked directly
and explicitly in the Higgs sector at the tree level, unlike in the MSSM,
where it is only radiatively induced into the Higgs sector
beyond the Born approximation. This is done by taking the Higgs sector
trilinear couplings to be complex parameters, hence we refer to this
version of the model  as
the complex NMSSM (cNMSSM) here. For non-zero 
CPV phases of these parameters, instead of the distinct CP-even and
CP-odd Higgs bosons, the model contains five CP-indefinite
neutral states. This CP mixing in the Higgs sector can get
additional contributions from the complex Higgs-sfermion-sfermion 
couplings as well as the phases in the neutralino-chargino sector,
as in the CPV MSSM.
Consequently, depending on the sizes of these
phases, the mass spectrum and production/decay rates of the Higgs states can
get considerably modified compared to the CPC 
case~\cite{Moretti:2013lya}, similarly to the
MSSM~\cite{Demir:1999hj,*Dedes:1999sj,*Dedes:1999zh,*Kane:2000aq,*Arhrib:2001pg,*Choi:2001pg,*Choi:2002zp,Ellis:2004fs,Hesselbach:2009gw,*Fritzsche:2011nr,*Chakraborty:2013si}.
The phenomenology of a single CPV Higgs boson near 125\,GeV in the cNMSSM
has been studied for a variety of possible underlying scenarios in~\cite{Moretti:2013lya,Graf:2012hh,Munir:2013dya}. The specific case of two mass-degenerate Higgs resonances near
125\,GeV within the cNMSSM has also been considered in~\cite{Moretti:2015bua},
where it was shown that these can give a better fit to the LHC Run-I data,
performed using the program HiggsSignals~\cite{Bechtle:2013xfa},
compared to a single resonance.

In all the above-mentioned analyses of two $\sim 125$\,GeV Higgs bosons in the NMSSM, it was assumed that each of them is produced on-shell and decays
subsequently into any of the observed final states.
However, for very strong mass degeneracy, it is possible that the two Higgs states produced in gluon fusion oscillate into each other before they decay. This could be perceived as being facilitated through quantum corrections of the propagator with two different mass eigenstates at the two ends. Such effects, coming into play beyond a Breit-Wigner (BW) resonance, in general contexts as well as in specific scenarios like the CPC MSSM, have been considered in some studies~\cite{Cacciapaglia:2009ic,*Fuchs:2014ola,*Chen:2016oib,*Fuchs:2016swt}. The specific case of the CPV MSSM with one-loop effects in the full propagator was treated in~\cite{Ellis:2004fs}. 

The purpose of the present work is to explore this possibility of quantum mechanical interference between two Higgs states near 125\,GeV in the NMSSM, both real and complex. We shall demonstrate here that the inclusion of such effects enhances the span of the model solutions mimicking the LHC observation. We shall then investigate possible ways to identify signatures of such a coupled system of Higgs bosons through a shape study of the profile of the resonance in the invariant mass distribution of the di-photon decay products. The analysis is carried out by first performing numerical scans of the model parameter space to identify specific Benchmark Points (BPs) where two of the Higgs bosons are degenerate with mass around 125\,GeV, within the uncertainty allowed by present LHC
measurements. 
For these BPs, we then calculate the cross section for the production of a di-photon pair with invariant mass near 125\,GeV via Higgs resonance(s) in the gluon fusion process at the LHC, using a Monte Carlo (MC) integration code developed in-house. This cross section is calculated assuming three different approaches: the full Higgs propagator matrix in the amplitude; only diagonal terms in the propagator matrix; for the two Higgs bosons individually without any mutual interference. A comparison of these three cross sections shows significant effects of interference, with the full propagator case deviating by up to about 38\% compared to the sum of the two individual Higgs boson contributions, along with a hint of a distortion in the line-shape of the differential distribution. 

The article is organised as follows. In the next section we will 
briefly revisit the Higgs sector of the NMSSM. In
section \ref{sec:cross-section} we will derive the analytical expression for the cross section that includes the full Higgs propagator matrix. In
section \ref{sec:numerical} we will provide details of our methodology for the numerical analysis. In section \ref{sec:results} we
will discuss the results of our analysis and in section
\ref{sec:concl} we will present our conclusions.

\section{The NMSSM Higgs sector}
\label{sec:cpvnmssm}

The NMSSM is defined by the superpotential 
\begin{equation}\label{eq:superp}
W_{\rm NMSSM} = h_u\, \widehat{Q} \cdot \widehat{H}_u\;
\widehat{U}^c_R \: + \: h_d\, \widehat{H}_d \cdot \widehat{Q}\;
\widehat{D}^c_R \: + \: h_e\, \widehat{H}_d \cdot \widehat{L}\;
\widehat{E}_R^c \: + \: 
\lambda \widehat{S} \widehat{H}_u \cdot \widehat{H}_d \: + \:
\frac{\kappa}{3}\ \widehat{S}^3\,,
\end{equation}
 where $y_u$, $y_d$ and $y_e$ are the quark and lepton Yukawa 
coupling constants, $\hat Q$ and $\hat L$ are the left-handed
quark and lepton doublet superfields, $\hat U$, $\hat D$
and $\hat E$ are the right-handed up-type, down-type and 
electron-type singlet superfields, respectively, 
and the charge conjugation is denoted by the superscript $c$. Furthermore,
$\hat H_u$ and $\hat H_d$ in the above superpotential are $SU(2)_L$ 
Higgs doublet 
superfields with opposite hypercharge, $Y = \pm 1$, as in the MSSM, 
and  $\hat S$ is a Higgs singlet superfield. Here,
$\lambda$ and $\kappa$ are dimensionless
trilinear coupling constants.

The fourth term on the right hand side of Eq.\,(\ref{eq:superp}) replaces the Higgs-higgsino mass term,
$\mu \hat H_u \hat H_d$, present in the MSSM superpotential. 
The Higgs singlet superfield acquires 
a non-zero vacuum expectation value (VEV), $v_s$, after EWSB. This $v_s$ is naturally of 
the order of the SUSY-breaking scale $M_{\rm SUSY}$
(herein operationally defined as $M_{\rm SUSY}^2 = \frac{m^2_{\tilde t_1} + m^2_{\tilde t_2}}{2}$, where $m_{\tilde t_1}$ and $m_{\tilde t_2}$ are the physical masses of the two stops), thus solving the $\mu$-problem of the MSSM by generating 
an effective $\mu$-term,
\begin{equation}\label{eq:3}
\mu_{\rm eff} \equiv \lambda \left < \hat S \right> = \lambda v_s\,.
\end{equation}
The absence of a  $\mu \hat H_u \hat H_d$ term, however, results in a $U(1)_{\rm PQ}$ symmetry, which is explicitly broken here by the last term in Eq.\,(\ref{eq:superp}), thus introducing instead a discrete $Z_3$ symmetry and making the NMSSM superpotential scale-invariant as well. 

The Higgs potential, derived from the above superpotential, 
is given as
\begin{eqnarray}\label{eq:potential}
V_0 &=& {| \lambda \left(H_u^+ H_d^- - H_u^0 H_d^0 \right) + \kappa S^2 |}^2
+ \left(m_{H_u}^2 + {| \mu + \lambda S |}^2 \right)
  \left({| H_u^0 |}^2 + {| H_u^+ |}^2 \right) \nonumber \\ 
&+& \left(m_{H_d}^2 + {| \mu + \lambda S |}^2 \right)
  \left({| H_d^0 |}^2 + {| H_d^- |}^2 \right)
+ \frac{g^2}{4} \left({| H_u^0 |}^2 + {| H_u^+ |}^2
- {| H_d^0 |}^2 - {| H_d^- |}^2 \right)^2 \nonumber \\
&+& \frac{g_2^2}{2} {| H_u^+ H_d^{0*} + H_u^0 H_d^{-*} |}^2
+ m_S^2 {| S |}^2 + \left[ \lambda A_{\lambda} \left(H_u^+ H_d^- - H_u^0 H_d^0 \right) S
+ \frac{1}{3} \kappa A_{\kappa} S^3 + {\rm h.c.} \right],
\end{eqnarray}
where $g_1$ and $g_2$ are the $U(1)_Y$ and $SU(2)_L$ gauge coupling 
constants, respectively, and 
$g^2 = \frac{g_1^2+g_2^2}{2}$. Here,
$A_{\lambda}$ and $A_{\kappa}$
are soft SUSY-breaking Higgs trilinear 
couplings, while $m_{H_d}$, $m_{H_u}$ and $m_S$ denote 
the soft Higgs masses. The fields $H_d$, $H_u$ and
$S$ are expanded about their respective VEVs, $v_d$, $v_u$ and $v_s$, as
\begin{equation}\label{eq:fields}
H_d^0 = 
\left( \begin{array}{c} \frac{1}{\sqrt 2}(v_d+H_{dR}+iH_{dI}) \\ H_d^- \end{array} \right),
H_u^0 = e^{i \phi_u}
\left( \begin{array}{c} H_d^+ \\ \frac{1}{\sqrt 2}(v_u+H_{uR}+iH_{uI}) \end{array} \right),
S^0 =  \frac{e^{i \phi_s}}{\sqrt 2}(v_s+S_R+iS_I).
\end{equation}
For correct EWSB, the $V_0$, rewritten in terms of these expanded fields, should
 have a minimum at non-vanishing $v_d$, $v_u$ and $v_s$, implying
\begin{equation}\label{eq:tadpole}
\left < \frac{\delta V_0}{\delta \theta} \right > = 0~~~ 
{\rm for}~~~ \theta = H_{dR}, H_{uR}, S_R, H_{dI}, H_{uI}, S_I\,,
\end{equation}
which leads to six `tadpole conditions' (see, e.g.,~\cite{Munir:2013dya}).  

Taking the second derivative of $V_0$ 
at the vacuum yields the tree-level $6 \times 6$ neutral Higgs mass matrix-squared, ${\cal M}^2_0$, in the basis ${\rm H}^T = (H_{dR},H_{uR},S_R,H_{dI},H_{uI},S_I)$.
 It can be expressed in the general form 
\begin{equation}\label{eq:matrix}
{\cal M}_0^2 =
\left( \begin{array}{c|c} {\cal M}_S^2  &  {\cal M}_{SP}^2 \\  
\\  \hline & \\
\left({\cal M}_{SP}^2\right)^T & {\cal M}_P^2 \end{array} \right)\,,
\end{equation}
where the $3 \times 3$ block ${\cal M}_S^2$ corresponds to the CP-even interaction states ($H_{dR}$, $H_{uR}$, $S_R$), the $3 \times 3$ block ${\cal M}_P^2$ to the CP-odd states ($H_{dI}$, $H_{uI}$, $S_I$), while ${\cal M}_{SP}^2$ is responsible for mixing between the CP-even and -odd states.

In the rNMSSM, where all the Higgs sector trilinear coupling parameters are real, ${\cal M}_{SP}^2$ is a null matrix. One can therefore simply rotate only the submatrix ${\cal M}_P^2$ to isolate the massless Nambu-Goldstone boson field, $G$, which can then be dropped to yield a $5 \times 5$ mass matrix ${\cal M}_0^{'2}$. This mass matrix receives higher order corrections, $\Delta {\cal M}^2$, from various sectors of the model, and thus gets modified as
\begin{equation}\label{eq:BLO}
{\cal M}_H^2 = {\cal M}_0^{'2} + \Delta {\cal M}^2\,.
\end{equation}
The most dominant of these corrections can be found in~\cite{Ellwanger:2009dp,Ellwanger:2005fh}. Some further two-loop corrections
have been calculated in~\cite{Degrassi:2009yq} and~\cite{Goodsell:2014pla}.
After the inclusion of these corrections, the submatrices ${\cal M}_S^2$ and ${\cal M}_P^{'2}$ are separately diagonalised to obtain the three CP-even mass eigenstates, $H_{1,2,3}$ (with $m_{H_1} < m_{H_2} < m_{H_3}$), and the two CP-odd physical Higgs bosons, $A_{1,2}$ (with $m_{A_1} < m_{A_2}$).

On the other hand, one can also invoke CP-violation explicitly by assuming $\lambda \equiv | \lambda | e^{i \phi_{\lambda}}$, $\kappa \equiv | \kappa | e^{i \phi_{\kappa}}$, $A_{\lambda} \equiv | A_{\lambda} | e^{i \phi_{A_{\lambda}}}$ and $A_{\kappa} \equiv | A_{\kappa} | e^{i \phi_{A_{\kappa}}}$. This leads to non-zero entries in the ${\cal M}_{SP}^2$ submatrix, implying that the CP-even and CP-odd interaction eigenstates also mix mutually. In this cNMSSM, the $G$ state is first separated out through a rotation of the entire ${\cal M}^2_0$ by ${\cal R}^G$,
\begin{equation}\label{eq:rotation1}
\left( H_{dR},H_{uR},S_R,H_I,S_I,G \right)^T = 
{\cal R}^G \left( H_{dR},H_{uR},S_R,H_{dI},H_{uI},S_I \right)^T\,,
\end{equation}
 and dropped before calculating the higher order corrections to the resulting ${\cal M}_0^{'2}$. The complete expressions for the tree-level ${\cal M}_0^{'2}$ and the dominant one-loop contributions to $\Delta {\cal M}^2$ 
from the (s)quark and gauge sectors in the cNMSSM were 
studied in~\cite{Ham:2001wt,*Funakubo:2004ka,Cheung:2010ba,Cheung:2011wn}
 in the renormalisation group equations-improved effective potential
approach. The corrections from the gaugino sector were added
in~\cite{Munir:2013dya} and, more inclusively, in~\cite{Domingo:2015qaa}. 
In the Feynman diagrammatic approach, the
complete one-loop Higgs mass matrix was derived in~\cite{Graf:2012hh} and
the $\mathcal{O}(\alpha_t \alpha_s)$ contributions to it were calculated
recently in~\cite{Muhlleitner:2014vsa}. 

The physical Higgs mass eigenstates of the cNMSSM are then obtained 
from the interaction states through another rotation by ${\cal R}^H$,
\begin{equation}\label{eq:rotation2}
\left( H_1,H_2,H_3,H_4,H_5 \right)^T = 
{\cal R}^H \left( H_{dR},H_{uR},S_R,H_I,S_I \right)^T\,,
\end{equation}
resulting in the diagonalised squared mass matrix,
\begin{equation}\label{eq:mdiag}
{\rm diag} \left( m_{H_1}^2,m_{H_2}^2,m_{H_3}^2,m_{H_4}^2,m_{H_5}^2 \right)
= {\cal R}^H \left[ {\cal R}^G {\cal M}_H^2 {\left({\cal R}^G \right)}^T \right]
{\left({\cal R}^H \right)}^T\,.
\end{equation}
Here $H_1$, $H_2$, $H_3$, $H_4$ and $H_5$ are the five 
neutral CP-indefinite Higgs bosons, ordered in terms of increasing mass.

Note here that only the physical phase combination
$\phi_{\lambda}-\phi_{\kappa}+\phi_u-2\phi_s$ appears in the cNMSSM
Higgs sector at the tree-level, as the other possible phase combinations
involving $\phi_{A_\lambda}$ and $\phi_{A_\kappa}$ are determined by it, up to a
twofold ambiguity, through the tadpole conditions.
Beyond this level, the CPV phases of the gaugino mass parameters, 
$M_{1,2,3}$, and of the soft trilinear couplings, $A_{\tilde f}$, of
the Higgs boson to the sfermions also get radiatively induced into
this sector.


\section{Di-photon production via gluon fusion}
\label{sec:cross-section}

We now turn our attention to the process under scrutiny, i.e.,  di-photon production from gluon fusion via Higgs states at the LHC.
The squared amplitude for the $gg \to H \to \gamma \gamma$ process,
with $H$ collectively denoting the five neutral CP-indefinite Higgs
bosons, can be written as~\cite{Spira:1995rr}
\begin{equation}\label{eq:ampsquare}
{\mid {\cal M} \mid}^2 = \sum_{\lambda , \sigma=\pm} {\cal M}_{P \lambda} {\cal M}_{P \lambda}^* \left|D_H(\hat s)\right|^2 
{\cal M}_{D \sigma} {\cal M}_{D \sigma}^*\,,
\end{equation}
where $\lambda, \sigma = \pm 1$ are the gluon and photon helicities,
respectively, and $D_H(\hat s)$ is the Higgs boson propagator, with
${\hat s}$ being the squared center-of-mass (CM) energy of the incoming
gluons. The amplitude for the production part is given as 
\begin{equation}\label{eq:prodamp}
{\cal M}_{P \lambda} = \sum_{i=1-5}{\cal M}_{P_i \lambda} =\sum_{i=1-5}\frac{\alpha_s m_{H_i}^2}{4 \pi v} 
\Bigl\{ S_i^g(m_{H_i}) + i \lambda P_i^g(m_{H_i}) \Bigr\}\,, 
\end{equation}
where the scalar and pseudoscalar form-factors are~\cite{Baglio:2013iia}
\begin{eqnarray}\label{eq:form}
S_i^g(m_{H_i}) &=& 2\sum_q g_{H_i qq}^S \tau_q \{ 1+(1-\tau_q) f(\tau_q) \}
- \sum_{\tilde q} 
\frac{m_Z^2 \tau_{\tilde q}}{m_{\tilde q}^2}
g_{H_i {\tilde q}{\tilde q}} \{1-\tau_{\tilde q} f(\tau_{\tilde q}) \}\,,\\
P_i^g(m_{H_i}) &=& 2\sum_q g_{H_i qq}^P \tau_q f(\tau_q)\,, 
\end{eqnarray}
with $\tau_x = 4m_x^2/m_{H_i}^2~(x=q,\tilde q)$ and the loop 
function being 
\begin{equation}\label{eq:func}
\begin{split}
f(\tau) = {\rm arcsin^2}(1/\sqrt\tau),~\tau \geq 1\,;\\
f(\tau) = -\frac{1}{4}\left\{{\rm ln}\left(\frac{1+\sqrt{1-\tau}}{1-\sqrt{1-\tau}}\right)
-i\pi\right\}^2,~\tau < 1\,.
\end{split}
\end{equation}
{Note that {$g^{S,P}_{H_i qq}$ and $g^{S}_{H_i {\tilde q}{\tilde q}}$}} in Eq.~(\ref{eq:form}) are the couplings of $H_i$ to quarks $q$ and squarks $\tilde q$, respectively, which depend on the elements of the Higgs mixing matrix, ${\cal R}^H$, noted in Eq.~(\ref{eq:mdiag}) above. The exact forms of these couplings can be found in~\cite{Cheung:2011wn}.

The amplitude for the decay part is similarly given as
\begin{equation}\label{eq:decayamp}
{\cal M}_{D \sigma} = \sum_{i=1-5} {\cal M}_{D_i \sigma}=\sum_{i=1-5}\frac{\alpha_{em} m_{H_i}^2}{4 \pi v} 
\Bigr\{ S_i^{\gamma}(m_{H_i}) + i \sigma P_i^{\gamma}(m_{H)i}) \Bigr\}\,, 
\end{equation}
with the form-factors being
\begin{eqnarray}\label{eq:decform}
S_i^\gamma(m_{H_i}) &=& 2\sum_f N_{cf} e_q^2 g_{H_i qq}^S \tau_q \{ 1+(1-\tau_q) f(\tau_q) \} - \sum_{\tilde f} N_{cf} e_{\tilde q}^2 \frac{M_Z^2}{M_{\tilde q}^2}
g_{H_i {\tilde q}{\tilde q}} \tau_{\tilde q} \{1-\tau_{\tilde q} f(\tau_{\tilde q}) \} \nonumber\\
&-& g_{H_i WW}\{2+3\tau_W+3\tau_W(2-\tau_W)f(\tau_W) \}
- \frac{M_Z^2}{2M_{H^\pm}^2} g_{H_i H^+ H^-} \tau_{H^\pm}\{1-\tau_{H^\pm}f(\tau_{H^\pm}) \} \nonumber \\
&+& 2\sum_{{\tilde \chi}_{j=1,2}^\pm} \frac{M_W}{M_{{\tilde \chi}_j^\pm}} 
g_{H_i {{\tilde \chi}_j^+}{{\tilde \chi}_j^-}}^S \tau_{{\tilde \chi}_j^\pm} 
\{ 1+(1-\tau_{{\tilde \chi}_j^\pm}) f(\tau_{{\tilde \chi}_j^\pm}) \}\,,\\ 
P_i^\gamma(m_{H_i}) &=& 2\sum_f N_{cf} e_q^2 g_{H_i qq}^P \tau_q f(\tau_q)
+ 2\sum_{{\tilde \chi}_{j=1,2}^\pm} \frac{M_W}{M_{{\tilde \chi}_j^\pm}} 
g_{H_i {{\tilde \chi}_j^+}{{\tilde \chi}_j^-}}^P \tau_{{\tilde \chi}_j^\pm} 
f(\tau_{{\tilde \chi}_j^\pm})\,, 
\end{eqnarray}
where $N_{cf} = 3,1$ is the colour factor for (s)quarks and 
charged (s)leptons, respectively, with $e_f$ being the corresponding
 electric charge.
Finally, the full propagator in Eq.~(\ref{eq:ampsquare}) is a $5\times
5$ matrix\footnote{Assuming negligible off-resonance self-energy
  transitions of any of the five Higgs bosons to the would-be Goldstone boson, $G$.}, given as 
\begin{equation}
\label{eq:propmat}
D_H(\hat s) = \hat s 
\left(\begin{array}{@{}ccccc@{}} 
{\rm m_{11}} 
& i{{\mathfrak{I}}{\rm m}\hat\Pi}_{12}(\hat s) 
& i{{\mathfrak{I}}{\rm m}\hat\Pi}_{13}(\hat s)
& i{{\mathfrak{I}}{\rm m}\hat\Pi}_{14}(\hat s) 
& i{{\mathfrak{I}}{\rm m}\hat\Pi}_{15}(\hat s) \\
i{{\mathfrak{I}}{\rm m}\hat\Pi}_{21}(\hat s) 
& {\rm m_{22}}
& i{{\mathfrak{I}}{\rm m}\hat\Pi}_{23}(\hat s)
& i{{\mathfrak{I}}{\rm m}\hat\Pi}_{24}(\hat s) 
& i{{\mathfrak{I}}{\rm m}\hat\Pi}_{25}(\hat s) \\
i{{\mathfrak{I}}{\rm m}\hat\Pi}_{31}(\hat s) 
& i{{\mathfrak{I}}{\rm m}\hat\Pi}_{32}(\hat s) 
& {\rm m_{33}}
& i{{\mathfrak{I}}{\rm m}\hat\Pi}_{34}(\hat s) 
& i{{\mathfrak{I}}{\rm m}\hat\Pi}_{35}(\hat s)  \\
i{{\mathfrak{I}}{\rm m}\hat\Pi}_{41}(\hat s) 
& i{{\mathfrak{I}}{\rm m}\hat\Pi}_{42}(\hat s) 
& i{{\mathfrak{I}}{\rm m}\hat\Pi}_{43}(\hat s) 
& {\rm m_{44}}
& i{{\mathfrak{I}}{\rm m}\hat\Pi}_{45}(\hat s)  \\
i{{\mathfrak{I}}{\rm m}\hat\Pi}_{51}(\hat s) 
& i{{\mathfrak{I}}{\rm m}\hat\Pi}_{52}(\hat s) 
& i{{\mathfrak{I}}{\rm m}\hat\Pi}_{53}(\hat s) 
& i{{\mathfrak{I}}{\rm m}\hat\Pi}_{54}(\hat s) 
& {\rm m_{55}}  \\
\end{array}\right)^{-1}\,, 
\end{equation}
where ${\rm m}_{ii} \equiv \hat s-m_{H_i}^2 +i{{\mathfrak{I}}{\rm m}\hat\Pi}_{ii}(\hat s)$, with ${{\mathfrak{I}}{\rm m}\hat\Pi}_{ij}(\hat s)$ being the absorptive parts of the Higgs self-energies, for $i,\,j=1 - 5$. The diagonal absorptive
parts are equivalent to the widths of the corresponding Higgs states, $\Gamma_{H_i}$. The explicit expressions for ${{\mathfrak{I}}{\rm m}\hat\Pi}_{ij}(\hat s)$ are given in the Appendix. Note here that in our numerical analysis we will only focus on $H_1$ and $H_2$ having masses very close to 125\,GeV. This implies that essentially the propagator matrix elements with $i,\,j=1 - 2$ are the only ones that contribute to the production of the di-photon pair with invariant mass near 125\,GeV, assuming that the remaining three Higgs bosons are relatively heavy. Moreover, in the case of the rNMSSM, since the CP-even-odd mixing terms in the Higgs mass matrix vanish, it is sufficient for our purpose to consider only the $3\times 3$ propagator matrix corresponding to the CP-even states instead of the complete one above. 

When the splitting between the Higgs boson
masses is much larger than the sizes of the absorptive parts in Eq.~(\ref{eq:propmat}), the
NWA can be applied to the $i$th Higgs boson propagator as
\begin{equation}\label{eq:Higgsprop}
\left|D_{ii}(\hat s)\right|^2={\left| \frac{1}{\hat s - m_{H_i}^2 + i m_{H_i} {\Gamma}_{H_i}} \right|^2} \to
\frac{\pi}{m_{H_i} {\Gamma}_{H_i}} \delta (\hat s - m_{H_i}^2)\,,
\end{equation}
so that the partonic cross section becomes
\begin{equation}\label{eq:partonXS}
\hat \sigma (gg \to H \to \gamma \gamma) =
\frac{1}{1024 \pi \hat s} 
\sum_{i=1-5}\left( \sum_{\lambda=\pm} {\left|{\cal M}_{P_i \lambda}\right|}^2 \times    
\frac{\pi}{m_{H_i} {\Gamma}_{H_i}} \delta (\hat s - m_{H_i}^2) \times 
\sum_{\sigma=\pm} {\left|{\cal M}_{D_i \sigma}\right|}^2 \right).
\end{equation}
The total cross-section for the process $pp \to H \to \gamma \gamma$ 
is then written as
\begin{eqnarray}\label{eq:totalXS}
\sigma_{pp}^{\gamma \gamma}&=&
\int_{0}^{1} dx_2 \int_{0}^{1} dx_1~~
\hat \sigma (gg \to H \to \gamma \gamma)~~ 
g(x_1) g(x_2) \nonumber \\
&=&\int_{0}^{1} dx_2 \int_{0}^{1} dx_1
\frac{g(x_1) g(x_2)}{1024 \pi \hat s} 
\sum_{i=1-5}\left( \sum_{\lambda=\pm} {\left|{\cal M}_{P_i \lambda}\right|}^2
\frac{\pi}{m_{H_i} {\Gamma}_{H_i}} \delta (\hat s - m_{H_i}^2)  
\sum_{\sigma=\pm} {\left|{\cal M}_{D_i \sigma}\right|}^2 \right),
\end{eqnarray}
where $g(x_1)$ and $g(x_2)$ are the parton distribution functions (PDFs) 
of the two gluons. By substituting $x_2$ in the above equation as
\begin{equation}\label{eq:subs}
\hat s = x_1 x_2 s \implies  x_1 x_2 = \frac{\hat s}{s} \equiv 
\tau \implies x_2 = \frac{\tau}{x_1} \implies dx_2 = \frac{d \tau}{x_1}\,,
\end{equation}
where $s$ is the total CM energy of the $pp$ system, and performing the integration over $d\tau$, one gets
\begin{eqnarray}\label{eq:finalXS}
\sigma_{pp}^{\gamma \gamma} &=& 
\int_{\frac{m_{H_i}^2}{s}}^{1} dx_1~\frac{1}{1024 s m_{H_i}^3 \Gamma_{H_i}}
\sum_{i=1-5}\left(\sum_{\lambda=\pm} {\left|{\cal M}_{P_i \lambda}\right|}^2
\sum_{\sigma=\pm} {\left|{\cal M}_{D_i \sigma}\right|}^2 \right)
\frac{g(x_1) g\left({\frac{m_{H_i}^2}{s~x_1}}\right)}{x_1}\,.
\end{eqnarray}

\begin{figure}[tbp]
\begin{center}
\includegraphics[scale=1.0]{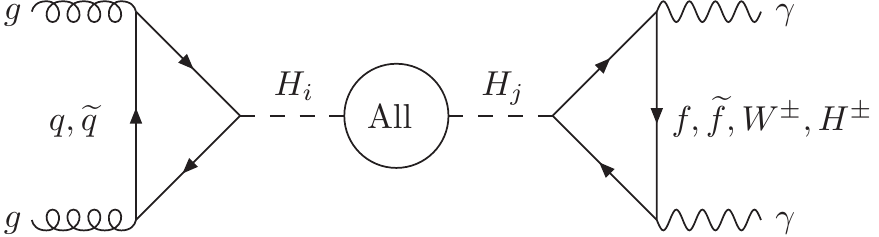}
\end{center}
\caption{Illustration of the effect of mixing in the propagator induced by quantum corrections.}
\label{fig:graph}
\end{figure}

In contrast, when two (or more) Higgs bosons of the model
are almost degenerate in
mass near a given ${\hat s}$, the sizes of the corresponding
absorptive parts can become comparable to their mass difference. As a
result, the $i$th Higgs state can undergo resonant transition to the
$j$th state through quantum corrections, as shown in Fig.~\ref{fig:graph}. In such a scenario, the NWA is no longer valid, and the total cross section is given as
\begin{equation}\label{eq:fullfinalXS}
\sigma_{pp}^{\gamma \gamma} =
\int_{0}^{1} d\tau \int_{\tau}^{1} \frac{dx_1}{x_1}
 \frac{g(x_1) g(\tau/x_1)}{1024 \pi \hat s^3} 
\sum_{i,j=1-5} 
\Bigl\{ 
\sum_{\lambda=\pm} {\left|{\cal M}_{P_i \lambda}\right|}^2
\left|D_{ij}(\hat s)\right|^2
\sum_{\sigma=\pm} {\left|{\cal M}_{D_j \sigma}\right|}^2
\Bigr\}\,,
\end{equation}
where $\left|D_{ij}(\hat s)\right|^2$ is the propagator matrix given in Eq.~(\ref{eq:propmat}).
From the above equation, one obtains the differential cross section
(recall that $\tau = \frac{\hat s}{s}$) as
\begin{equation}\label{eq:diffXShats}
\frac{d\sigma_{pp}^{\gamma \gamma}}{d\sqrt{\hat s}} =
\int_{\tau}^{1} \frac{2\sqrt{\hat s}}{s} \frac{dx_1}{x_1}
\frac{g(x_1) g(\hat s/s x_1)}{1024 \pi \hat s^3} 
\sum_{i,j=1-5} 
\Bigl\{ 
\sum_{\lambda=\pm} {\left|{\cal M}_{P_i \lambda}\right|}^2
\left|D_{ij}(\hat s)\right|^2
\sum_{\sigma=\pm} {\left|{\cal M}_{D_j \sigma}\right|}^2
\Bigr\}\,.
\end{equation}

\section{Numerical analysis}
\label{sec:numerical}

We first performed numerical scanning of the parameter space of the
NMSSM, requiring $H_1$ and $H_2$ to lie within the $123\,{\rm GeV} - 127\,{\rm GeV}$ range\footnote{The extended range
  of Higgs boson masses around the actual measured experimental value
  of $\sim 125$\,GeV is to allow for up to $\pm 2$\,GeV uncertainty
  from unknown higher order corrections in their model prediction.}. Our first scan corresponded to the rNMSSM, wherein
sufficient mass degeneracy near 125\,GeV between the two lightest scalars can generally be obtained for large values of
the couplings $\lambda$ and $\kappa$ and a relatively small
$\tan\beta$, which results in maximal mixing between the doublet- and
singlet-like states, as noted in some earlier
studies~\cite{Gunion:2012gc,*King:2012tr,*Gherghetta:2012gb}.
In the rNMSSM, while it is also possible for $A_1$ 
to lie near 125\,GeV~\cite{Munir:2013wka}, it does not mix with the SM-like $H_1$ when the coupling parameters are all real. Therefore, the corresponding off-diagonal absorptive parts in the propagator matrix given in Eq.~(\ref{eq:propmat}) are zero. When the complex
phases are turned on though, all the Higgs states become CP-indefinite, and any of the off-diagonal terms in the full $5 \times 5$ propagator matrix can be non-zero and contribute to the interference effects.
Therefore, either one of the scalar-singlet-like and 
pseudoscalar-singlet-like states can have strong mass-degeneracy with the $\sim 125$\,GeV SM-like state and interfere with it.    

As stated earlier, at the tree level, only the phase combination 
$\phi_{\lambda}-\phi_{\kappa}+\phi_u-2\phi_s$ appears in the Higgs sector of the cNMSSM. Furthermore, several studies~\cite{Graf:2012hh,Cheung:2010ba,Cheung:2011wn} have shown that, out of all the individual phases, including those appearing beyond the Born approximation, the phase $\phi_\kappa$ is virtually unconstrained by the measurements of the fermion Electric Dipole Moments (EDMs). Therefore, after setting all the other phases to $0^\circ$, we performed two separate parameter space scans of the cNMSSM also, with the value of $\phi_\kappa$ fixed to
$3^\circ$ in one and to $10^\circ$ in the other. In Tab. \ref{tab:scan} we list the scanned ranges of the free parameters (input at the EW scale), which assume the following universality conditions:
\begin{gather}
M_0 \equiv M_{Q_{1,2,3}} = M_{U_{1,2,3}} = M_{D_{1,2,3}} =
M_{L_{1,2,3}} = M_{E_{1,2,3}}\,; \nonumber \\
M_{1/2} \equiv 2M_1 = M_2 = \frac{1}{3}M_3\,;~~~
A_0 \equiv A_{\tilde{t}} = A_{\tilde{b}} = A_{\tilde{\tau}}\,, \nonumber
\end{gather}
where 
$M_{Q_{1,2,3}},\,M_{U_{1,2,3}},\,M_{D_{1,2,3}} ,\,M_{L_{1,2,3}}$ and $M_{E_{1,2,3}}$ are the soft masses of the sfermions, $M_{1,2,3}$ those of the gauginos and $A_{\tilde{t},\tilde{b},\tilde{\tau}}$ the soft trilinear couplings. These ranges are consistent across the three scans and correspond to the parameter space region that was noted to yield maximally mass-degenerate $H_1$ and $H_2$ in a previous study~\cite{Moretti:2015bua}, where more details about the scanning methodology can also be found. It was additionally pointed out in that study that for larger values of $\phi_\kappa$ it gets increasingly difficult to obtain both $H_1$ and $H_2$ near 125\,GeV in the cNMSSM. 

\begin{table}[tbp]
\centering\begin{tabular}{|c|c|}
\hline
Parameter & Scanned range  \\
\hline
$M_0$\,(GeV) 	& 800 -- 2000 \\
$M_{1/2}$\,(GeV)  & 100 -- 500	\\
$A_0$\, (GeV)  & $-3000$ -- 0\\
$\tan\beta$ 		& 2 -- 8 \\
$\lambda$ 		& 0.58 -- 0.7 \\
$\kappa$ 		& 0.3 -- 0.6 \\
$\mu_{\rm eff}$\,(GeV) 	& 100 -- 200 \\
$A_\lambda$\,(GeV)  	& 200 -- 1000 \\
$A_\kappa$\,(GeV)  	& $-300$ -- 0\\
\hline
\end{tabular}
\caption{\label{tab:scan} NMSSM parameters and their scanned ranges.}
\end{table}

For each parameter space input point generated by the scanning
algorithm, the masses as well as branching ratios (BRs) of the Higgs
bosons were calculated with the public code
NMSSMCALC v2.00~\cite{Baglio:2013iia}. The Supersymmetric Les Houches
Accord~\cite{Skands:2003cj,*Allanach:2008qq} output file produced
by NMSSMCALC for a scanned point was then passed to HiggsBounds v4.3.1~\cite{Bechtle:2008jh,*Bechtle:2011sb,*Bechtle:2013gu,*Bechtle:2013wla} to check for the consistency of each Higgs boson with the direct Higgs search results from LEP, Tevatron and LHC. We further made sure that a point only got through the scan if it satisfied the limits from measurements of the EDMs, computed intrinsically by NMSSMCALC. Finally, the CMS and ATLAS collaborations have performed measurements of the total width of the \hobs\ by analysing its off-shell production and subsequent decays in the $ZZ$ and $W^+ W^-$ channels~\cite{Aad:2015xua,Khachatryan:2015mma,*Khachatryan:2016ctc}. The most recent observed 95\% confidence level upper limit for the two channels combined is 13\,MeV. Therefore, we also require each of the $H_1$ and $H_2$ in a given scan to observe this constraint, unless stated otherwise for exceptional scenarios, which may well be plausible, as such a limit presumes an underlying BW resonance for the signal~\cite{Caola:2013yja,Englert:2014aca}, which is not the case here. 

Next, from the points collected in each scan, we selected BPs satisfying certain specific criteria, which will be explained later. In order to perform the numerical calculation of the cross sections for these BPs, we
implemented the expressions given in
Eqs.~(\ref{eq:fullfinalXS}) and (\ref{eq:diffXShats}) in a locally
developed {\tt fortran} program. 
This code is linked to the LAPACK package v3.6.0~\cite{lapack} for propagator matrix inversion, as well as to a locally modified version of the VEGAS routine~\cite{Lepage:1977sw}
to perform the 2-dimensional numerical integration. As a test of the
reliability of our results, for a given model point, we calculated
the cross section in the NWA for each of the two Higgs bosons with our 
code and compared it with 
the gluon fusion cross section computed using the publicly available code
SusHi v1.6.0~\cite{Harlander:2012pb,*Liebler:2015bka,*Harlander:2016hcx}
multiplied by its di-photon BRs obtained from NMSSMCALC. We found that the two results agreed within 5\% or better in all cases. The various Higgs boson
couplings for a given parameter space
point, needed for the calculation of the absorptive parts of the
propagator matrix as well as of the production and decay form-factors in our code, were also obtained from NMSSMCALC.

Note that our program calculates the total cross section only at
the Leading Order (LO), since implementing  Higher Order (HO) corrections, e.g., as
included in SusHi, is a highly involved task beyond the scope
of this work, which is aimed at comparing the effects of including the
full propagator against the simplest approach of two separate BWs on the total cross section. In fact, since these HO corrections apply only to the production
process, they should have exactly the same impact in both approaches, hence including their effect is simply tantamount to rescaling
the cross section by a `$k$ factor' (defined as $k_{\rm HO}\equiv \sigma_{\rm HO} / \sigma_{\rm LO}$, with HO implying the perturbative order at which the cross section is to be evaluated), which can be obtained from a
dedicated public tool. For a few test points corresponding to the rNMSSM, using SusHi, we thus also estimated the Next-to-Next-to-LO (NNLO) factor, $k_{\rm NNLO}$, by calculating the gluon-fusion cross section at both LO and NNLO in QCD. We found this $k_{\rm NNLO}$ to always lie very close to 3 for both $H_1$ and $H_2$. However, since SusHi is not yet compatible with the cNMSSM, the $k$ factor cannot be estimated exactly for the points corresponding to the cNMSSM. Therefore, for an approximate evaluation of the NNLO cross section in our discussion below, we will multiply the LO cross section obtained with our tool for a given point, both in the real and the complex NMSSM, with a constant $k_{\rm NNLO}=3$. (In fact, for an almost decoupled SUSY sector, as is generally the case here, the $k$ factor is indeed a constant for a SM-like $\sim 125$\,GeV Higgs boson, which can be obtained from, e.g., \cite{Heinemeyer:2013tqa}.)

Finally, we used the CT10~\cite{Lai:2010vv} PDF set for gluons in our cross section calculation, with renormalisation/factorisation scale set to the \hobs\ mass, i.e., 125\,GeV. We have, however, verified that the gross features of our analysis are independent of the choice of the PDF set as well as of the fixed numerical inputs for the SM and NMSSM parameters, for which the default NMSSMCALC values were retained.


\section{Results and discussion}
\label{sec:results}

\begin{figure}[h!]
\centering\includegraphics[width=81mm]{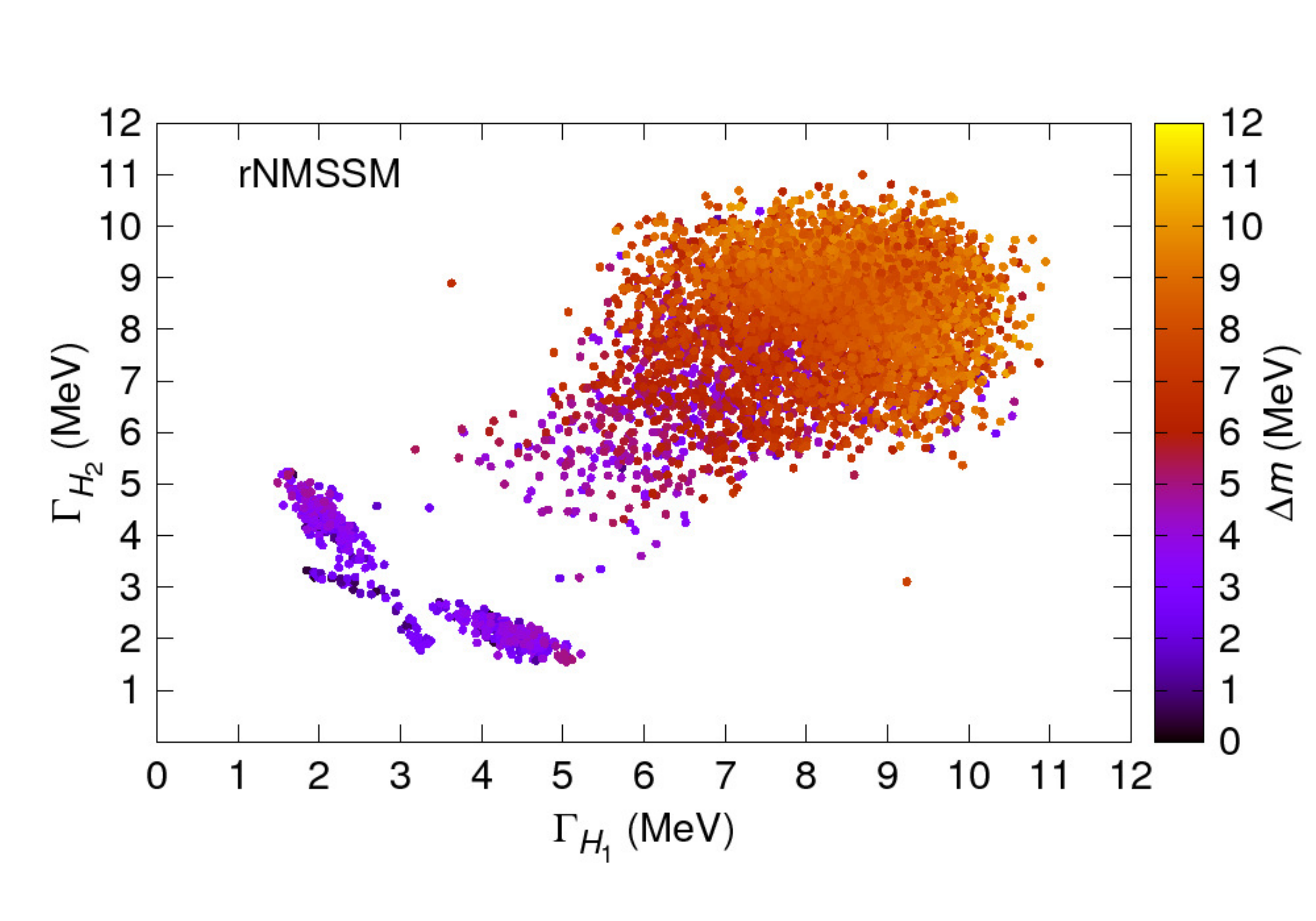}\\
\hspace*{-0.5cm}\includegraphics[width=81mm]{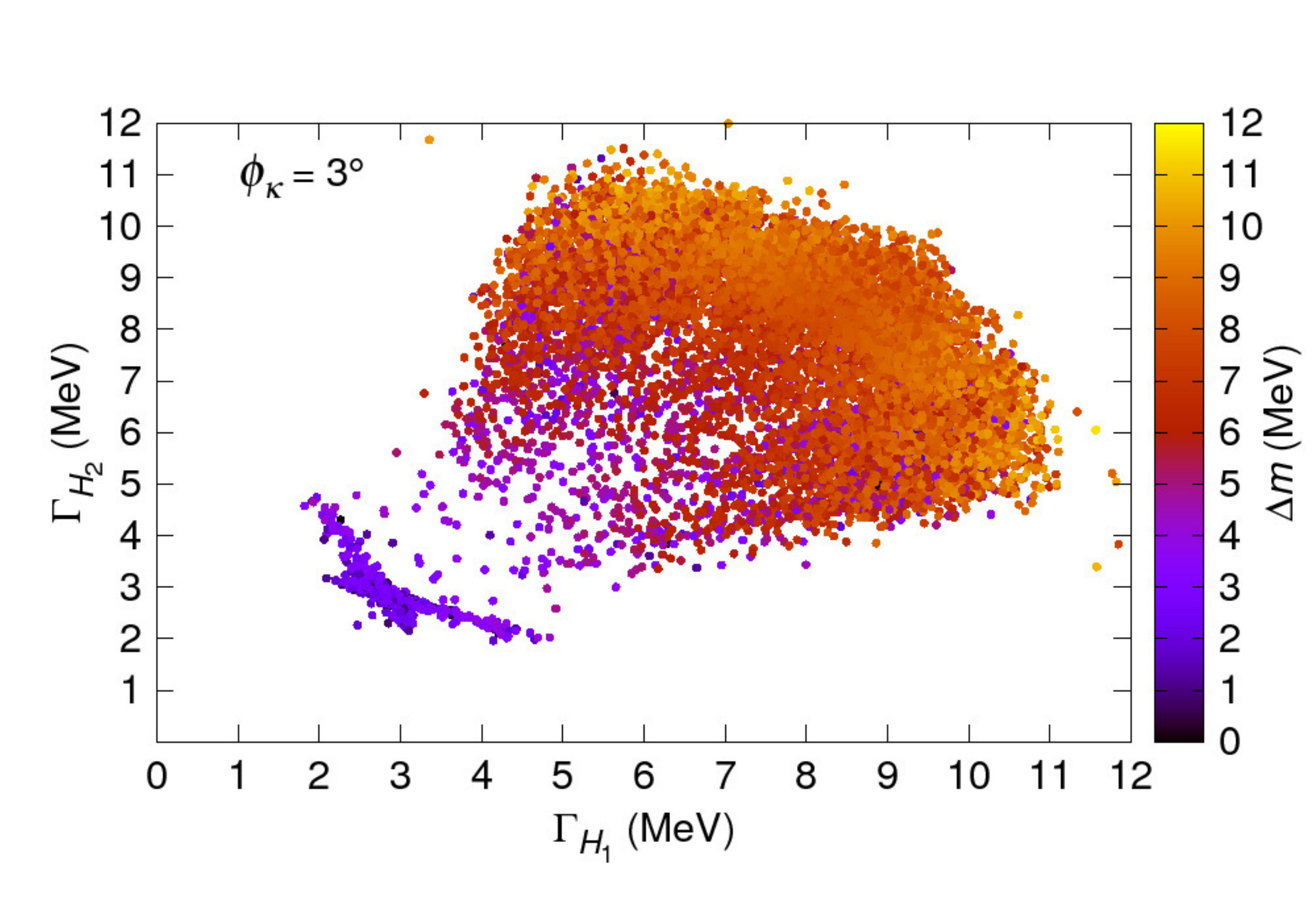}
\includegraphics[width=81mm]{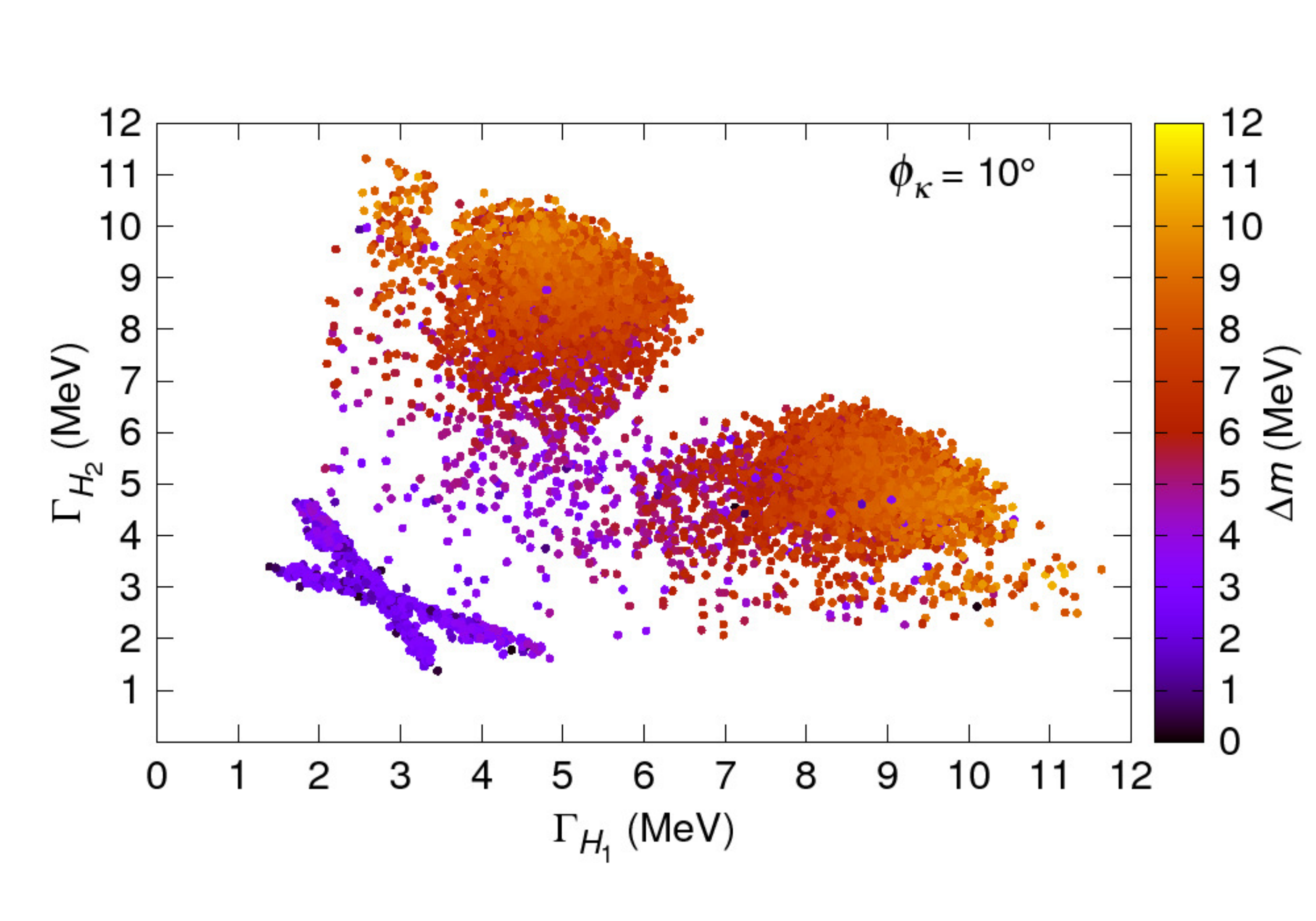}
\caption{\label{fig:scatter} Points obtained from the parameter 
scans of the rNMSSM (top) and of the cNMSSM with 
$\phi_\kappa = 3^\circ$ (bottom left) and with 
$\phi_\kappa = 10^\circ$ (bottom right). For all the points 
shown, $\Delta m$ (colour map) is always smaller than 
$\Gamma_{H_1}$ (x-axis) and/or $\Gamma_{H_2}$ (y-axis).}
\end{figure}

In Fig.~\ref{fig:scatter} we show those points obtained from the three scans for which $\Delta m\equiv m_{H_2}-m_{H_1}$ is smaller than one (or both) of the widths, $\Gamma_{H_1}$ and $\Gamma_{H_2}$, of the two lightest Higgs bosons. The top panel corresponds to the rNMSSM, while the bottom panels to the cNMSSM with $\phi_\kappa = 3^\circ$ (left) and with $\phi_\kappa = 10^\circ$ (right). We note in the figure that, given the parameter space in Tab. \ref{tab:scan}, for the vast majority of points, $\Gamma_{H_1}$ and $\Gamma_{H_2}$ tend to lie within 3--4\,MeV of each other in the rNMSSM. For $\phi_\kappa = 3^\circ$, the size of splitting between $\Gamma_{H_1}$ and $\Gamma_{H_2}$ can range from very small to fairly large across the points collected. However, for $\phi_\kappa = 10^\circ$, no points appear along the diagonal for $\Gamma_{H_1/H_2} > 6$\,MeV in the bottom right frame, as the splitting between these two widths starts growing beyond this value. 

From each of these scans we selected a few BPs to study the cross section for the production of a di-photon pair with an invariant mass, ${\cal M}_{\gamma\gamma}$, near 125\,GeV via resonant Higgs boson(s) in gluon fusion at the LHC with $\sqrt{s}=14$\,TeV.
More specifically, we studied the distributions, with respect to the partonic CM energy $\sqrt{\hat s}$ (which is the same as ${\cal M}_{\gamma\gamma}$ at LO), of the differential cross section, calculated such that: \\
Case 1: the $m_{11}$ and $m_{22}$ terms in the propagator matrix, Eq.~(\ref{eq:propmat}), each contribute alternatively to two amplitudes, which are squared and then summed, implying two independent Higgs boson resonances; \\
Case 2: both $m_{11}$ and $m_{22}$ contribute to the amplitude which is then squared, thus allowing for interference between $H_1$ and $H_2$ but without any mixing effects; \\
Case 3: besides $m_{11}$ and $m_{22}$, the off-diagonal terms $i{{\mathfrak{I}}{\rm m}\hat\Pi}_{12}$ and $i{{\mathfrak{I}}{\rm m}\hat\Pi}_{21}$ also contribute to the amplitude before squaring, leading to additional interference effects arising from the mixing of the two Higgs states.

The distributions obtained for the Cases 1, 2 and 3 (colour-coded in red, green and blue, respectively) are plotted in Fig.~\ref{fig:BPs-real} for each of the three selected BPs corresponding to the rNMSSM, with the integrated cross section for each curve also given in the legend. For these distributions, a ${\cal M}_{\gamma\gamma}$ bin width of 2\,MeV has been used. As noted above, $\Gamma_{H_1}\sim \Gamma_{H_2}$ for most of the points in the rNMSSM. Therefore, in order to illustrate the dependence of the interference effects on the mass difference and relative widths of the two Higgs bosons, we selected  BP1 such that $\Delta m \sim\Gamma_{H_1/H_2} $, BP2 such that $\Delta m < \Gamma_{H_1/H_2}$ and BP3 with $\Delta m \ll \Gamma_{H_1/H_2}$. One sees, going from the top panel to the bottom right one in the figure, that these effects are always positive and grow larger as $\Delta m$ decreases compared to $\Gamma_{H_1/H_2}$, as expected. Also, the interference effects due to the mixing terms in the propagators matrix (Case 3) are notably larger than those due only to the diagonal terms (Case 2) for each of the three BPs. The deviation in the total cross section with the full propagator matrix compared to the Case 1 for  BP3 at an inclusive level is about 38\%, clearly indicating that the interference effects can be quite sizeable. We point out here that although the BP3 represents maximal enhancement of these effects among all the points collected in our scans, it is possible that they can be even slightly larger for certain other parameter combinations in the vicinity of this BP. {Note also that the total integrated cross section (obtained at NNLO in QCD, as mentioned earlier) is generally consistent with the fiducial one, as estimated for the SM-like Higgs boson near 125\,GeV~\cite{deFlorian:2016spz}, or measured by the ATLAS and CMS collaborations for $h_{\rm obs}$ at $\sqrt{s}=13$\,TeV.\footnote{We also point out here that a considerable discrepancy exists between the ATLAS measurement, which reads $43.2\pm 14.9\pm 4.9$\,fb~\cite{ATLAS:2016nke}, and the CMS one, $69^{+18}_{-22}$\,fb~\cite{CMS:2016ixj}, besides a fairly large error in each of these itself. This renders an accurate estimation of the total cross section of little  significance here and justifies our use of an approximate constant NNLO $k$ factor.}

The values of the input parameters for all the selected BPs can be found in Tab.~\ref{table:BPinput}, and the masses and widths of $H_1$ and $H_2$ as well as the total cross sections corresponding to the three Cases for each of the BPs are given in Tab.~\ref{tab:XSections}.

\begin{figure}[h!]
\centering\includegraphics[width=82mm]{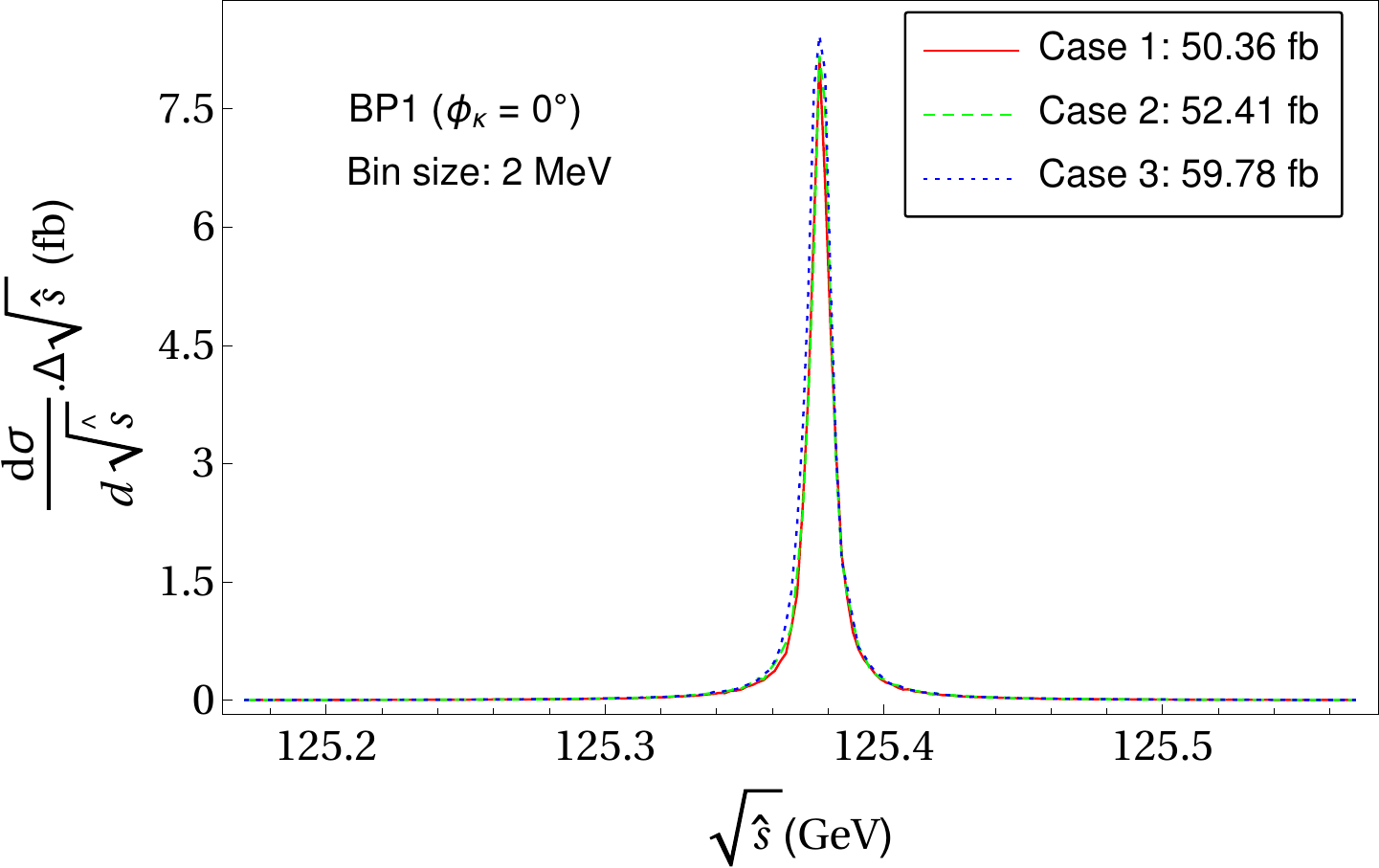}\\
\hspace*{-0.2cm}\includegraphics[width=82mm]{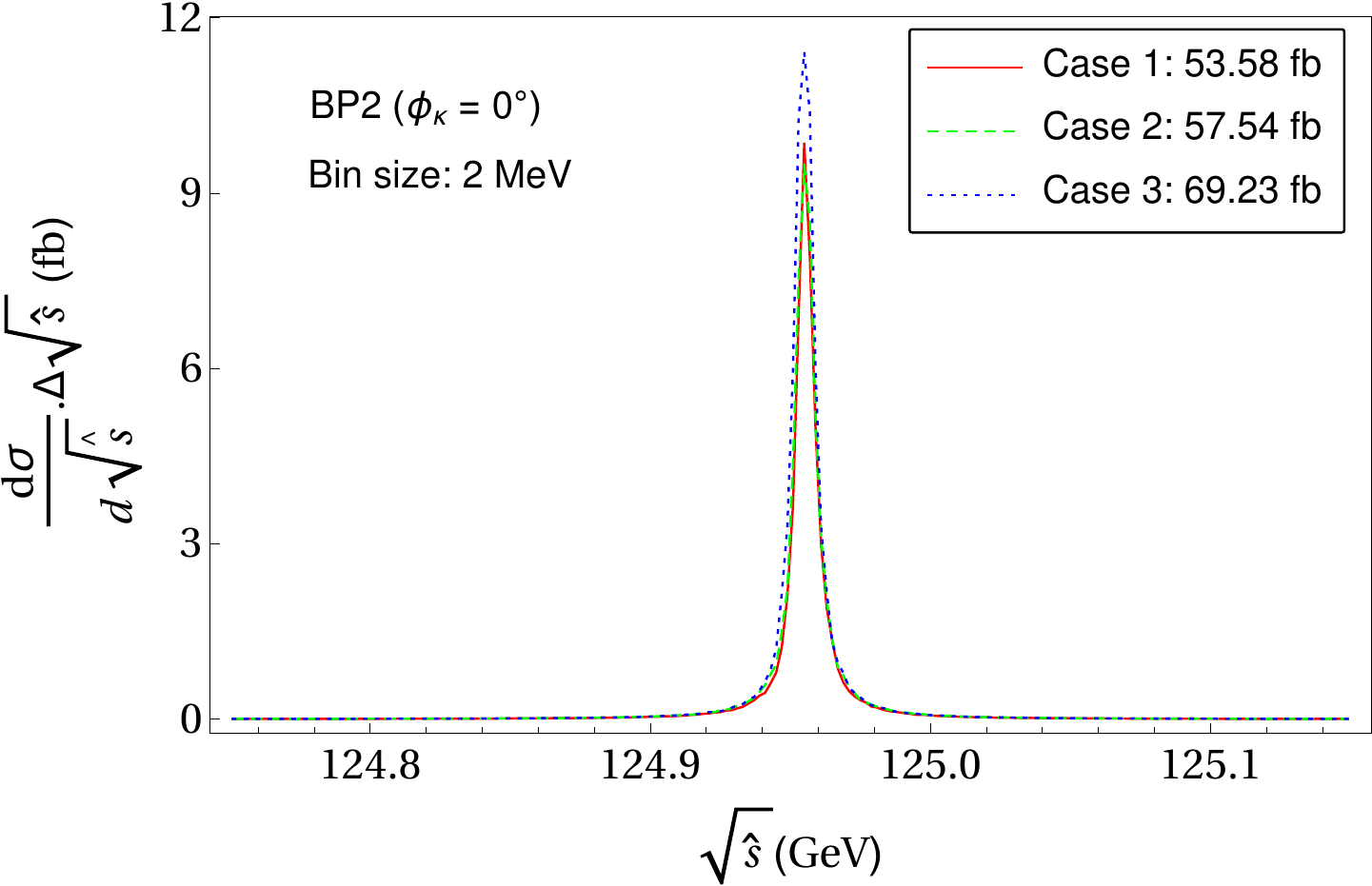}
\includegraphics[width=82mm]{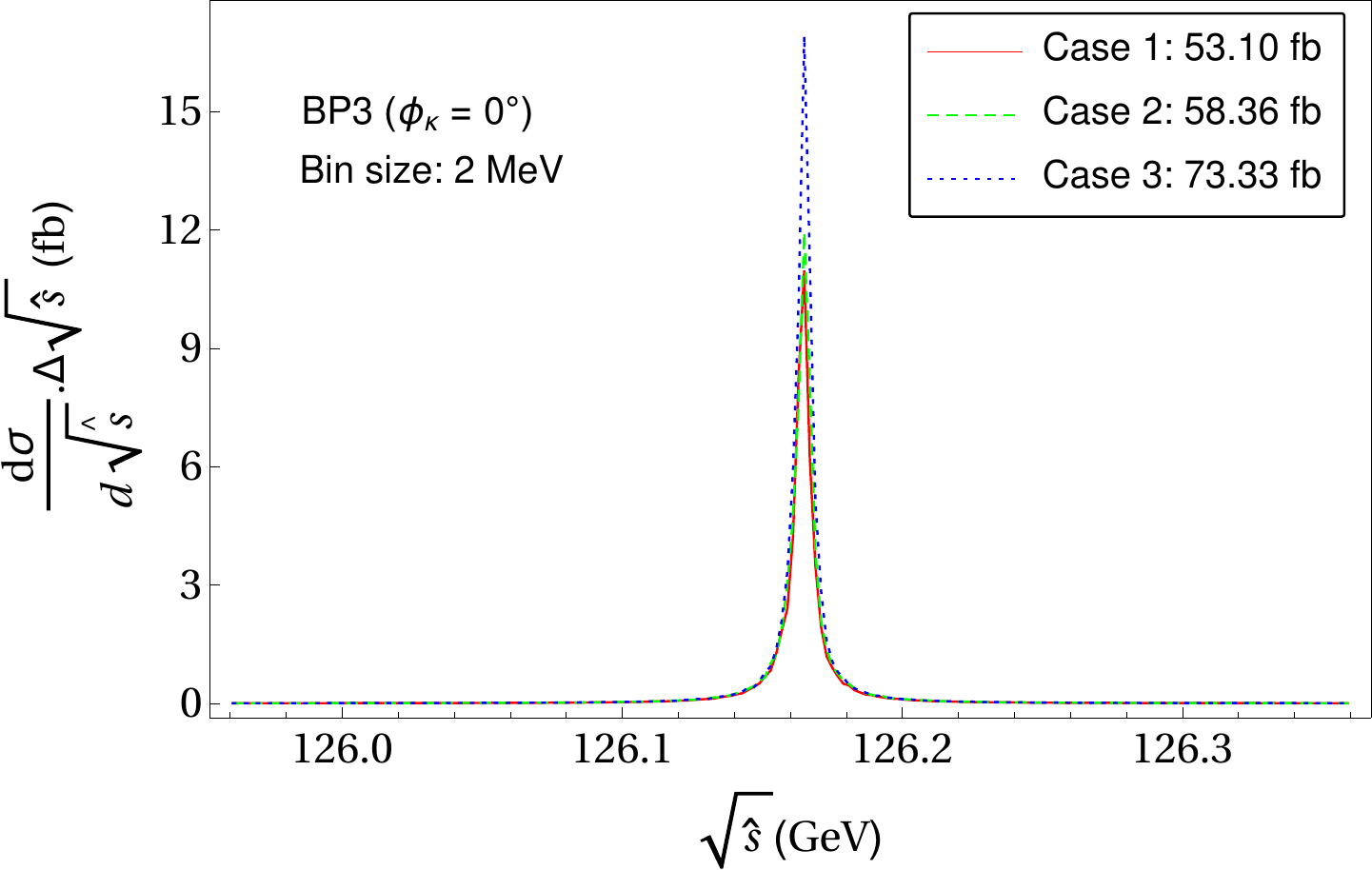}
\caption{\label{fig:BPs-real} Distribution of the differential cross section as a function of the di-photon invariant mass (assumed equal to $\sqrt{\hat{s}}$) for the three benchmark points in the rNMSSM. The red, green and blue curves correspond to the Cases 1, 2 and 3, respectively, discussed in the text.}
\end{figure}

\begin{table}[h]
\centering
\begin{tabular}{c|cccccccccc}
\hline
BP & $\phi_\kappa$ & $M_0$ & $M_{1/2}$ & $A_0$ & tan$\beta$ 
& $\lambda$ & 
$\kappa$ & 
$A_\lambda$ & 
$A_\kappa$ 
& $\mu_{\rm eff}$ \\
\hline\hline
1 & \multirow{3}{*}{$0^\circ$} & 1380.9 & 458.51 & $-2946.2$ & 4.39 & 0.6970 & 0.4594 
& 423.23 & $-5.271$ & 113.60\\ 
2 & & 1598.3 & 471.51 & $-2875.0$ & 4.34 & 0.6907 & 0.4823 
& 402.53 & $-17.117$ & 110.86\\ 
3 & & 1498.2 & 379.87 & $-2822.4$ 
& 3.91 & 0.6969 & 0.4538 & 385.05 & $-16.566$ & 117.92\\ 
\hline
4 & \multirow{4}{*}{$3^\circ$} & 1366.6 & 426.35 & $-2694.3$ & 3.92 & 0.6878 & 0.4657 
& 361.11 & $-13.780$ & 112.79\\ 
5 & & 1476.6 & 363.81 & $-2969.1$ & 4.67 & 0.6725 & 0.4304 & 485.87 & $-35.335$ & 120.41\\ 
6 & & 1427.1 & 249.93 & $-2918.1$ & 4.53 & 0.6852 & 0.3360 & 610.69 & $-26.038$ & 147.10\\
7 & & 1350.2 & 23.24 & $-2727.6$ & 4.50 & 0.6630 & 0.3053 & 618.04 & $-13.900$ & 148.83\\ 
\hline
8 & \multirow{5}{*}{$10^\circ$} & 1270.6 & 176.67 & $-2218.0$ & 3.96 & 0.6781 & 0.4501 
& 538.70 & $-263.98$ & 168.65\\
9 & & 1491.9 & 167.11 & $-2728.0$ & 5.22 & 0.6920 & 0.4599 
& 797.56 & $-291.36$ & 175.84\\
10 & & 1378.0 & 173.35 & $-2291.7$ & 3.99 & 0.6877 & 0.4483 
& 564.66 & $-266.73$ & 172.87\\  
11 & & 1416.6 & 170.40 & $-2741.2$ 
& 4.45 & 0.6684 & 0.3853 & 687.11 & $-221.00$ & 177.72\\
12 & & 1429.0 & 168.46 & $-2821.6$ & 4.71 & 0.6562 & 0.4303 
& 689.40 & $-276.65$ & 173.02\\
\hline
\end{tabular}
\caption{Values of the input parameters for all the selected BPs. All dimensionful parameters are in units of GeV.}
\def\baselinestretch{1.0}
\label{table:BPinput}
\end{table}

The enhancement in the interference effects for a larger difference between $\Delta m$ and $\Gamma_{H_1}\sim \Gamma_{H_2}$ is further confirmed by the distributions shown in the top panels of Fig.~\ref{fig:BPs-pk03} for BP4 and BP5, in the cNMSSM with $\phi_\kappa=3^\circ$. For both these BPs, the interference is again constructive, as in the rNMSSM. It is, however, also possible for the interference to be destructive in this scenario. This is the case for BP6, shown in the bottom left panel of the figure, for which the cumulative cross section is smaller for Case 2 compared to Case 1. Turning on the mixing terms in the propagator matrix further contributes negatively to lower the cross section, although the overall effect is hardly per cent level for this particular parameter space point.

\begin{figure}[t!]
\includegraphics[width=82mm]{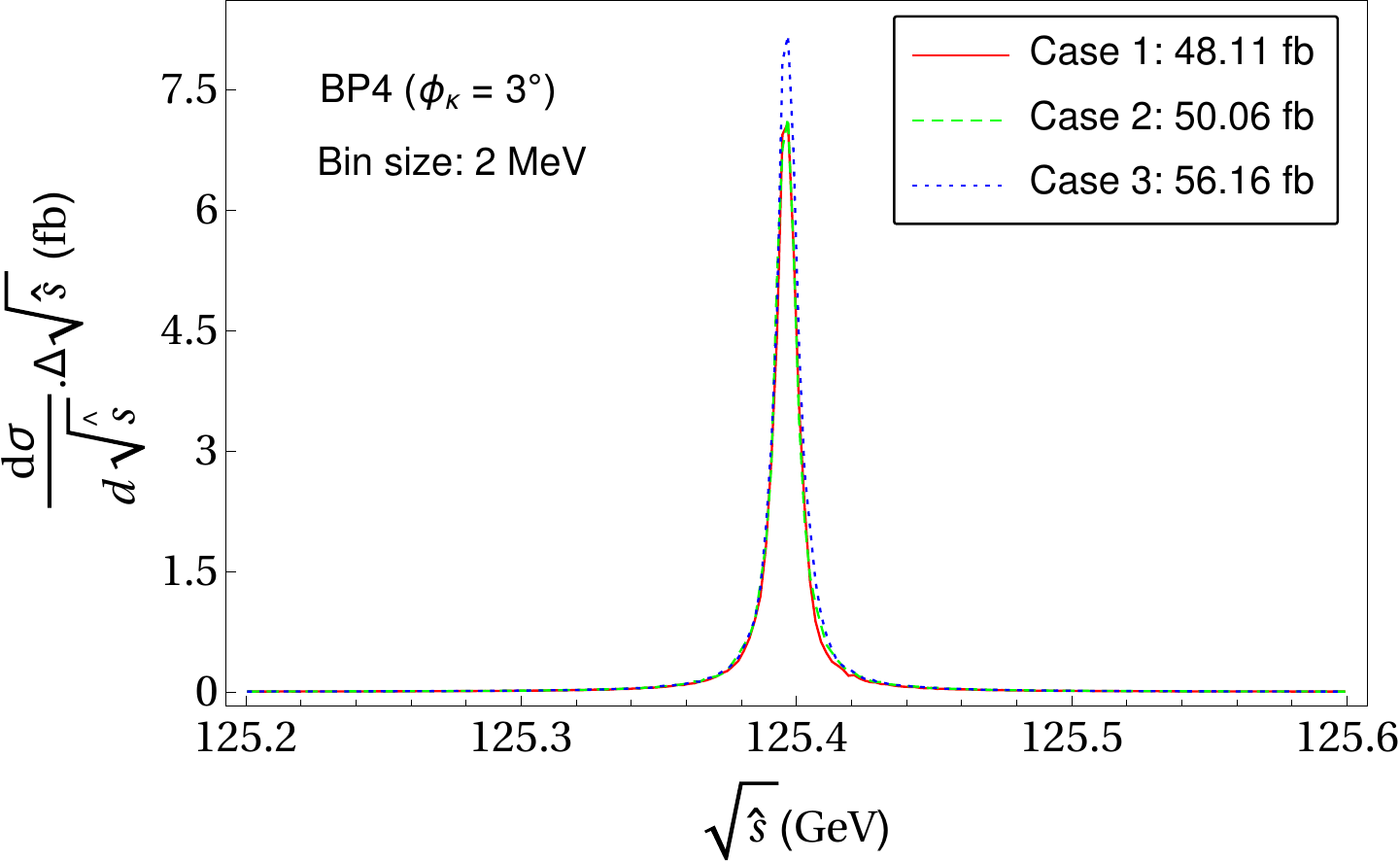}\hspace*{0.2cm}\includegraphics[width=78mm]{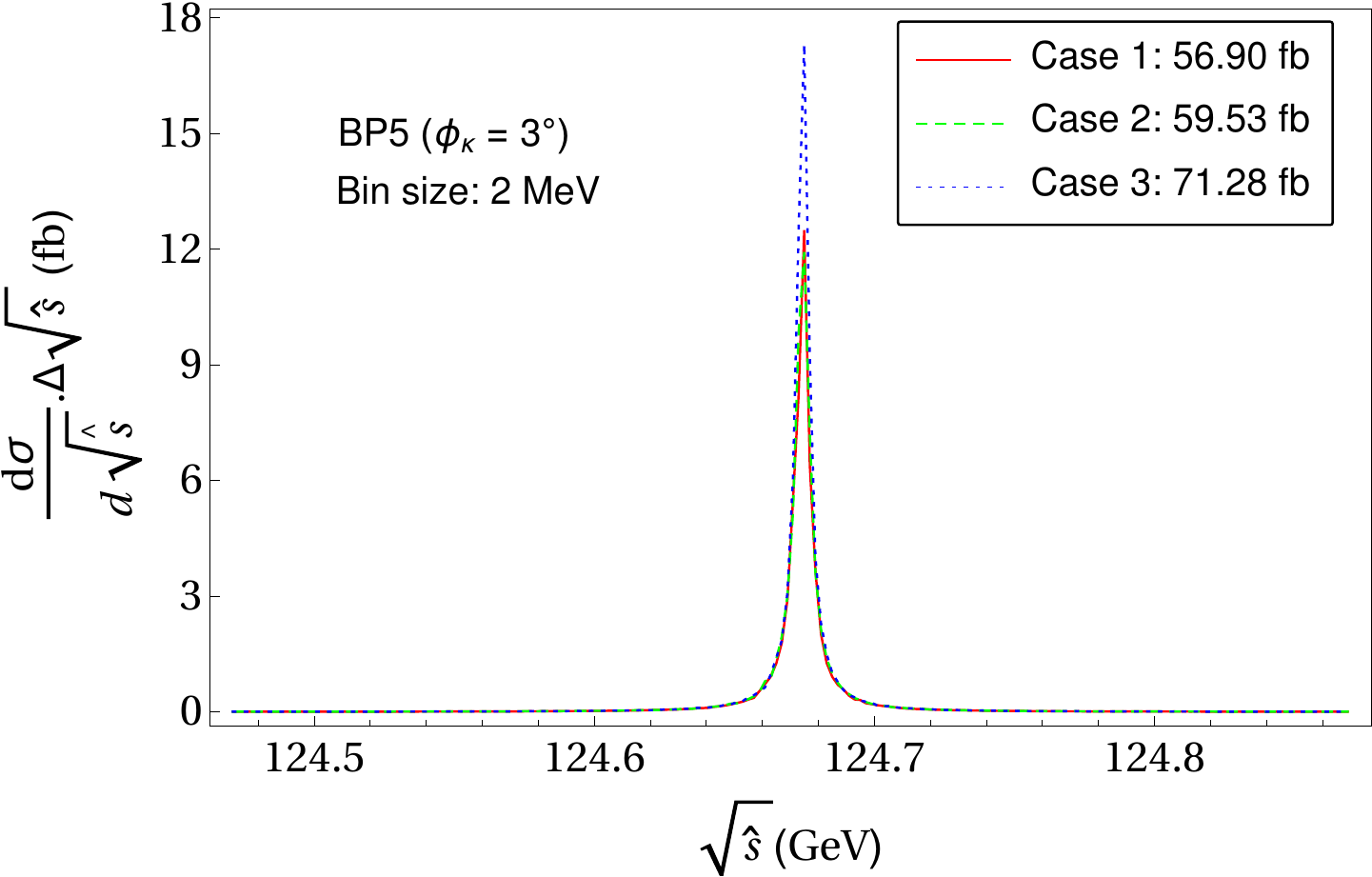}\\
\includegraphics[width=80mm]{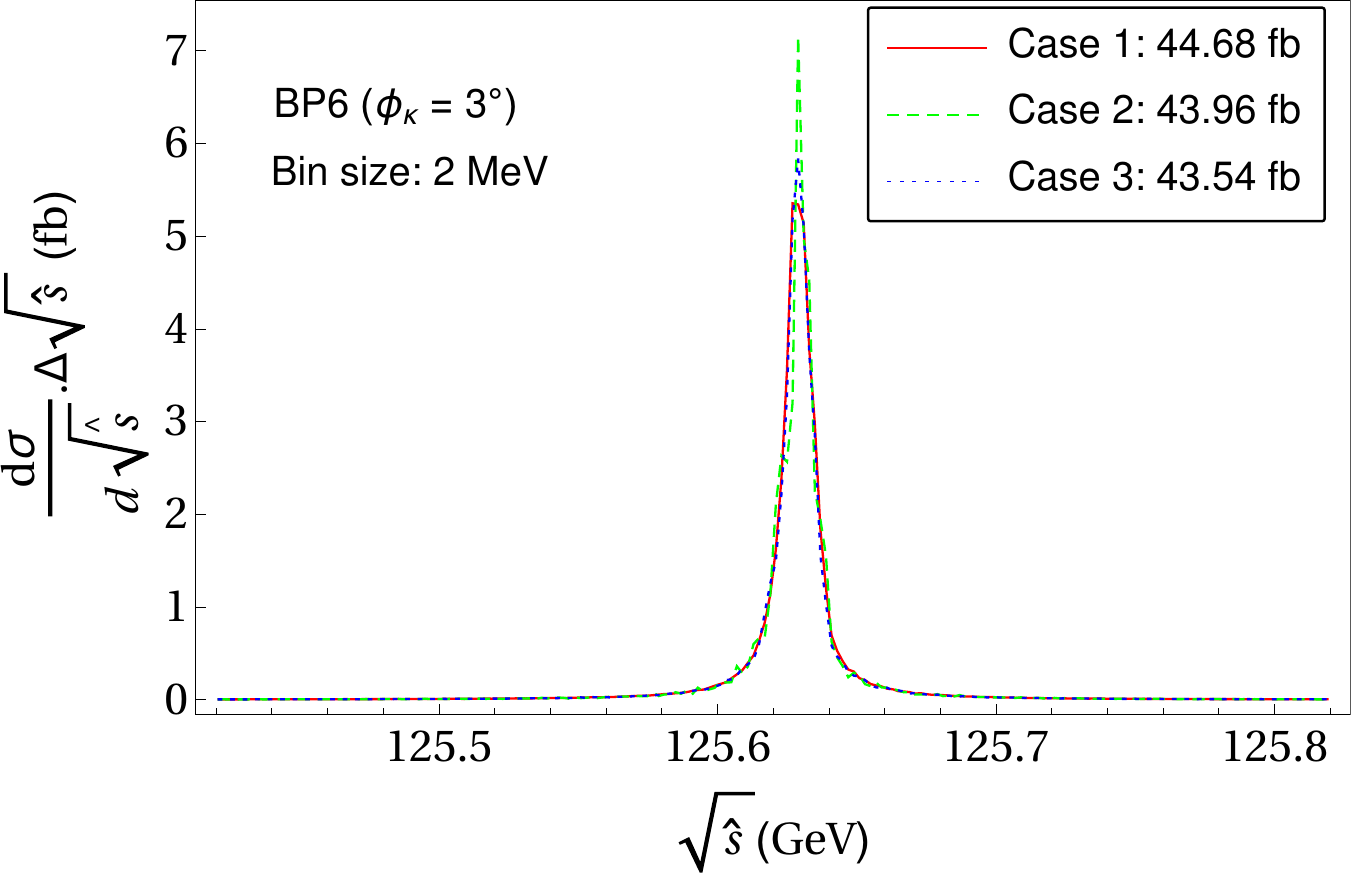}\includegraphics[width=85mm]{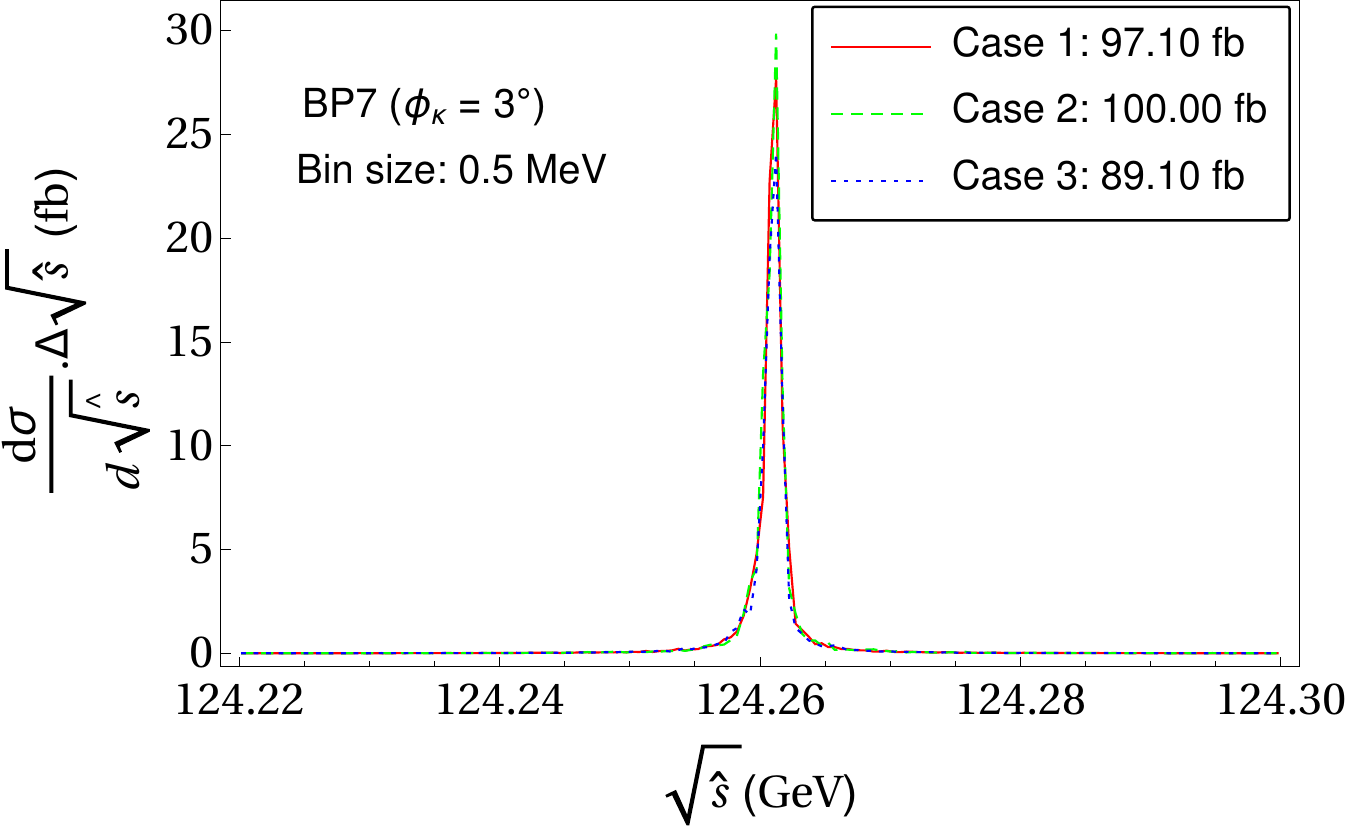}\\
\caption{\label{fig:BPs-pk03} As in Fig.~\ref{fig:BPs-real}, for the BPs corresponding to the cNMSSM with $\phi_\kappa = 3^\circ$.}
\end{figure}
\begin{table}[t]
\hspace*{-.8cm}
\centering
\begin{tabular}{c|cccccccc}
\hline
BP & $m_{H_1}$ & $m_{H_2}$ & $\Delta m_H$ 
& $\Gamma_{H_1}$ & $\Gamma_{H_2}$ 
& \multicolumn{3}{c}{$\sigma_{pp}^{\gamma \gamma}$ (fb)} \\
& (GeV) &  (GeV)&  (MeV) &  (MeV) &  (MeV) & Case 1 & Case 2 & Case 3 \\
\hline\hline
1 & 125.3688 & 125.3782 & 9.4 & 10.7 & 9.7 &
50.36 & 52.41 & 59.78\\
2 & 124.9498 & 124.9562 & 6.4 & 10.1 & 9.1 &
53.58 & 57.54 & 69.23\\
3 & 126.1641 & 126.1667 & 2.6 & 10.1 & 9.3 & 
53.10 & 58.36 & 73.33\\
\hline
4 & 125.3960 & 125.4052 & 9.2 & 9.6 & 9.5 & 
48.11 & 50.06 & 56.16\\
5 & 124.6742 & 124.6757 & 1.5 & 9.1 & 8.4 & 
56.90 & 59.53 & 71.28\\
6 & 125.6285 & 125.6393 & 10.8 & 11.1 & 5.9 & 
44.68 & 43.96 & 43.54 \\
7 & 124.2607 & 124.2625 & 1.8 & 2.5 & 2.3 & 
97.10 & 100.00 & 89.10\\
\hline
8 & 124.9873 & 124.9968 & 9.5 & 10.3 & 3.0 & 
46.94 & 48.38 & 48.89\\
9 & 124.9669 & 124.9742 & 7.3 & 10.6 & 3.0 & 
45.22 & 46.54 & 47.31\\
10 & 125.1874 & 125.1924 & 5.0 & 10.3 & 2.9 & 
46.56 & 49.55 & 50.62\\
11 & 125.1826 & 125.1828 & 2.0 & 10.1 & 2.6 & 
50.14 & 51.42 & 52.22\\
12 & 124.7542 & 124.7604 & 6.2 & 10.3 & 2.7 & 
47.96 & 47.12 & 49.03\\
\hline
\end{tabular}
\caption{The masses and total widths of $H_1$ and $H_2$ in the selected BPs. Also listed for each BP is the cross section for the $pp\to H \to \gamma\gamma$ process calculated in the three different ways considered.}
\label{tab:XSections}
\end{table}

\begin{figure}[t!]
\includegraphics[width=83mm]{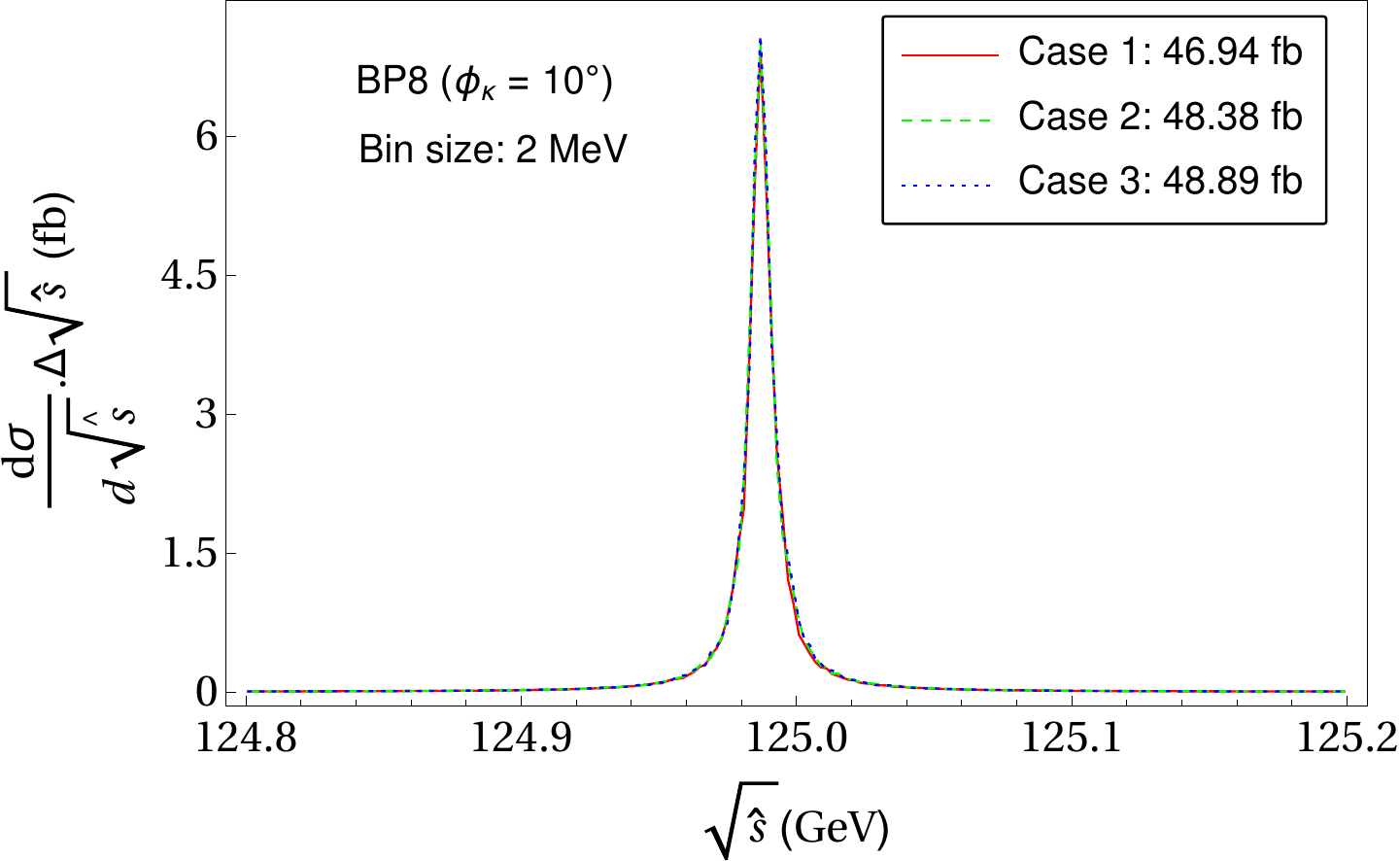}
\hspace*{-0.2cm}\includegraphics[width=81mm]{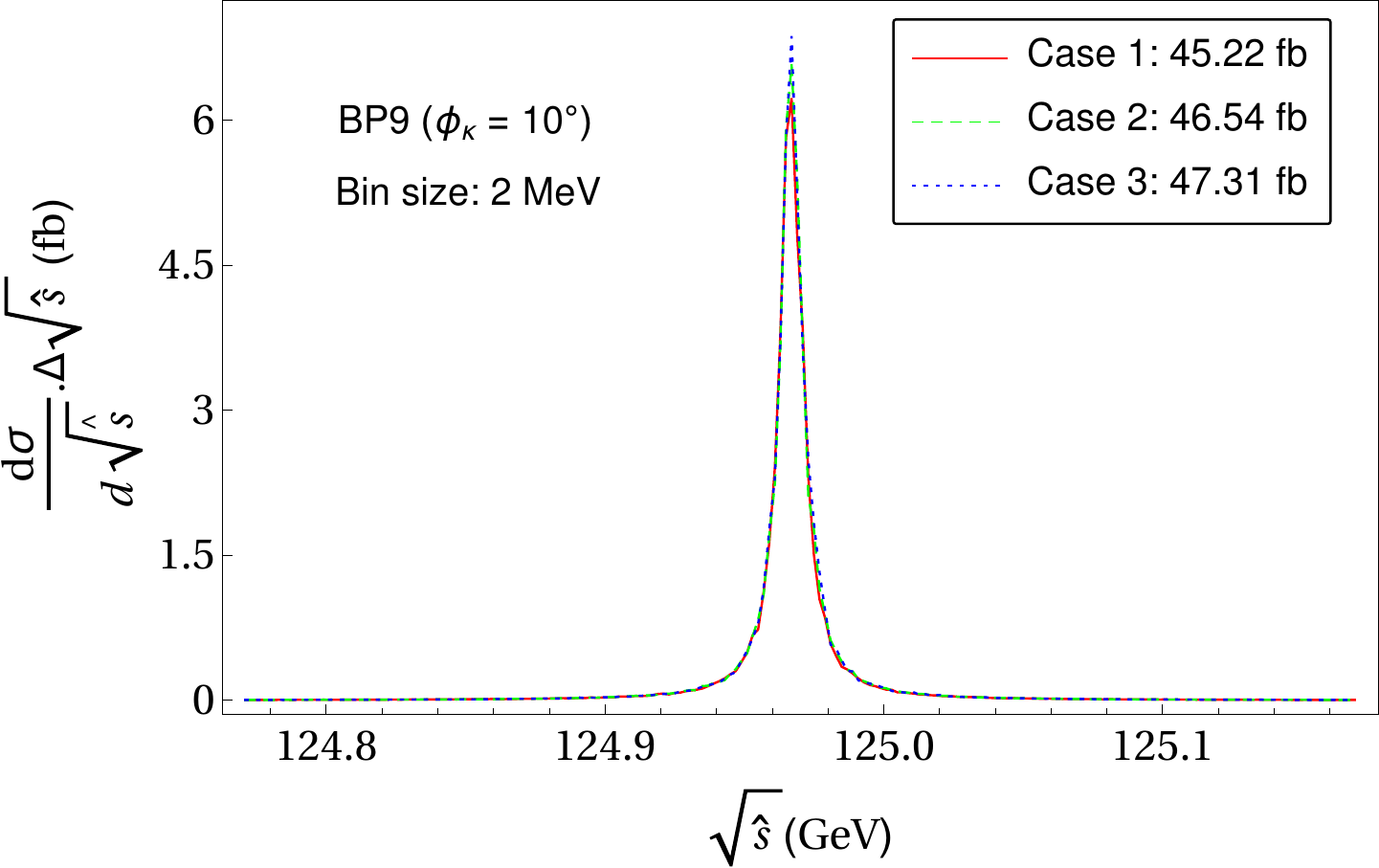}\\
\includegraphics[width=81mm]{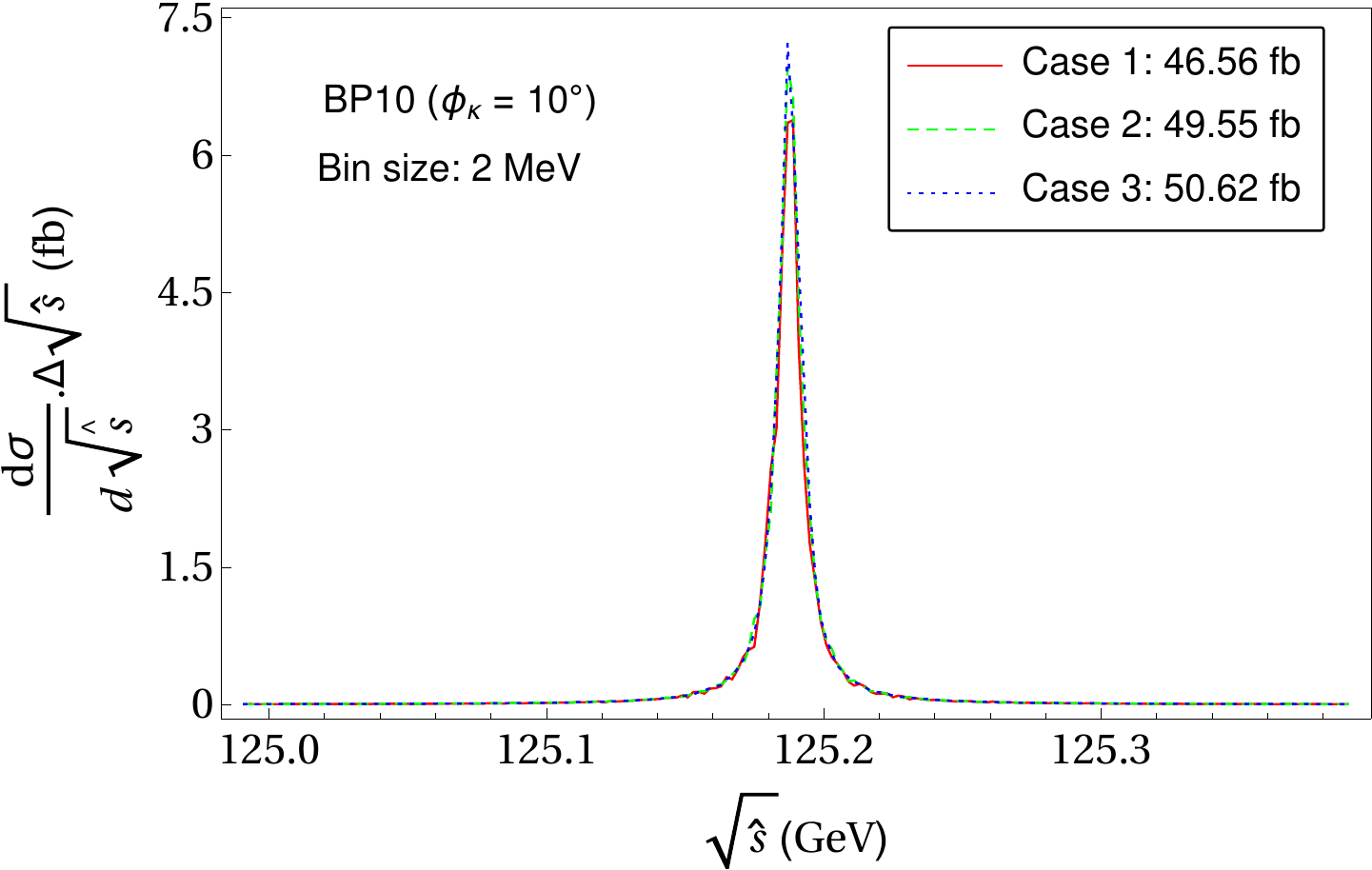}
\includegraphics[width=81mm]{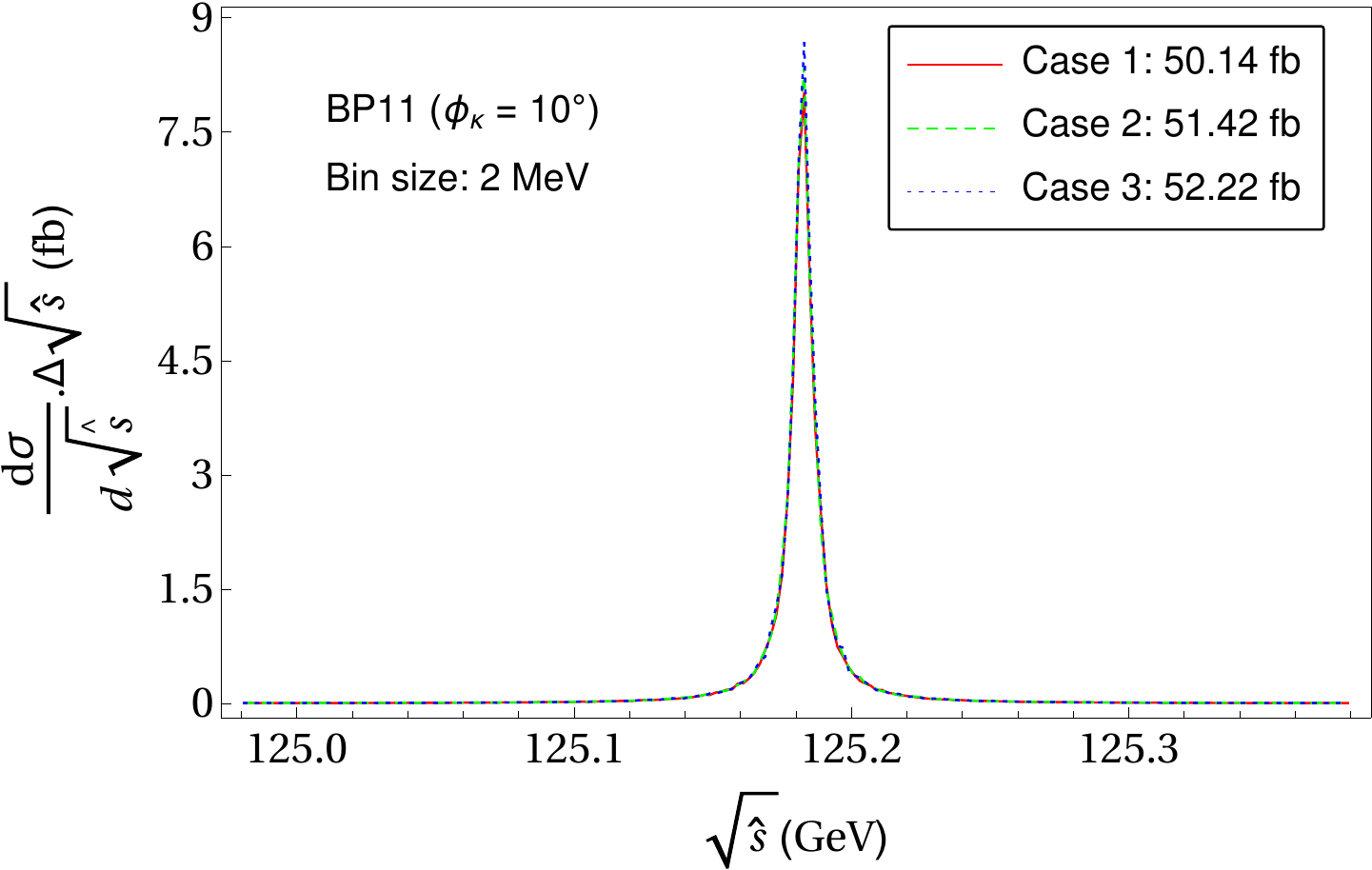}\\
\centering\includegraphics[width=82mm]{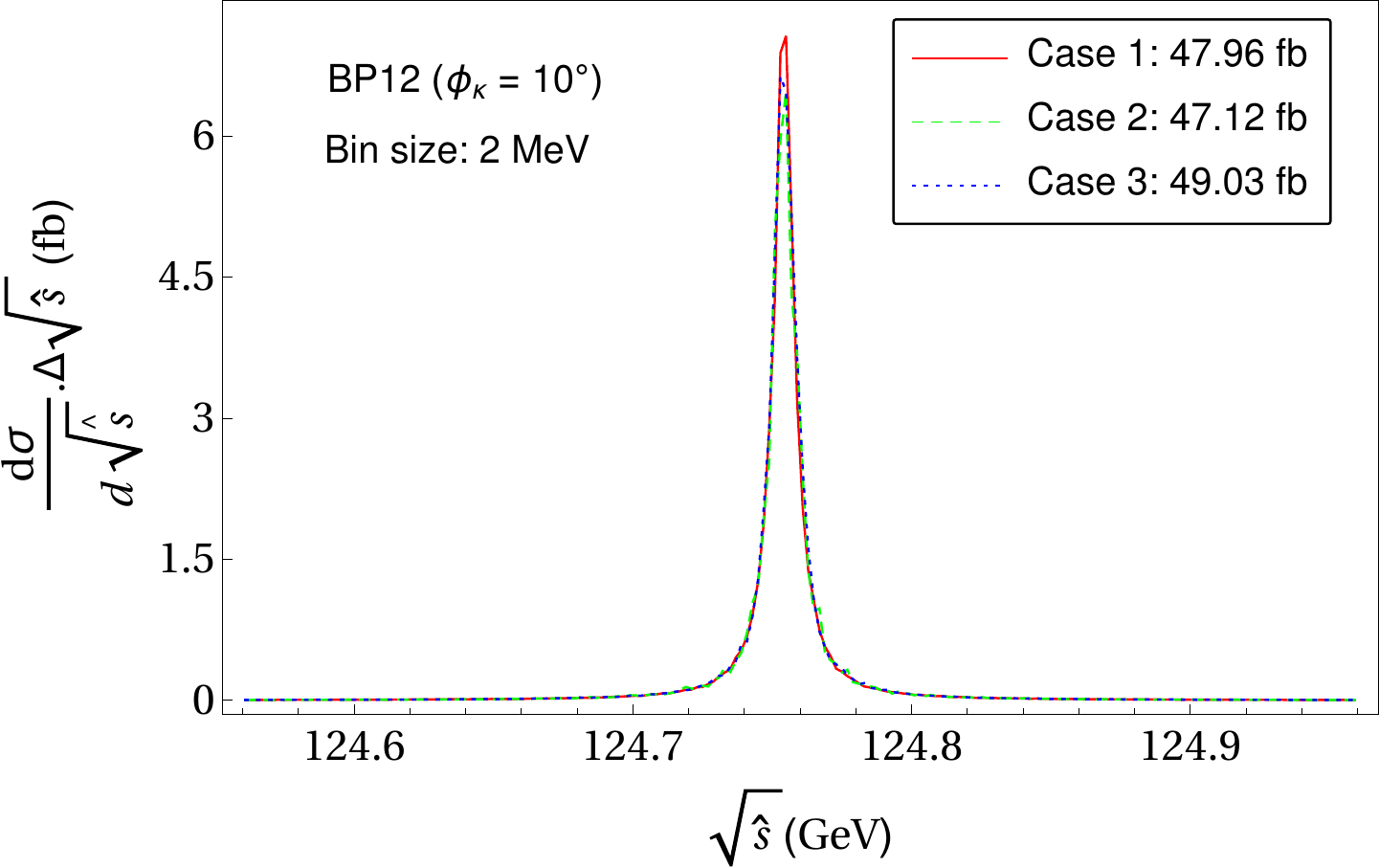}
\caption{\label{fig:BPs-pk10} As in Fig.~\ref{fig:BPs-real}, for the BPs corresponding to the cNMSSM with $\phi_\kappa = 10^\circ$.}
\end{figure}

For BPs 4--6 above, the $H_1$ and $H_2$ are scalar-like, which is the case for almost all the points obtained in the scan for this scenario. We note here that the singlet-like Higgs boson near 125\,GeV is classified as scalar (pseudoscalar)-like if its coupling to the gauge bosons are significantly larger (smaller) than those of the singlet-like $H_3$, which itself also lies fairly close in mass.\footnote{Evidently, both $H_1$ and $H_2$ cannot be singlet-like, since in that case both of them will have highly reduced couplings to the SM particles and resultantly the di-photon production cross section will be extremely suppressed. Moreover, $H_3$ in such a scenario ought to have SM-like couplings and would therefore be ruled out by the collider limits tested against using HiggsBounds.} While one of the main reasons for considering the cNMSSM was to explore the possibility of interference of a $\sim 125$\,GeV pseudoscalar-like Higgs boson with the SM-like one, only a handful of such points were found by our scan, wherein the widths of the $H_1$ and $H_2$ are always relatively small. 

BP7 in Fig.~\ref{fig:BPs-pk03} illustrates the scenario with a pseudoscalar-like $H_2$, with its total width, as well as that of the SM-like $H_1$, being smaller than 3\,MeV. An intriguing feature of this point is that, while the overall interference effect is positive for Case 2, enhancing the cross section compared to  Case 1, it becomes negative when the complete propagator matrix is included. The destructive interference here is much stronger than in  BP6 above and has the desirable consequence of bringing the total cross section down to a level consistent with the LHC measurements noted earlier. The reason why the cross section for  Case 1 for this point is considerably higher than that seen for the other BPs so far can be ascribed to the fact that the $H_1$ is much more SM-like here compared to the points where the scalar-like $H_2$ gets closer in mass to it owing to large singlet-doublet mixing.

In Fig.~\ref{fig:BPs-pk10} we display the distributions for the five BPs selected in the cNMSSM with $\phi_\kappa=10^\circ$. As noted from the corresponding scatter plot above, this value of $\phi_\kappa$ allows for a much larger splitting between $\Gamma_{H_1}$ and $\Gamma_{H_2}$, compared particularly to the rNMSSM. This makes it possible to test the impact of interference when  $\Delta m$, instead of being smaller than both the widths, lies somewhere in between them. Thus, for each of these points $\Gamma_{H_{1}}<3$\,MeV and $\Gamma_{H_{2}}>10$\,MeV, with $\Delta m$ varying from about 9.5\,MeV for BP8 to 2\,MeV for BP11. While the behaviour of the interference is similar to what has been observed earlier, i.e., it grows larger as the gap between $\Delta m$ and $\Gamma_{H_2}$ increases, its overall size remains comparatively small, reaching about 9\% for BP10, for which $\Delta m$ is only slightly larger than $\Gamma_{H_1}$. However, when $\Delta m$ is lowered below $\Gamma_{H_1}$ also, as in BP11, the interference effects get reduced instead of continuing to grow. BP12 is another example of mutually opposite contributions to the interference effects from the diagonal and off-diagonal elements of the Higgs propagator matrix. As opposed to BP7 though, here the negative interference comes from the diagonal elements in the propagator matrix, while the mixing effects contribute positively to again raise the cross section slightly for Case 3. 

A closer look at the curves for BP6 and BP10 above reveals very small kinks near $m_{H_1}$ in addition to tall peaks near $m_{H_2}$. These kinks result from the large splitting between the $\Gamma_{H_1}$ and $\Gamma_{H_2}$, coupled with the fact that $\Delta m$, while still being sufficiently smaller than the $\Gamma_{H_2}$ to cause notable interference effects, is larger than the bin size of 2\,MeV adopted for plotting the distribution. Thus, for these points not only does the values of the inclusive (i.e., integrated) cross section change between the respective Cases 1 and 3, but also the shape of the distribution for the exclusive (i.e., differential) one. However, a bin width of 2\,MeV is about three orders of magnitude smaller than the current experimental ${\cal M}_{\gamma\gamma}$ resolution of around 1\,GeV~\cite{Khachatryan:2016vau}. Evidently, any differences between their shapes corresponding to each of the three Cases for a given BP, which could prove crucial for mutually distinguishing them and hence unraveling the interference effects, would not appear had a realistic bin width of 1\,GeV been used. It is nevertheless interesting to study whether these differences persist to some extent once the differential distributions are convolved with a Gaussian distribution emulating detector effects. We performed the convolution using the {\tt ListConvolve} function~\cite{mathematica} in {\tt Mathematica} by supplying the differential distribution data for a point as a list, as well as specifying the mean and width of the Gaussian.

In the top frames of Fig.~\ref{fig:BP10} we display the result of the convolution of the distributions corresponding to Cases 1 and 3 for BP10 with a Gaussian of width (resolution) 1\,GeV, which is also used as the size of the ${\cal M}_{\gamma\gamma}$ bins for first reproducing these distributions with our MC  integrator. We use two prospective integrated luminosities at the LHC: 300\,fb$^{-1}$ (left), which is expected by the end of the machine run in standard configuration; and 1000\,fb$^{-1}$ (right), foreseen for the High-Luminosity (HL) LHC option~\cite{Gianotti:2002xx}.\footnote{The higher luminosity only serves to reduce the sizes of the error bars in the figure, which refer only to the statistical error in fact, as we are not able to account for the systematic one.} It is worth appreciating in the figures that, while the kinks have expectedly disappeared and the distributions are smoother, there exists some marginal scope at the LHC to distinguish at the differential level the simplistic scenario generally assumed (Case 1) and the one rigorously predicted (Case 3), as the difference in the heights of the red and blue curves is not constant for all the bins in $\sqrt{\hat{s}}$. However, evidently this requires knowledge of additional model inputs (e.g., the Higgs boson couplings) as such difference could also be generated by a different point in the parameter space. This difference is even more pronounced in the bottom panels of the figure, which correspond to the convolution with a Gaussian of resolution 300\,MeV (clearly an unrealistic value at present, yet potentially within the reach of a detector upgrade). In fact, the histograms referring to these two predictions could eventually be statistically separable, again though, with the aforementioned caveat that one ought to have established additional model parameters elsewhere.

\begin{figure}[t!]
\includegraphics[width=80mm]{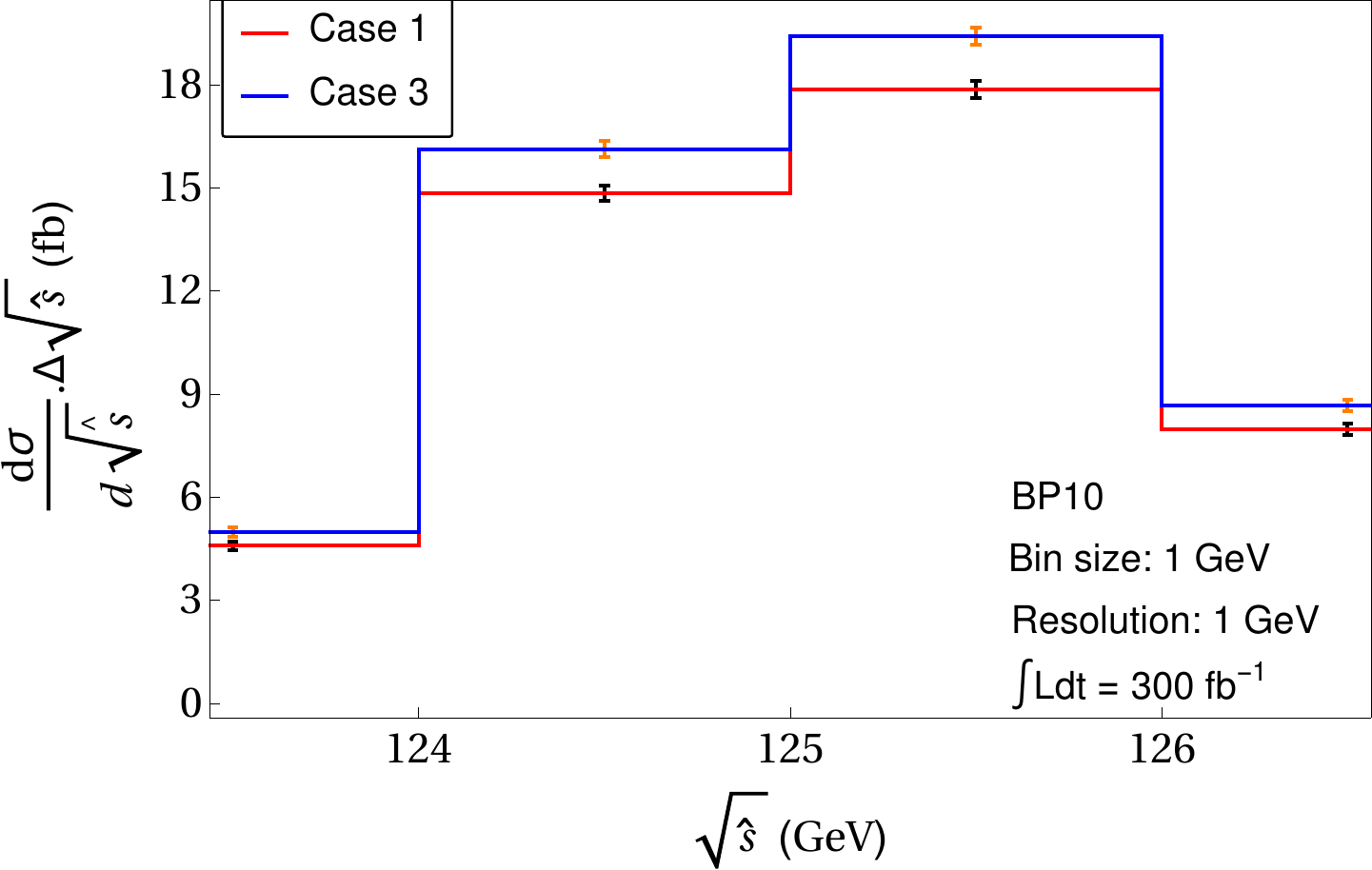}
\includegraphics[width=80mm]{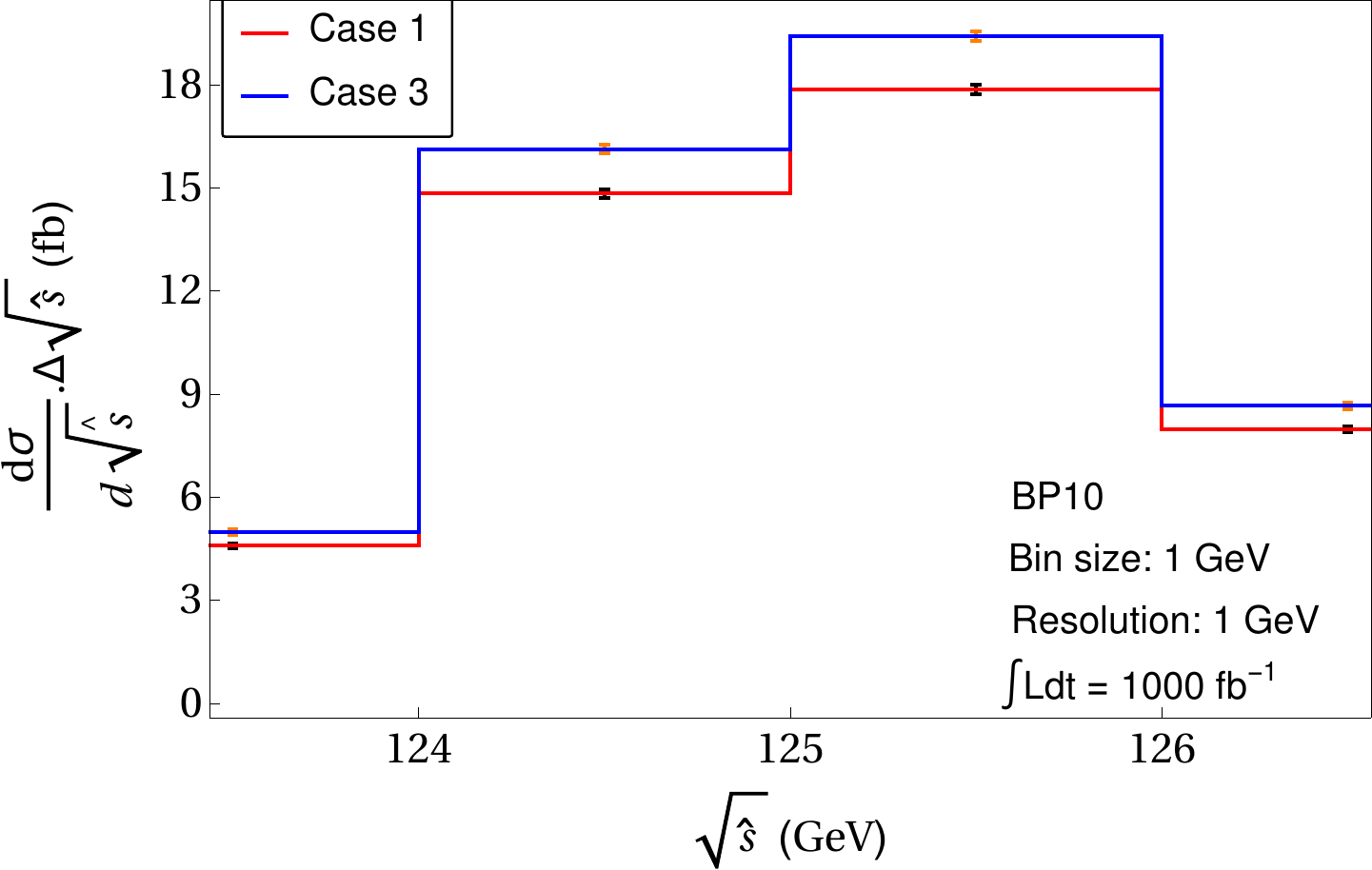}\\
\includegraphics[width=80mm]{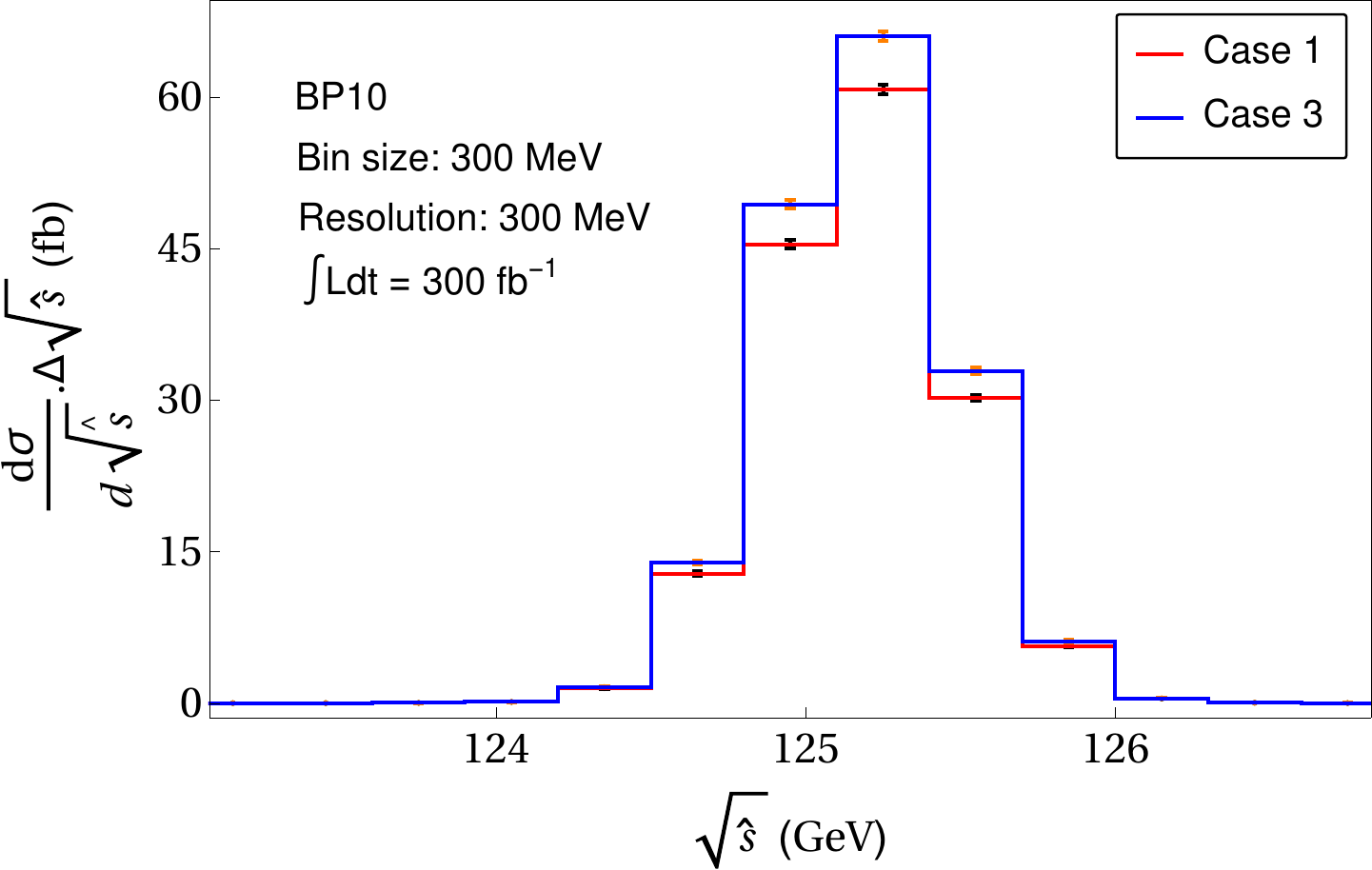}
\includegraphics[width=80mm]{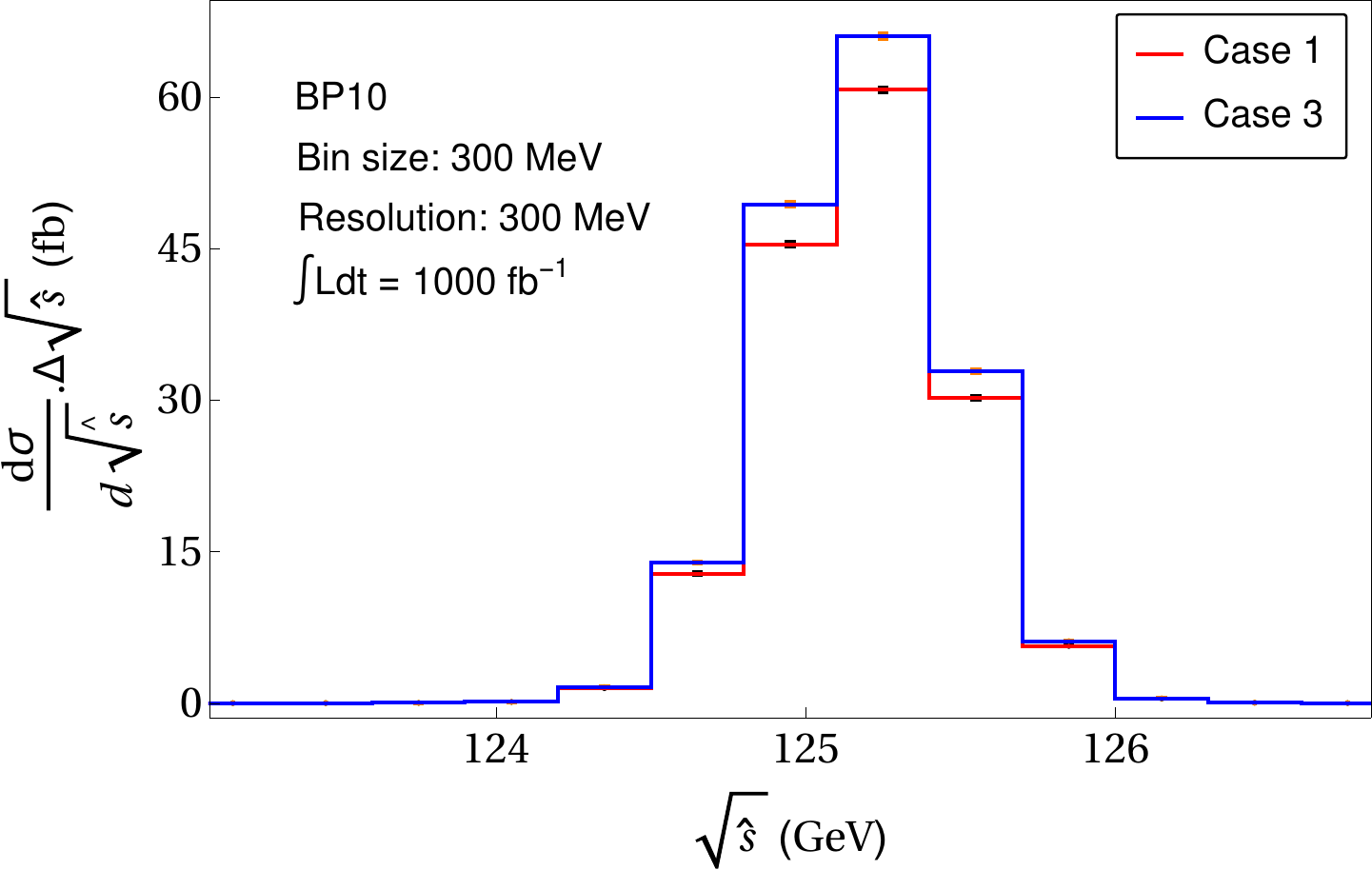}\\
\caption{\label{fig:BP10} Convolution of the distributions
1 and 3 for BP10 with Gaussians of width
1\,GeV (top) and 300\,MeV (bottom). An integrated luminosity of 300\,fb$^{-1}$ is assumed in the left panels and of 1000\,fb$^{-1}$ in the right panels.}
\end{figure}

The difficulty to separate the Cases 1 and 3 for BP10 (as it would be for all other BPs) with present machine and detector conditions at the LHC is ultimately related to the enforcement of the $\Gamma_{H_1/H_2}< 13$\,MeV constraint from off-shell Higgs measurements in our selection of the BPs. We therefore consider a Test Point (TP) 1, where this constraint is dismissed and only the milder $\Gamma_{h_{\rm obs}}< 41$\,MeV constraint, as obtained from a global fit to the on-shell Higgs boson signal strength measurements~\cite{CMS:2016ilx}, is imposed. This is all the more important in light of the fact that some critiques have been drawn about the model-independence of such a measurement, see, e.g.,~\cite{Cacciapaglia:2014rla} and~\cite{Logan:2014ppa}, or its stability against theoretical uncertainties~\cite{Englert:2015zra,*Englert:2015bwa}\footnote{Recall also that the mentioned experimental measurement of the (off-shell) SM-like Higgs width suffers from a small signal yield and large backgrounds.}. The top right frame of Fig.~\ref{fig:TP1} shows the convoluted distributions 1 and 3 for TP1 with a Gaussian of 1\,GeV width, illustrating again the fact that also in this case there exists some scope in separating the Cases 1 and 3, even for current di-photon mass resolutions, so long that sufficient luminosity is accrued. The bottom frames of the figure illustrate that this scope gets further enriched if the mass resolution is improved to 300\,MeV.

\begin{figure}[t!]
\includegraphics[width=82mm]{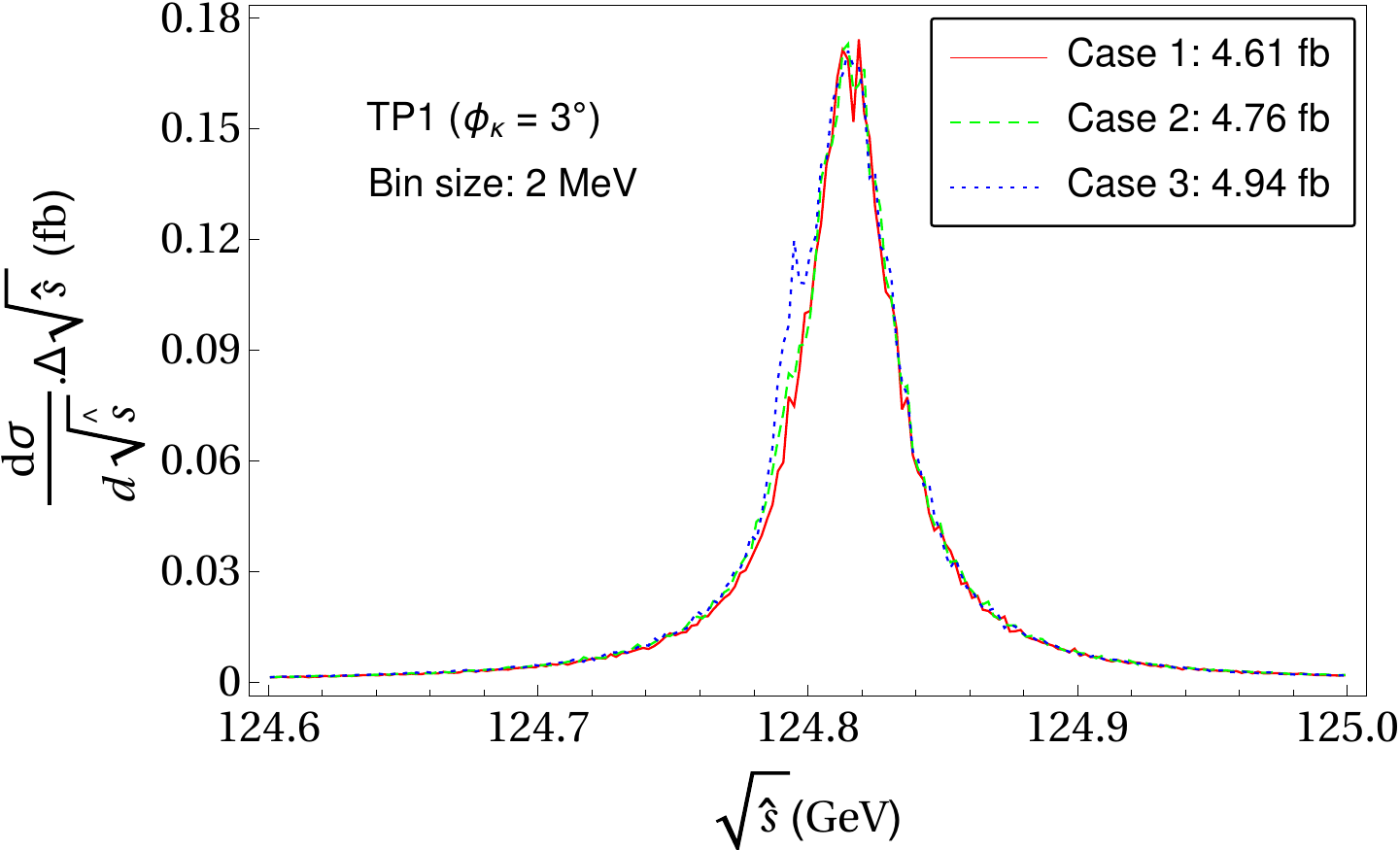}
\includegraphics[width=80mm]{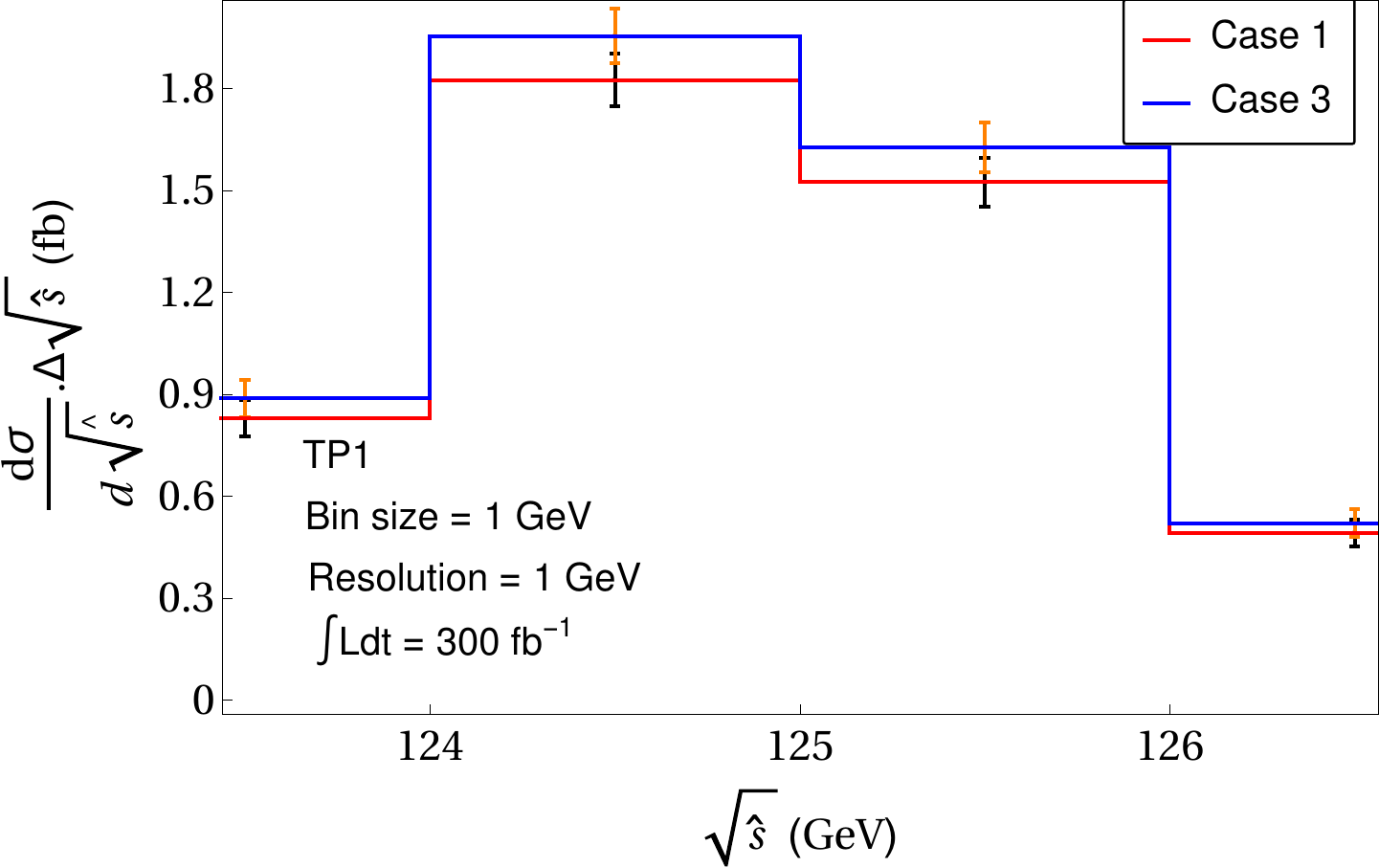}\\
\includegraphics[width=80mm]{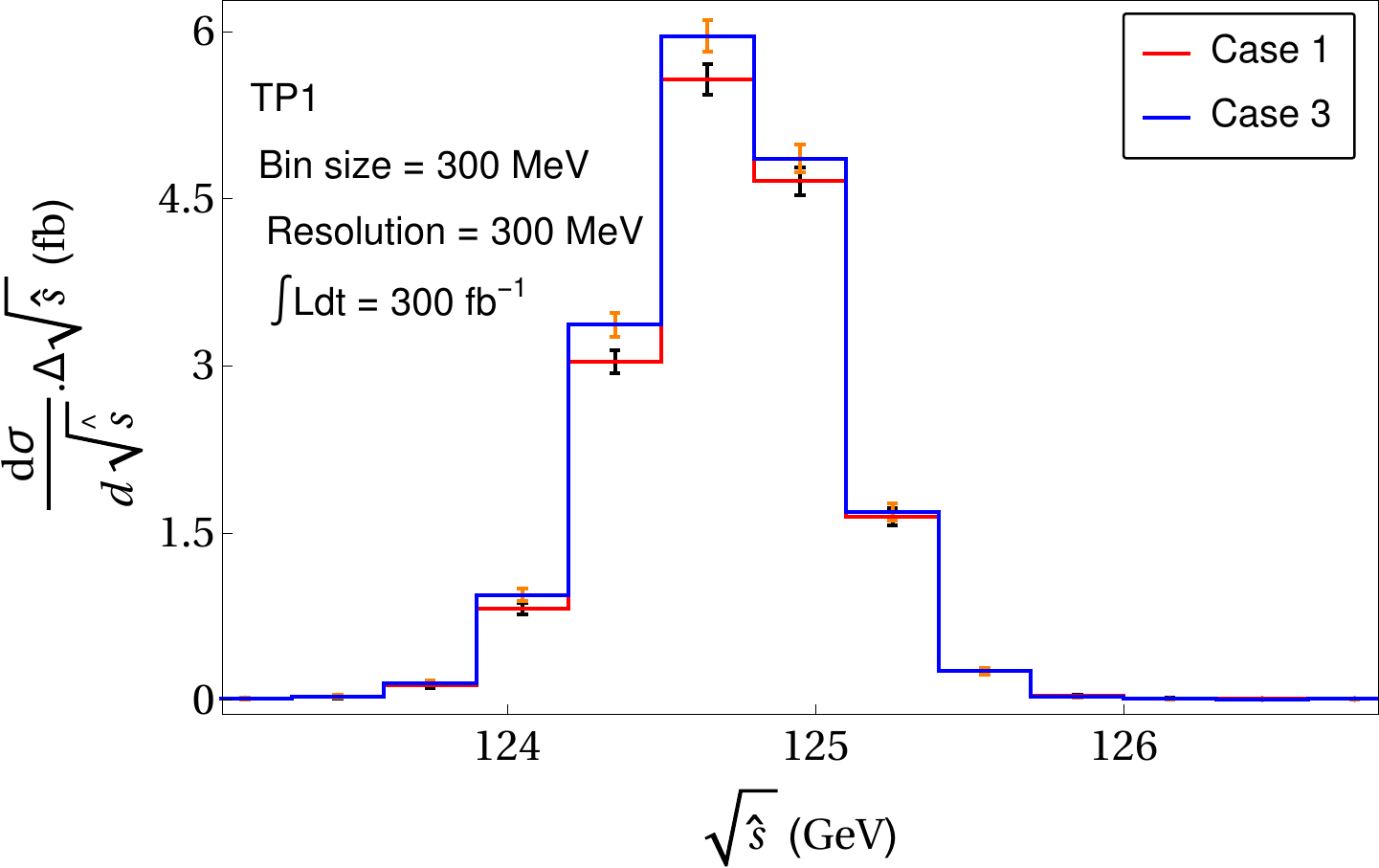}
\hspace*{0.2cm}\includegraphics[width=80mm]{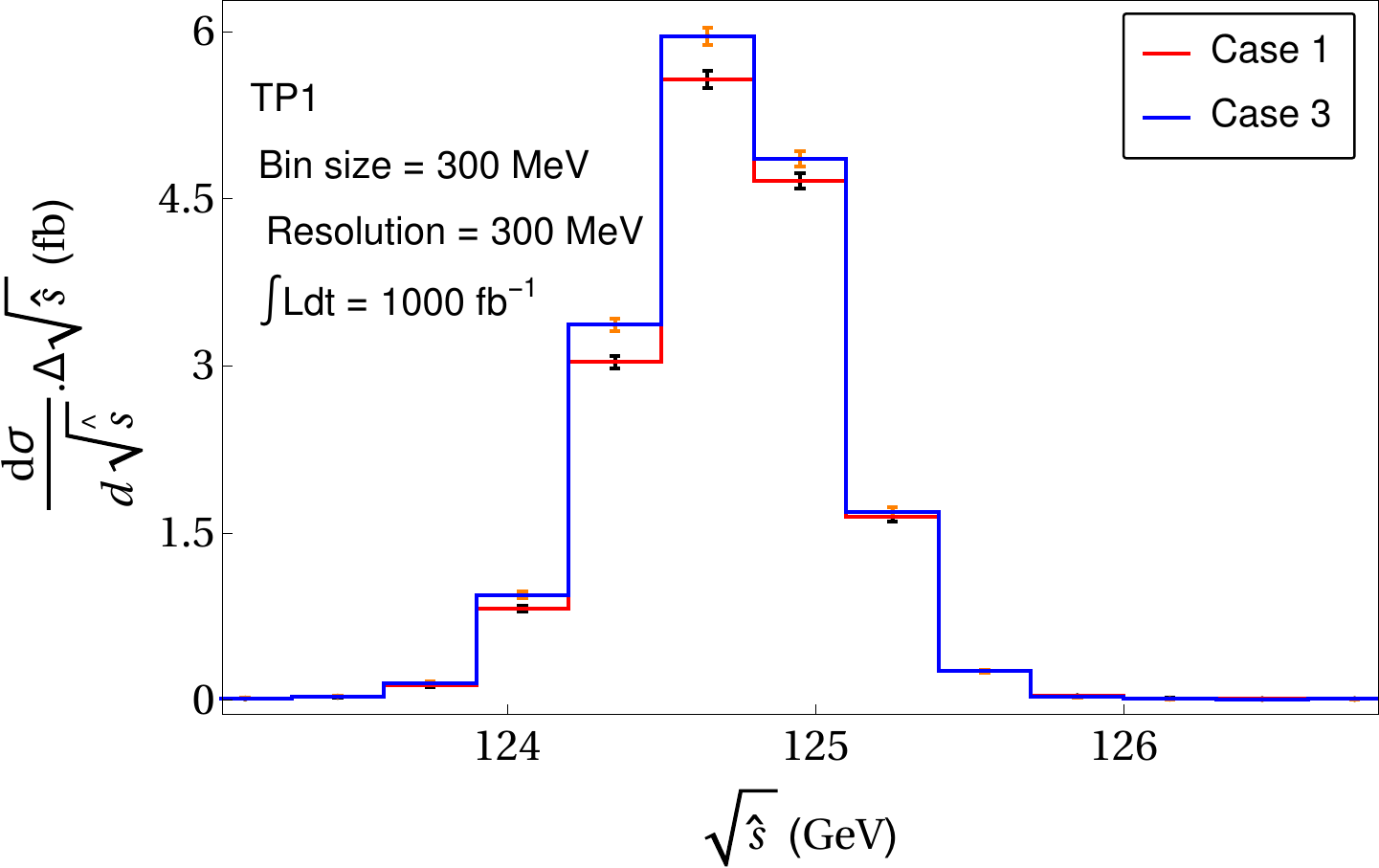}\\
\caption{\label{fig:TP1} Top: The differential distributions for TP1 without convolution (left) and after convolution with a 
Gaussian of width 1\,GeV for an integrated luminosity of $300~{\rm fb^{-1}}$ (right). Bottom: TP1 distributions after convolution with a Gaussian of width 300\,MeV for an integrated luminosity of $300~{\rm fb^{-1}}$ (left) and $1000~{\rm fb^{-1}}$ (right). }
\end{figure}

In fact, one could ignore constraints on the total width altogether in order to estimate the minimal mass splitting that could be potentially detectable. While this exercise may appear academic (i.e., to dismiss a crucial experimental constraint), it is worth noting that the current
procedures adopted to extract the Higgs boson properties inevitably work with the underlying assumption that only one resonance is produced around 125\,GeV. This implies that the allowed intrinsic widths of a pairs of degenerate Higgs states need not relate directly to the currently fitted value. 

In Figs.~\ref{fig:TP2} and \ref{fig:TP3} we thus display the distributions of Cases 1 and 3 for two more TPs, 2 and 3, respectively, convolved, again, with Gaussians of 1\,GeV and 300\,MeV widths for two prospective integrated luminosities. In both these TPs, the $H_1$ is very wide, $\mathcal{O}$(100)\,MeV, while $\Gamma_{H_2}$ is a few 10s of MeVs, as seen in Tab.~\ref{table:crosssection-TP}.\footnote{The input parameters for the three TPs are provided in Tab.~\ref{table:TPinput}.} But since $\Gamma_{H_1} - \Delta m$ is only about 70\,MeV for TP2, while it is larger than half of $\Gamma_{H_1}$ for TP3, the interference effects are highly enhanced for the latter (about 46\%) compared to the former ($\sim 30$\%). {These figures more effectively bring home the point that a very large $\Gamma_{H_1}$ (as noticeable in the top-left frames) does not impact significantly the quality of the fit to what, in the
end, looks like a single object shape (as visible in the other three frames)}. Though, clearly, the difference between the Cases 1 and 3 is much more pronounced here than for TP1 (and all the BPs). This difference may potentially  be established experimentally within the next few years, more likely so the wider (one of) the Higgs states. 
\begin{figure}[tbp]
\includegraphics[width=82mm]{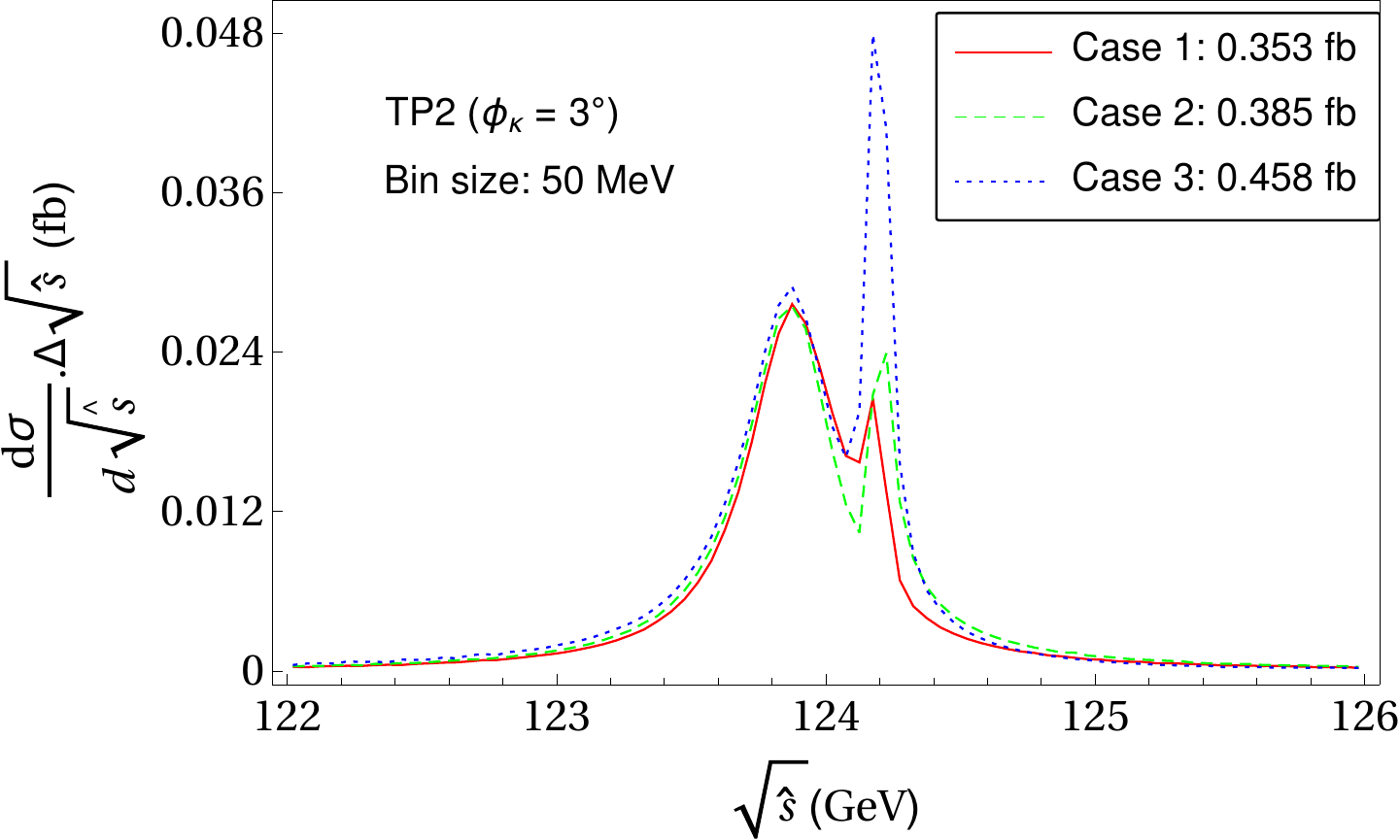}
\includegraphics[width=80mm]{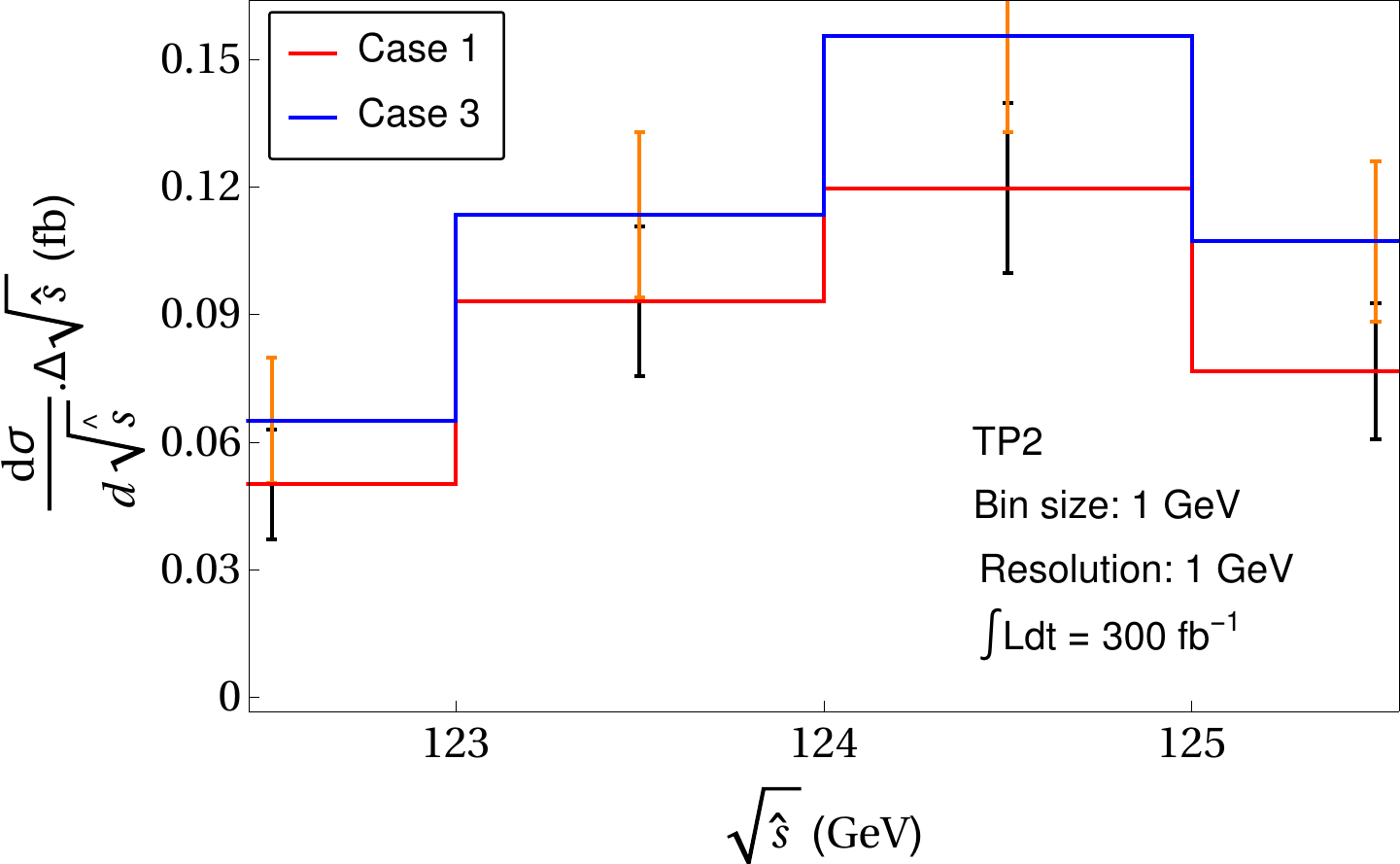}\\
\includegraphics[width=80mm]{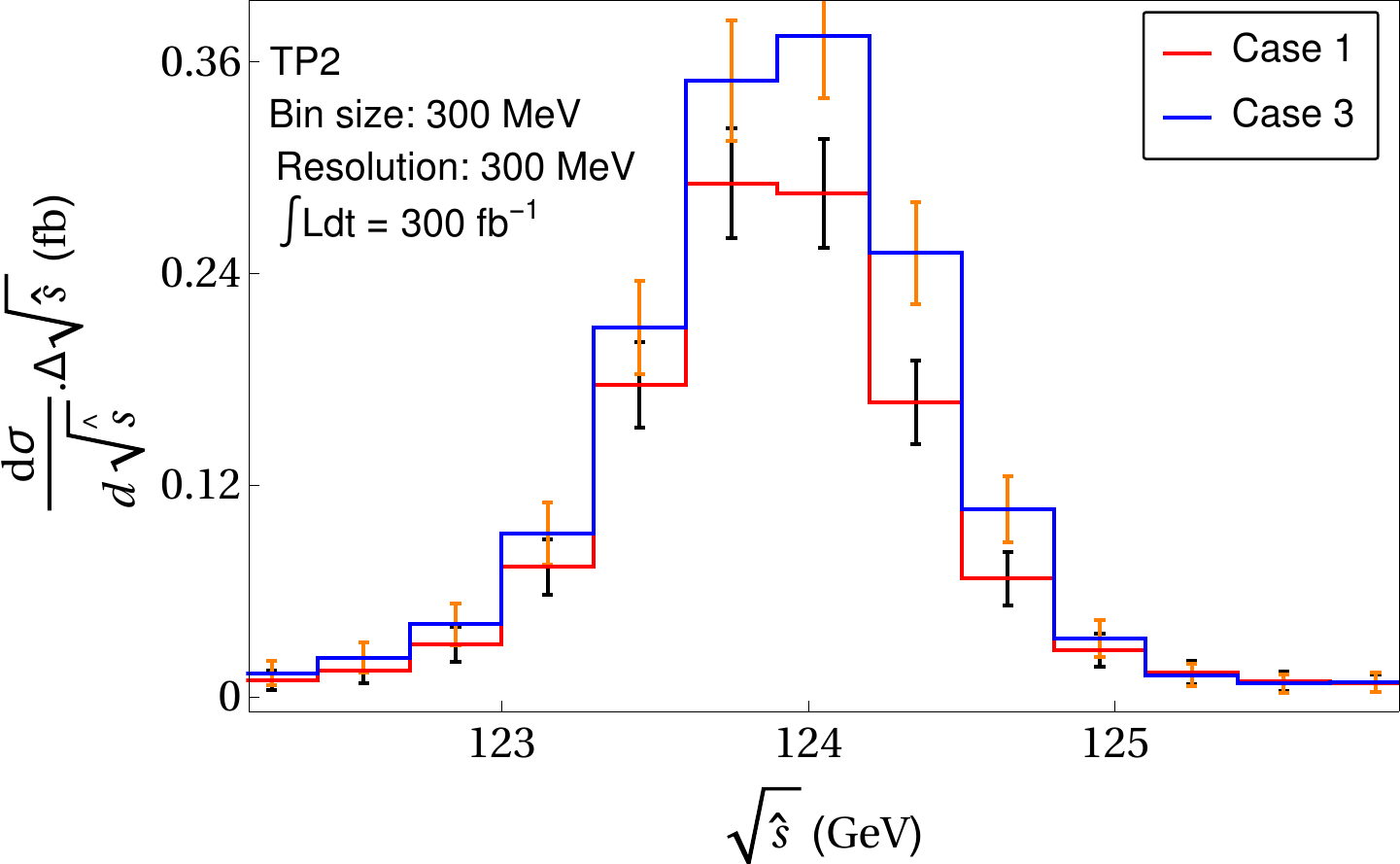}
\hspace*{0.2cm}\includegraphics[width=80mm]{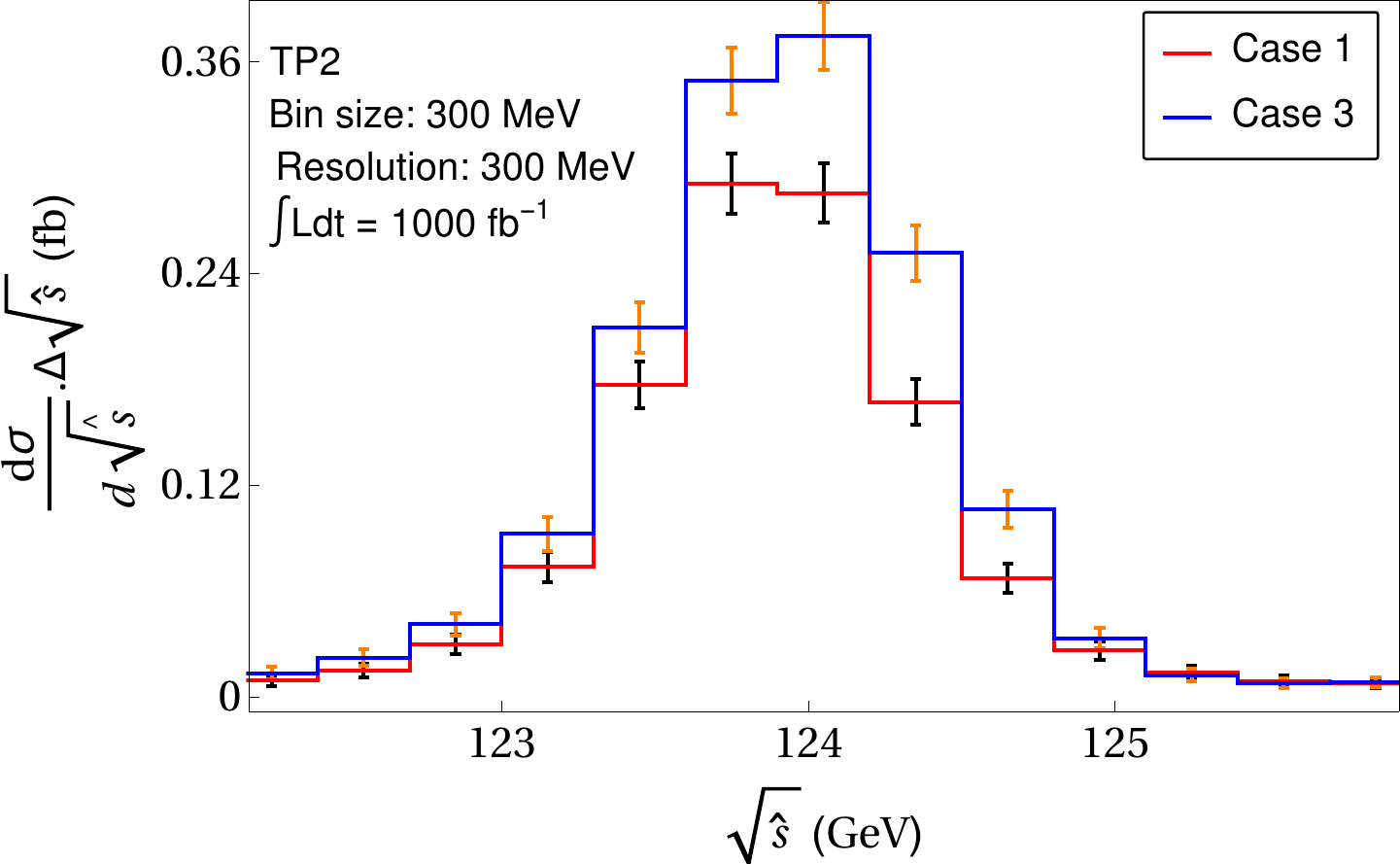}\\
\caption{\label{fig:TP2} As in Fig.~\ref{fig:TP1}, for the TP2.}
\end{figure}

\begin{figure}[tbp]
\includegraphics[width=82mm]{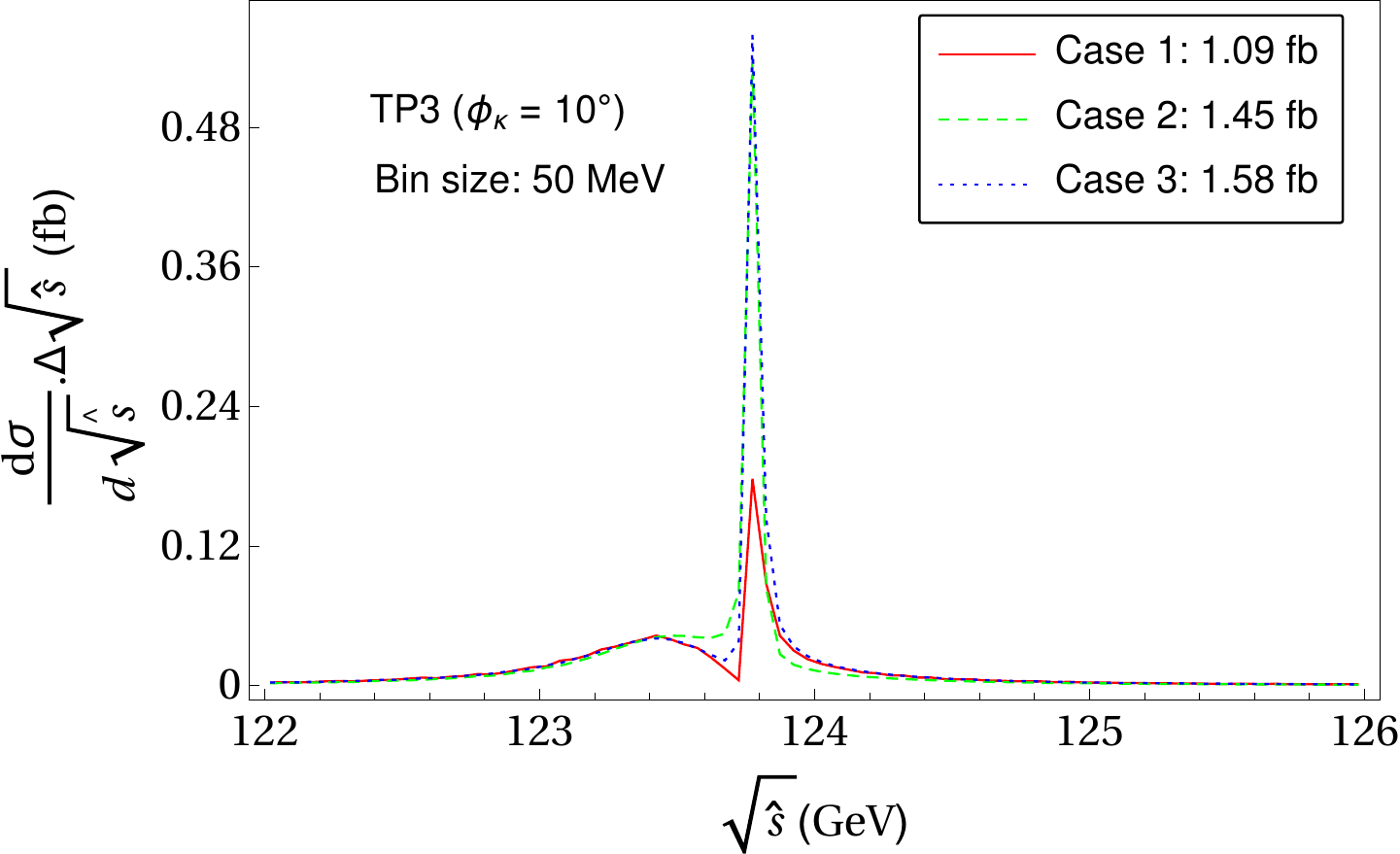}
\includegraphics[width=80mm]{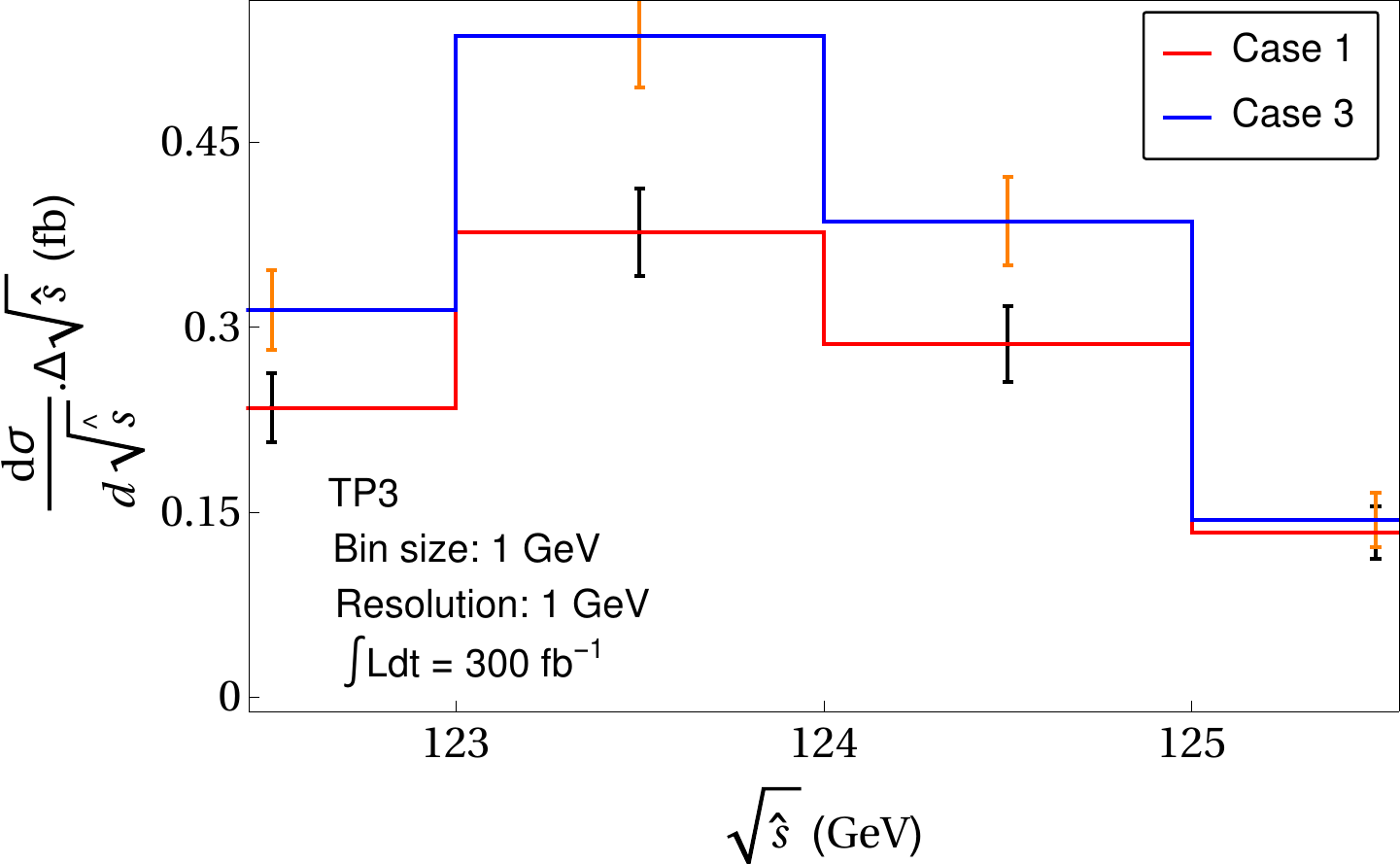}\\
\includegraphics[width=80mm]{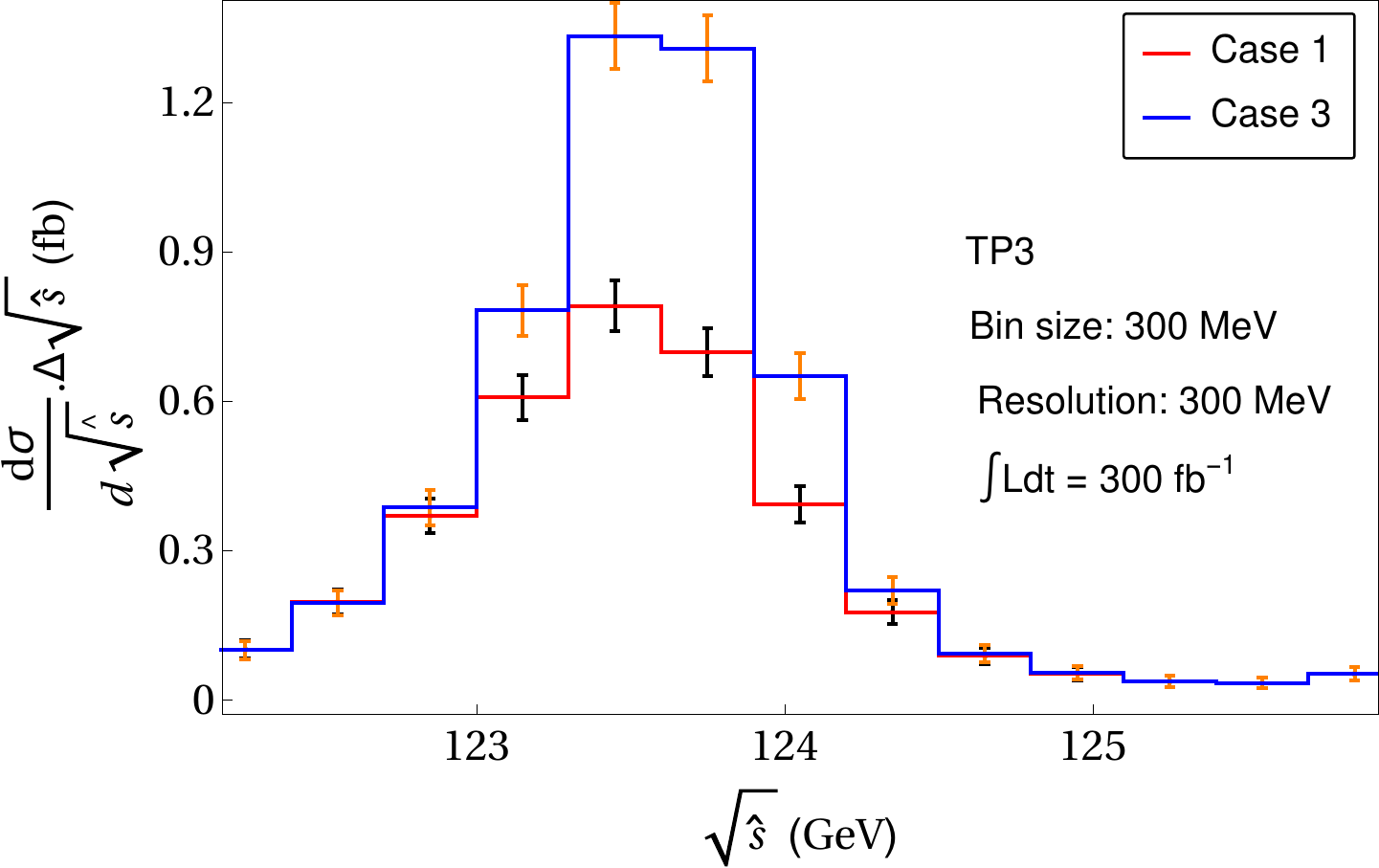}
\hspace*{0.2cm}\includegraphics[width=80mm]{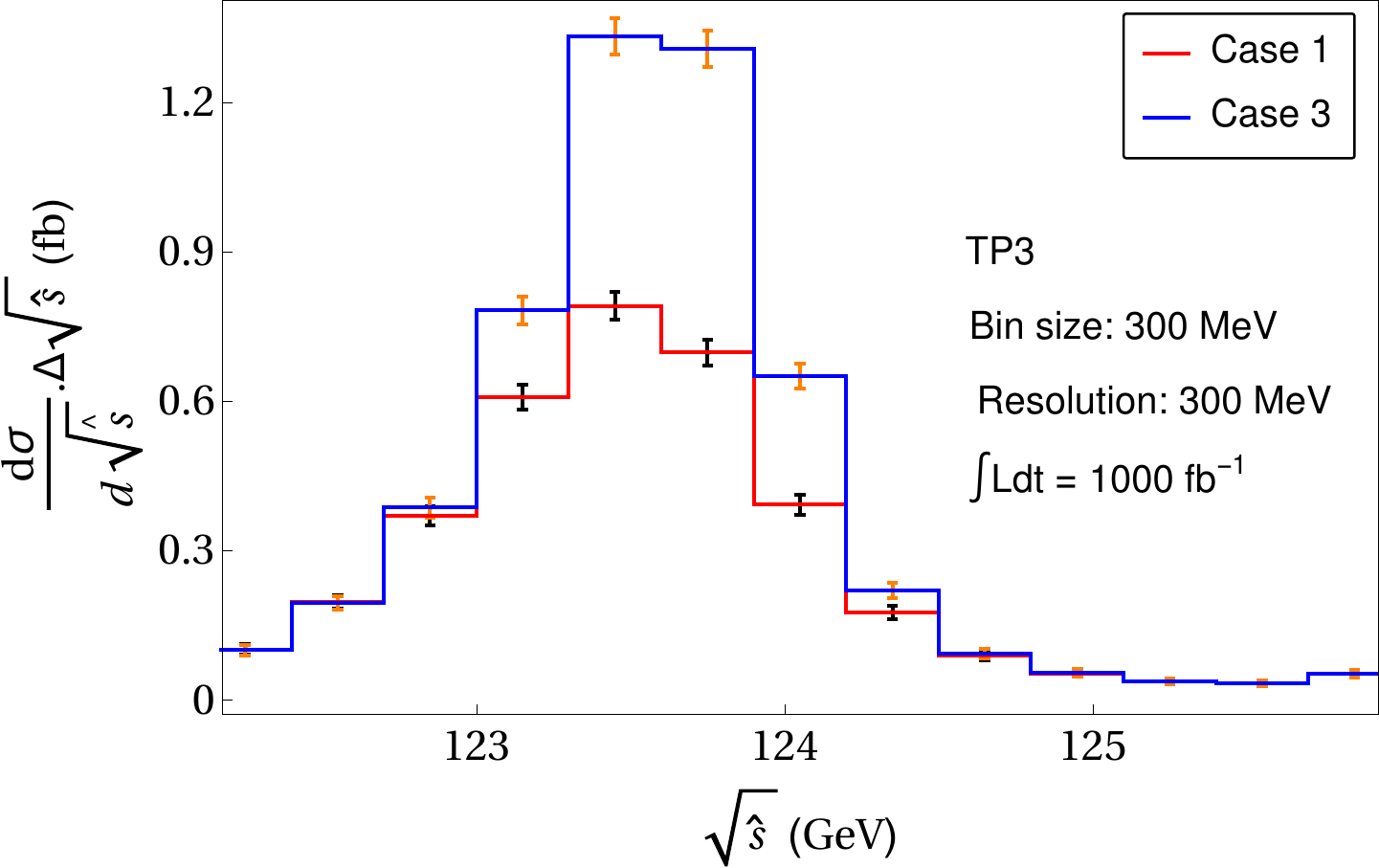}\\
\caption{\label{fig:TP3} As in Fig.~\ref{fig:TP1}, for the TP3.}
\end{figure}

\begin{table}[t!]
\hspace*{-.8cm}
\centering
\begin{tabular}{c|cccccccc}
\hline
TP & $m_{H_1}$ & $m_{H_2}$ & $\Delta m_H$ 
& $\Gamma_{H_1}$ & $\Gamma_{H_2}$ 
& \multicolumn{3}{c}{$\sigma_{pp}^{\gamma \gamma}$ (fb)} \\
& (GeV) &  (GeV)&  (MeV) &  (MeV) &  (MeV) & Case 1 & Case 2 & Case 3 \\
\hline\hline
1 & 124.7928 & 124.8158 & 23 & 10.8 & 38.3 
& 4.61 & 4.76 & 4.94\\
2 & 123.8696 & 124.1991 & 329.5 & 400.2 & 73.5 
& 0.353 & 0.385 & 0.458\\
\hline
3 & 123.4590 & 123.7876 & 328.6 & 704.9 & 39.2
& 1.09 & 1.45 & 1.58\\
\hline
\end{tabular}
\caption{Higgs boson masses and widths as well as the $pp\to H \to \gamma\gamma$ cross sections corresponding to the three Cases for the three selected TPs.}
\label{table:crosssection-TP}
\end{table}

\begin{table}[t]
\centering
\begin{tabular}{c|cccccccccc}
\hline
TP & $\phi_\kappa$ & $M_0$ & $M_{1/2}$ & $A_0$ & tan$\beta$ 
& $\lambda$ & 
$\kappa$ & 
$A_\lambda$ & 
$A_\kappa$ 
& $\mu_{\rm eff}$ \\
\hline\hline
1 & \multirow{2}{*}{$3^\circ$}& 1438.0 & 255.53 & $-2859.2$ & 4.80 & 0.6935 & 0.3287 
& 653.21 & $-6.399$ & 145.77\\
2 & & 1405.7 & 154.63 & $-2706.5$ & 5.63 & 0.6861 & 0.4602 
& 546.59 & $1.555$ & 108.58\\
\hline
3 & $10^\circ$ & 1895.2 & 115.14 & $-835.20$ & 1.76 & 0.6524 & 0.5752 
& 74.865 & $-120.70$ & 105.95\\ 
\hline
\end{tabular}
\caption{Values of the input parameters 
for the three TPs considered. All dimensionful parameters are in units of GeV.}
\def\baselinestretch{1.0}
\label{table:TPinput}
\end{table}


\section{\label{sec:concl}Conclusions}

In summary, we have scrutinised in detail the proton-proton to di-photon process, through which a 125\,GeV resonance consistent with the SM Higgs boson has been discovered at the LHC. Indeed, this is the signature for which the Higgs mass resolution is highest amongst all those accessible at the CERN collider.
Measurements of its cross section, at both the inclusive and exclusive level, however, do not exclude the possibility of non-SM explanations. Amongst these, particularly intriguing are those invoking two Higgs bosons produced via gluon fusion, with such a small mass difference that they cannot be resolved by the experimental apparatus. This scenario can emerge only in non-minimal realisations of SUSY, such as the NMSSM, wherein (unlike the MSSM) two coexisting Higgs bosons can contribute to the 125\,GeV signal (in $\gamma\gamma$ as well as other final states). In this case, an accurate treatment is required of the propagation of the two states, which not only goes beyond the NWA but also allows for full interference between these. Hence, we have studied the quantitative impact of interference between two Higgs states near 125\,GeV, with and without mixing effects, relative to the simplistic approach where the two resonant objects are treated independently of each other. For a full treatment, including the possibility of complex couplings as well, we have considered both real and complex NMSSM.
 
Our analysis involved scanning of the parameter space of the model for finding possible solutions consistent not only with the LHC exclusion limits on the additional Higgs bosons but also with the constraints from EDM measurements. These scans further collected only model solutions yielding two Higgs bosons with masses lying within the uncertainty of the measurements of the 125\,GeV resonance. This was followed by a dedicated computation,  performed with the help of a locally developed MC  program, producing both integrated and differential cross sections for the full process $pp(gg)\to H_1,H_2\to\gamma\gamma$. We have found that the aforementioned interference effects can be sizeable, with some of the selected BPs providing a difference of around 40\% in inclusive rates between the standard approach consisting in treating the two resonances as separate BW functions and the full propagator one including all non-trivial quantum effects. 

We then considered the possibility of a shape analysis of the emerging profile, which could reveal the presence of multiple resonances, assuming realistic, current and prospective, di-photon mass resolutions of the LHC detectors. {This revealed some long-term potential to see the difference between the generally exploited simplistic case of assuming two separate resonances and the one where the two nearly mass-degenerate states interfere due to the inclusion of the complete propagator matrix in the amplitude calculation.} These differences are in fact more visible with a smaller di-photon mass resolution and a larger data sample, both of which can only be achieved with upgraded detectors and/or machine. Finally, in attempting to distinguish the two approaches, we have also noted a tension in the underlying dynamics. Any distortion
effect of a single BW shape can only be exploited when the mass difference is sufficiently larger than the assumed width of the bins (which should naturally be consistent with the available experimental mass resolution) in the distribution of the differential cross section. However, a larger mass difference leads to smaller interference effects. 

\section*{Acknowledgments}

S.~Moretti is supported in part through the NExT Institute. We thank Korea Institute for Advanced Study for providing computing resources (Linux Cluster System at KIAS Center for Advanced Computation) for this work. 

\appendix
\numberwithin{equation}{section}

\section{Appendix}

The absorptive part of the Higgs propagator matrix can be written as
\begin{equation}
\label{eq:all}
\begin{split}
{{\mathfrak{I}}{\rm m}\hat\Pi}_{ij}(s)=
{{\mathfrak{I}}{\rm m}\hat\Pi}_{ij}^{ff}(s)
+{{\mathfrak{I}}{\rm m}\hat\Pi}_{ij}^{VV}(s)
+{{\mathfrak{I}}{\rm m}\hat\Pi}_{ij}^{HV}(s)
+{{\mathfrak{I}}{\rm m}\hat\Pi}_{ij}^{HH}(s)
+{{\mathfrak{I}}{\rm m}\hat\Pi}_{ij}^{\tilde f \tilde f}(s)\,.
\end{split}
\end{equation}
%
We reproduce here the expressions for the individual contributions 
from~\cite{Ellis:2004fs}, where those to vector bosons as well as
associated Higgs and vector boson pairs were derived using the Pinch
Technique~\cite{Cornwall:1981zr,*Cornwall:1989gv,*Papavassiliou:1989zd,*Papavassiliou:1994pr,Degrassi:1992ue,*Hashimoto:1994ct,*Watson:1994tn,*Binosi:2009qm}, which ensures their linear dependence on $s$. These two contributions are given as
\begin{eqnarray}
\label{eq:VV}
{{\mathfrak{I}}{\rm m}\hat\Pi}_{ij}^{VV}(s)&=&
\frac{g^2 g_{H_iVV} g_{H_jVV} \delta_V \beta_V}{128\pi m_W^2}
\Bigl\{
-4m_V^2\Bigl(2 s-3m_V^2\Bigr)
+2m_V^2\Bigl(m_{H_i}^2+m_{H_j}^2\Bigr) \nonumber \\
&+&m_{H_i}^2m_{H_j}^2
\Bigr\}~
\Theta\Bigl(s-4m_V^2\Bigr)\,,
\end{eqnarray}
with $\beta_V=\sqrt{1-4\kappa_V}$ and $\delta_W=2,~\delta_Z=1$, and
\begin{eqnarray}
\label{eq:HV}
{{\mathfrak{I}}{\rm m}\hat\Pi}_{ij}^{HV}(s)&=&
\frac{g^2}{64\pi m_W^2}
\sum_{k=1-5} g_{H_i H_k Z} g_{H_j H_k Z}~ 
\lambda^{1/2}\Bigl(1,\kappa_Z,\kappa_{H_k}\Bigr)
\Bigl\{
-4 s m_Z^2+ \Bigl(m_Z^2-m_{H_k}^2\Bigr)^2 \nonumber \\
&+&\Bigl(m_Z^2-m_{H_k}^2\Bigr)\Bigl(m_{H_i}^2+m_{H_j}^2\Bigr)
+m_{H_i}^2 m_{H_j}^2
\Bigr\}~
\Theta\Bigl(s-(m_Z+m_{H_k})^2\Bigr) \nonumber \\
&+&\frac{g^2}{32\pi m_W^2}{\mathfrak{R}}{\rm e}
\Bigl(g_{H_i H^+ W^-} g_{H_j H^+ W^-}^*\Bigr)~
\lambda^{1/2}\Bigl(1,\kappa_W,\kappa_{H^\pm}\Bigr)
\Bigl\{
-4\hat s m_W^2+ \Bigl(m_W^2-m_{H^\pm}^2\Bigr)^2 \nonumber \\
&+&\Bigl(m_W^2-m_{H^\pm}^2\Bigr)\Bigl(m_{H_i}^2+m_{H_j}^2\Bigr)
+m_{H_i}^2 m_{H_j}^2
\Bigr\}~
\Theta\Bigl(s-(m_W+m_{H^\pm})^2\Bigr)\,. 
\end{eqnarray}

The contribution from loops of Higgs boson pairs reads
\begin{equation}
\label{eq:HH}
{{\mathfrak{I}}{\rm m}\hat\Pi}_{ij}^{HH}(s)=
\frac{v^2}{16\pi}\sum_{k\geq l=1-5} 
\frac{S_{ij;kl}}{1+\delta_{kl}}
g_{H_i H_k H_l} g_{H_j H_k H_l}~
\lambda^{1/2}\Bigl(1,\kappa_{H_k},\kappa_{H_l}\Bigr)~
\Theta\Bigl(s-(m_{H_k}+m_{H_{H_l}})^2\Bigr)\,,
\end{equation}
where the symmetry factor $S_{ij;kl}$ is equal to 2 
for $i=j$ and $k\neq l$, or $i\neq j$ and $k=l$, to 4 for 
$i=j$ and $k=l$, and to 1 otherwise.

Finally, the loops of 
fermions (omitting the QCD $K$-factors) and sfermions give 
\begin{eqnarray}
\label{eq:ff}
{{\mathfrak{I}}{\rm m}\hat\Pi}_{ij}^{ff}(s)&=&
\frac{s}{8\pi} \sum_{f,f^\prime}
g_f^2 \Delta_{ff^\prime} N_{cf} 
~\lambda^{1/2}\Bigl(1,\kappa_f,\kappa_{f^\prime}\Bigr)\Bigl\{
\Bigl(1-\kappa_f-\kappa_{f^\prime}\Bigl)
\Bigl(g_{H_i f \bar f^\prime}^S g_{H_j f \bar f^\prime}^{S*}
+g_{H_i f \bar f^\prime}^P g_{H_j f \bar f^\prime}^{P*}\Bigr) \nonumber \\
&-&2\sqrt{\kappa_f \kappa_{f^\prime}}
\Bigl(g_{H_i f \bar f^\prime}^S g_{H_j f \bar f^\prime}^{S*}
-g_{H_i f \bar f^\prime}^P g_{H_j f \bar f^\prime}^{P*}\Bigr)
\Bigr\}
~\Theta\Bigl(s-(m_f+m_{f^\prime})^2 \Bigr)\,,
\end{eqnarray}
where $g_f=\frac{gm_f}{2m_W}$, $\Delta_{ff^\prime}=\delta_{ff^\prime}~(f,f^\prime=t,b,c,s,\tau,\mu)$, 
$\frac{4}{1+\delta_{ff^\prime}}~(f,f^\prime=\tilde\chi_{1,2,3,4,5}^0)$,
1 $(f,f^\prime=\tilde\chi_{1,2}^+)$, $\lambda(x,y,z)=x^2+y^2+z^2-2(xy+yz+zx)$
and $\kappa_x=\frac{m_x^2}{\hat s}$, and
\begin{equation}
\label{eq:sfsf}
{{\mathfrak{I}}{\rm m}\hat\Pi}_{ij}^{\tilde f \tilde f}(s)=
\frac{v^2}{16\pi}
\sum_{\tilde f} \sum_{k,l=1-2} N_{cf}
g_{H_i \tilde f_k \tilde f_l^*} g_{H_j f_k \tilde f_l^*}^*~
\lambda^{1/2}\Bigl(1,\kappa_{\tilde f_k},\kappa_{\tilde f_l}\Bigr)~
\Theta\Bigl(s-(m_{\tilde f_k}+m_{\tilde f_l})^2\Bigr)\,.
\end{equation}


\begin{mcitethebibliography}{10}

\bibitem{Aad:2012tfa}
{\bfseries ATLAS Collaboration}, G.~Aad {\em et~al.}, ``{Observation of a new
  particle in the search for the Standard Model Higgs boson with the ATLAS
  detector at the LHC},''
  \href{http://dx.doi.org/10.1016/j.physletb.2012.08.020}{{\em Phys.Lett.}
  {\bfseries B716} (2012) 1--29},
\href{http://arxiv.org/abs/1207.7214}{{\ttfamily arXiv:1207.7214 [hep-ex]}}.
\mciteBstWouldAddEndPunctfalse
\mciteSetBstMidEndSepPunct{\mcitedefaultmidpunct}
{}{\mcitedefaultseppunct}\relax
\EndOfBibitem
\bibitem{Chatrchyan:2012xdj}
{\bfseries CMS Collaboration}, S.~Chatrchyan {\em et~al.}, ``{Observation of a
  new boson at a mass of 125 GeV with the CMS experiment at the LHC},''
  \href{http://dx.doi.org/10.1016/j.physletb.2012.08.021}{{\em Phys. Lett.}
  {\bfseries B716} (2012) 30--61},
\href{http://arxiv.org/abs/1207.7235}{{\ttfamily arXiv:1207.7235 [hep-ex]}}.
\mciteBstWouldAddEndPunctfalse
\mciteSetBstMidEndSepPunct{\mcitedefaultmidpunct}
{}{\mcitedefaultseppunct}\relax
\EndOfBibitem
\bibitem{Kim:1983dt}
J.~E. Kim and H.~P. Nilles, ``{The mu Problem and the Strong CP Problem},''
\href{http://dx.doi.org/10.1016/0370-2693(84)91890-2}{{\em Phys. Lett.}
  {\bfseries B138} (1984) 150--154}.
\mciteBstWouldAddEndPunctfalse
\mciteSetBstMidEndSepPunct{\mcitedefaultmidpunct}
{}{\mcitedefaultseppunct}\relax
\EndOfBibitem
\bibitem{Nir:1995bu}
Y.~Nir, ``{Gauge unification, Yukawa hierarchy and the mu problem},''
  \href{http://dx.doi.org/10.1016/0370-2693(95)00619-V}{{\em Phys. Lett.}
  {\bfseries B354} (1995) 107--110},
\href{http://arxiv.org/abs/hep-ph/9504312}{{\ttfamily arXiv:hep-ph/9504312
  }}.
\mciteBstWouldAddEndPunctfalse
\mciteSetBstMidEndSepPunct{\mcitedefaultmidpunct}
{}{\mcitedefaultseppunct}\relax
\EndOfBibitem
\bibitem{Fayet:1974pd}
P.~Fayet, ``{Supergauge Invariant Extension of the Higgs Mechanism and a Model
  for the electron and Its Neutrino},''
\href{http://dx.doi.org/10.1016/0550-3213(75)90636-7}{{\em Nucl.Phys.}
  {\bfseries B90} (1975) 104--124}
\mciteBstWouldAddEndPunctfalse
\mciteSetBstMidEndSepPunct{\mcitedefaultmidpunct}
{}{\mcitedefaultseppunct}\relax
\EndOfBibitem
\bibitem{Ellis:1988er}
J.~R. Ellis, J.~Gunion, H.~E. Haber, L.~Roszkowski and F.~Zwirner, ``{Higgs
  Bosons in a Nonminimal Supersymmetric Model},''
\href{http://dx.doi.org/10.1103/PhysRevD.39.844}{{\em Phys.Rev.} {\bfseries
  D39} (1989) 844}
\mciteBstWouldAddEndPunctfalse
\mciteSetBstMidEndSepPunct{\mcitedefaultmidpunct}
{}{\mcitedefaultseppunct}\relax
\EndOfBibitem
\bibitem{Durand:1988rg}
L.~Durand and J.~L. Lopez, ``{Upper Bounds on Higgs and Top Quark Masses in the
  Flipped SU(5) x U(1) Superstring Model},''
\href{http://dx.doi.org/10.1016/0370-2693(89)90079-8}{{\em Phys.Lett.}
  {\bfseries B217} (1989) 463}
\mciteBstWouldAddEndPunctfalse
\mciteSetBstMidEndSepPunct{\mcitedefaultmidpunct}
{}{\mcitedefaultseppunct}\relax
\EndOfBibitem
\bibitem{Drees:1988fc}
M.~Drees, ``{Supersymmetric Models with Extended Higgs Sector},''
\href{http://dx.doi.org/10.1142/S0217751X89001448}{{\em Int.J.Mod.Phys.}
  {\bfseries A4} (1989) 3635}.
\mciteBstWouldAddEndPunctfalse
\mciteSetBstMidEndSepPunct{\mcitedefaultmidpunct}
{}{\mcitedefaultseppunct}\relax
\EndOfBibitem
\bibitem{Ellwanger:2009dp}
U.~Ellwanger, C.~Hugonie and A.~M. Teixeira, ``{The Next-to-Minimal
  Supersymmetric Standard Model},''
  \href{http://dx.doi.org/10.1016/j.physrep.2010.07.001}{{\em Phys.Rept.}
  {\bfseries 496} (2010) 1--77},
\href{http://arxiv.org/abs/0910.1785}{{\ttfamily arXiv:0910.1785 [hep-ph]}}.
\mciteBstWouldAddEndPunctfalse
\mciteSetBstMidEndSepPunct{\mcitedefaultmidpunct}
{}{\mcitedefaultseppunct}\relax
\EndOfBibitem
\bibitem{Maniatis:2009re}
M.~Maniatis, ``{The Next-to-Minimal Supersymmetric extension of the Standard
  Model reviewed},'' \href{http://dx.doi.org/10.1142/S0217751X10049827}{{\em
  Int.J.Mod.Phys.} {\bfseries A25} (2010) 3505--3602},
\href{http://arxiv.org/abs/0906.0777}{{\ttfamily arXiv:0906.0777 [hep-ph]}}.
\mciteBstWouldAddEndPunctfalse
\mciteSetBstMidEndSepPunct{\mcitedefaultmidpunct}
{}{\mcitedefaultseppunct}\relax
\EndOfBibitem
\bibitem{Djouadi:2005gj}
A.~Djouadi, ``{The Anatomy of electro-weak symmetry breaking. II. The Higgs
  bosons in the minimal supersymmetric model},''
  \href{http://dx.doi.org/10.1016/j.physrep.2007.10.005}{{\em Phys. Rept.}
  {\bfseries 459} (2008) 1--241},
\href{http://arxiv.org/abs/hep-ph/0503173}{{\ttfamily arXiv:hep-ph/0503173}}.
\mciteBstWouldAddEndPunctfalse
\mciteSetBstMidEndSepPunct{\mcitedefaultmidpunct}
{}{\mcitedefaultseppunct}\relax
\EndOfBibitem
\bibitem{Gunion:2012gc}
J.~F. Gunion, Y.~Jiang and S.~Kraml, ``{Could two NMSSM Higgs bosons be
  present near 125 GeV?},''
  \href{http://dx.doi.org/10.1103/PhysRevD.86.071702}{{\em Phys.Rev.}
  {\bfseries D86} (2012) 071702},
\href{http://arxiv.org/abs/1207.1545}{{\ttfamily arXiv:1207.1545 [hep-ph]}}
\mciteBstWouldAddEndPunctfalse
\mciteSetBstMidEndSepPunct{\mcitedefaultmidpunct}
{}{\mcitedefaultseppunct}\relax
\EndOfBibitem
\bibitem{King:2012tr}
S.~King, M.~Muhlleitner, R.~Nevzorov and K.~Walz, ``{Natural NMSSM Higgs
  Bosons},'' \href{http://dx.doi.org/10.1016/j.nuclphysb.2013.01.020}{{\em
  Nucl.Phys.} {\bfseries B870} (2013) 323--352},
\href{http://arxiv.org/abs/1211.5074}{{\ttfamily arXiv:1211.5074 [hep-ph]}}
\mciteBstWouldAddEndPunctfalse
\mciteSetBstMidEndSepPunct{\mcitedefaultmidpunct}
{}{\mcitedefaultseppunct}\relax
\EndOfBibitem
\bibitem{Gherghetta:2012gb}
T.~Gherghetta, B.~von Harling, A.~D. Medina, and M.~A. Schmidt, ``{The
  Scale-Invariant NMSSM and the 126 GeV Higgs Boson},''
  \href{http://dx.doi.org/10.1007/JHEP02(2013)032}{{\em JHEP} {\bfseries 1302}
  (2013) 032},
\href{http://arxiv.org/abs/1212.5243}{{\ttfamily arXiv:1212.5243 [hep-ph]}}.
\mciteBstWouldAddEndPunctfalse
\mciteSetBstMidEndSepPunct{\mcitedefaultmidpunct}
{}{\mcitedefaultseppunct}\relax
\EndOfBibitem
\bibitem{Wu:2015nba}
L.~Wu, J.~M. Yang, C.-P. Yuan and M.~Zhang, ``{Higgs self-coupling in the MSSM
  and NMSSM after the LHC Run 1},''
  \href{http://dx.doi.org/10.1016/j.physletb.2015.06.020}{{\em Phys. Lett.}
  {\bfseries B747} (2015) 378--389},
\href{http://arxiv.org/abs/1504.06932}{{\ttfamily arXiv:1504.06932 [hep-ph]}}
\mciteBstWouldAddEndPunctfalse
\mciteSetBstMidEndSepPunct{\mcitedefaultmidpunct}
{}{\mcitedefaultseppunct}\relax
\EndOfBibitem
\bibitem{Domingo:2015eea}
F.~Domingo and G.~Weiglein, ``{NMSSM interpretations of the observed Higgs
  signal},'' \href{http://dx.doi.org/10.1007/JHEP04(2016)095}{{\em JHEP}
  {\bfseries 04} (2016) 095},
\href{http://arxiv.org/abs/1509.07283}{{\ttfamily arXiv:1509.07283 [hep-ph]}}.
\mciteBstWouldAddEndPunctfalse
\mciteSetBstMidEndSepPunct{\mcitedefaultmidpunct}
{}{\mcitedefaultseppunct}\relax
\EndOfBibitem
\bibitem{Munir:2013wka}
S.~Munir, L.~Roszkowski and S.~Trojanowski, ``{Simultaneous enhancement in
  gamma.gamma, b.b-bar and tau+.tau- rates in the NMSSM with nearly degenerate
  scalar and pseudoscalar Higgs bosons},'' {\em Phys. Rev.} {\bfseries D88}
  (2013) 055017,
\href{http://arxiv.org/abs/1305.0591}{{\ttfamily arXiv:1305.0591 [hep-ph]}}.
\mciteBstWouldAddEndPunctfalse
\mciteSetBstMidEndSepPunct{\mcitedefaultmidpunct}
{}{\mcitedefaultseppunct}\relax
\EndOfBibitem
\bibitem{Moretti:2013lya}
S.~Moretti, S.~Munir and P.~Poulose, ``{125 GeV Higgs Boson signal within the
  complex NMSSM},'' \href{http://dx.doi.org/10.1103/PhysRevD.89.015022}{{\em
  Phys.Rev.} {\bfseries D89} no.~1, (2014) 015022},
\href{http://arxiv.org/abs/1305.0166}{{\ttfamily arXiv:1305.0166 [hep-ph]}}.
\mciteBstWouldAddEndPunctfalse
\mciteSetBstMidEndSepPunct{\mcitedefaultmidpunct}
{}{\mcitedefaultseppunct}\relax
\EndOfBibitem
\bibitem{Demir:1999hj}
D.~A. Demir, ``{Effects of the supersymmetric phases on the neutral Higgs
  sector},'' \href{http://dx.doi.org/10.1103/PhysRevD.60.055006}{{\em
  Phys.Rev.} {\bfseries D60} (1999) 055006},
\href{http://arxiv.org/abs/hep-ph/9901389}{{\ttfamily arXiv:hep-ph/9901389}}
\mciteBstWouldAddEndPunctfalse
\mciteSetBstMidEndSepPunct{\mcitedefaultmidpunct}
{}{\mcitedefaultseppunct}\relax
\EndOfBibitem
\bibitem{Dedes:1999sj}
A.~Dedes and S.~Moretti, ``{Effect of large supersymmetric phases on Higgs
  production.},'' \href{http://dx.doi.org/10.1103/PhysRevLett.84.22}{{\em
  Phys.Rev.Lett.} {\bfseries 84} (2000) 22--25},
\href{http://arxiv.org/abs/hep-ph/9908516}{{\ttfamily arXiv:hep-ph/9908516}}
\mciteBstWouldAddEndPunctfalse
\mciteSetBstMidEndSepPunct{\mcitedefaultmidpunct}
{}{\mcitedefaultseppunct}\relax
\EndOfBibitem
\bibitem{Dedes:1999zh}
A.~Dedes and S.~Moretti, ``{Effects of CP violating phases on Higgs boson
  production at hadron colliders in the minimal supersymmetric standard
  model},'' \href{http://dx.doi.org/10.1016/S0550-3213(00)00144-9}{{\em
  Nucl.Phys.} {\bfseries B576} (2000) 29--55},
\href{http://arxiv.org/abs/hep-ph/9909418}{{\ttfamily arXiv:hep-ph/9909418}}
\mciteBstWouldAddEndPunctfalse
\mciteSetBstMidEndSepPunct{\mcitedefaultmidpunct}
{}{\mcitedefaultseppunct}\relax
\EndOfBibitem
\bibitem{Kane:2000aq}
G.~L. Kane and L.-T. Wang, ``{Implications of supersymmetry phases for Higgs
  boson signals and limits},''
  \href{http://dx.doi.org/10.1016/S0370-2693(00)00871-6}{{\em Phys.Lett.}
  {\bfseries B488} (2000) 383--389},
\href{http://arxiv.org/abs/hep-ph/0003198}{{\ttfamily arXiv:hep-ph/0003198}}
\mciteBstWouldAddEndPunctfalse
\mciteSetBstMidEndSepPunct{\mcitedefaultmidpunct}
{}{\mcitedefaultseppunct}\relax
\EndOfBibitem
\bibitem{Arhrib:2001pg}
A.~Arhrib, D.~K. Ghosh and O.~C. Kong, ``{Observing CP violating MSSM Higgs
  bosons at hadron colliders?},''
  \href{http://dx.doi.org/10.1016/S0370-2693(02)01912-3}{{\em Phys.Lett.}
  {\bfseries B537} (2002) 217--226},
\href{http://arxiv.org/abs/hep-ph/0112039}{{\ttfamily arXiv:hep-ph/0112039}}
\mciteBstWouldAddEndPunctfalse
\mciteSetBstMidEndSepPunct{\mcitedefaultmidpunct}
{}{\mcitedefaultseppunct}\relax
\EndOfBibitem
\bibitem{Choi:2001pg}
S.~Choi, K.~Hagiwara and J.~S. Lee, ``{Higgs boson decays in the minimal
  supersymmetric standard model with radiative Higgs sector CP violation},''
  \href{http://dx.doi.org/10.1103/PhysRevD.64.032004}{{\em Phys.Rev.}
  {\bfseries D64} (2001) 032004},
\href{http://arxiv.org/abs/hep-ph/0103294}{{\ttfamily arXiv:hep-ph/0103294}}
\mciteBstWouldAddEndPunctfalse
\mciteSetBstMidEndSepPunct{\mcitedefaultmidpunct}
{}{\mcitedefaultseppunct}\relax
\EndOfBibitem
\bibitem{Choi:2002zp}
S.~Choi, M.~Drees, J.~S. Lee and J.~Song, ``{Supersymmetric Higgs boson decays
  in the MSSM with explicit CP violation},''
  \href{http://dx.doi.org/10.1007/s10052-002-0997-8}{{\em Eur.Phys.J.}
  {\bfseries C25} (2002) 307--313},
\href{http://arxiv.org/abs/hep-ph/0204200}{{\ttfamily arXiv:hep-ph/0204200}}.
\mciteBstWouldAddEndPunctfalse
\mciteSetBstMidEndSepPunct{\mcitedefaultmidpunct}
{}{\mcitedefaultseppunct}\relax
\EndOfBibitem
\bibitem{Ellis:2004fs}
J.~R. Ellis, J.~S. Lee and A.~Pilaftsis, ``{CERN LHC signatures of resonant CP
  violation in a minimal supersymmetric Higgs sector},''
  \href{http://dx.doi.org/10.1103/PhysRevD.70.075010}{{\em Phys.Rev.}
  {\bfseries D70} (2004) 075010},
\href{http://arxiv.org/abs/hep-ph/0404167}{{\ttfamily arXiv:hep-ph/0404167}}.
\mciteBstWouldAddEndPunctfalse
\mciteSetBstMidEndSepPunct{\mcitedefaultmidpunct}
{}{\mcitedefaultseppunct}\relax
\EndOfBibitem
\bibitem{Hesselbach:2009gw}
S.~Hesselbach, S.~Moretti, S.~Munir and P.~Poulose, ``{Explicit CP violation
  in the MSSM through $gg\rightarrow H \rightarrow \gamma \gamma$},''
  \href{http://dx.doi.org/10.1103/PhysRevD.82.074004}{{\em Phys.Rev.}
  {\bfseries D82} (2010) 074004},
\href{http://arxiv.org/abs/0903.0747}{{\ttfamily arXiv:0903.0747 [hep-ph]}}
\mciteBstWouldAddEndPunctfalse
\mciteSetBstMidEndSepPunct{\mcitedefaultmidpunct}
{}{\mcitedefaultseppunct}\relax
\EndOfBibitem
\bibitem{Fritzsche:2011nr}
T.~Fritzsche, S.~Heinemeyer, H.~Rzehak and C.~Schappacher, ``{Heavy Scalar Top
  Quark Decays in the Complex MSSM: A Full One-Loop Analysis},''
  \href{http://dx.doi.org/10.1103/PhysRevD.86.035014}{{\em Phys. Rev.}
  {\bfseries D86} (2012) 035014},
\href{http://arxiv.org/abs/1111.7289}{{\ttfamily arXiv:1111.7289 [hep-ph]}}
\mciteBstWouldAddEndPunctfalse
\mciteSetBstMidEndSepPunct{\mcitedefaultmidpunct}
{}{\mcitedefaultseppunct}\relax
\EndOfBibitem
\bibitem{Chakraborty:2013si}
A.~Chakraborty, B.~Das, J.~L. Diaz-Cruz, D.~K. Ghosh, S.~Moretti {\em et~al.},
  ``{125 GeV Higgs signal at the LHC in the $CP$-violating MSSM},''
  \href{http://dx.doi.org/10.1103/PhysRevD.90.055005}{{\em Phys.Rev.}
  {\bfseries D90} no.~5, (2014) 055005},
\href{http://arxiv.org/abs/1301.2745}{{\ttfamily arXiv:1301.2745 [hep-ph]}}.
\mciteBstWouldAddEndPunctfalse
\mciteSetBstMidEndSepPunct{\mcitedefaultmidpunct}
{}{\mcitedefaultseppunct}\relax
\EndOfBibitem
\bibitem{Graf:2012hh}
T.~Graf, R.~Grober, M.~Muhlleitner, H.~Rzehak and K.~Walz, ``{Higgs Boson
  Masses in the Complex NMSSM at One-Loop Level},''
  \href{http://dx.doi.org/10.1007/JHEP10(2012)122}{{\em JHEP} {\bfseries 1210}
  (2012) 122},
\href{http://arxiv.org/abs/1206.6806}{{\ttfamily arXiv:1206.6806 [hep-ph]}}.
\mciteBstWouldAddEndPunctfalse
\mciteSetBstMidEndSepPunct{\mcitedefaultmidpunct}
{}{\mcitedefaultseppunct}\relax
\EndOfBibitem
\bibitem{Munir:2013dya}
S.~Munir, ``{Novel Higgs-to-125 GeV Higgs boson decays in the complex NMSSM},''
  \href{http://dx.doi.org/10.1103/PhysRevD.89.095013}{{\em Phys.Rev.}
  {\bfseries D89} (2014) 095013},
\href{http://arxiv.org/abs/1310.8129}{{\ttfamily arXiv:1310.8129 [hep-ph]}}.
\mciteBstWouldAddEndPunctfalse
\mciteSetBstMidEndSepPunct{\mcitedefaultmidpunct}
{}{\mcitedefaultseppunct}\relax
\EndOfBibitem
\bibitem{Moretti:2015bua}
S.~Moretti and S.~Munir, ``{Two Higgs Bosons near 125 GeV in the Complex NMSSM
  and the LHC Run I Data},'' \href{http://dx.doi.org/10.1155/2015/509847}{{\em
  Adv. High Energy Phys.} {\bfseries 2015} (2015) 509847},
\href{http://arxiv.org/abs/1505.00545}{{\ttfamily arXiv:1505.00545 [hep-ph]}}.
\mciteBstWouldAddEndPunctfalse
\mciteSetBstMidEndSepPunct{\mcitedefaultmidpunct}
{}{\mcitedefaultseppunct}\relax
\EndOfBibitem
\bibitem{Bechtle:2013xfa}
P.~Bechtle, S.~Heinemeyer, O.~Stal, T.~Stefaniak and G.~Weiglein,
  ``{$HiggsSignals$: Confronting arbitrary Higgs sectors with measurements at
  the Tevatron and the LHC},''
  \href{http://dx.doi.org/10.1140/epjc/s10052-013-2711-4}{{\em Eur.Phys.J.}
  {\bfseries C74} no.~2, (2014) 2711},
\href{http://arxiv.org/abs/1305.1933}{{\ttfamily arXiv:1305.1933 [hep-ph]}}.
\mciteBstWouldAddEndPunctfalse
\mciteSetBstMidEndSepPunct{\mcitedefaultmidpunct}
{}{\mcitedefaultseppunct}\relax
\EndOfBibitem
\bibitem{Cacciapaglia:2009ic}
G.~Cacciapaglia, A.~Deandrea and S.~De~Curtis, ``{Nearby resonances beyond the
  Breit-Wigner approximation},''
  \href{http://dx.doi.org/10.1016/j.physletb.2009.10.090}{{\em Phys. Lett.}
  {\bfseries B682} (2009) 43--49},
\href{http://arxiv.org/abs/0906.3417}{{\ttfamily arXiv:0906.3417 [hep-ph]}}
\mciteBstWouldAddEndPunctfalse
\mciteSetBstMidEndSepPunct{\mcitedefaultmidpunct}
{}{\mcitedefaultseppunct}\relax
\EndOfBibitem
\bibitem{Fuchs:2014ola}
E.~Fuchs, S.~Thewes and G.~Weiglein, ``{Interference effects in BSM processes
  with a generalised narrow-width approximation},''
  \href{http://dx.doi.org/10.1140/epjc/s10052-015-3472-z}{{\em Eur. Phys. J.}
  {\bfseries C75} (2015) 254},
\href{http://arxiv.org/abs/1411.4652}{{\ttfamily arXiv:1411.4652 [hep-ph]}}
\mciteBstWouldAddEndPunctfalse
\mciteSetBstMidEndSepPunct{\mcitedefaultmidpunct}
{}{\mcitedefaultseppunct}\relax
\EndOfBibitem
\bibitem{Chen:2016oib}
N.~Chen and Z.~Liu, ``{Degenerate Higgs bosons: hiding a second Higgs at 125
  GeV},''
\href{http://arxiv.org/abs/1607.02154}{{\ttfamily arXiv:1607.02154 [hep-ph]}}
\mciteBstWouldAddEndPunctfalse
\mciteSetBstMidEndSepPunct{\mcitedefaultmidpunct}
{}{\mcitedefaultseppunct}\relax
\EndOfBibitem
\bibitem{Fuchs:2016swt}
E.~Fuchs and G.~Weiglein, ``{Breit-Wigner approximation for propagators of
  mixed unstable states},''
\href{http://arxiv.org/abs/1610.06193}{{\ttfamily arXiv:1610.06193 [hep-ph]}}.
\mciteBstWouldAddEndPunctfalse
\mciteSetBstMidEndSepPunct{\mcitedefaultmidpunct}
{}{\mcitedefaultseppunct}\relax
\EndOfBibitem
\bibitem{Ellwanger:2005fh}
U.~Ellwanger and C.~Hugonie, ``{Yukawa induced radiative corrections to the
  lightest Higgs boson mass in the NMSSM},''
  \href{http://dx.doi.org/10.1016/j.physletb.2005.07.039}{{\em Phys. Lett.}
  {\bfseries B623} (2005) 93--103},
\href{http://arxiv.org/abs/hep-ph/0504269}{{\ttfamily arXiv:hep-ph/0504269
  [hep-ph]}}.
\mciteBstWouldAddEndPunctfalse
\mciteSetBstMidEndSepPunct{\mcitedefaultmidpunct}
{}{\mcitedefaultseppunct}\relax
\EndOfBibitem
\bibitem{Degrassi:2009yq}
G.~Degrassi and P.~Slavich, ``{On the radiative corrections to the neutral
  Higgs boson masses in the NMSSM},''
  \href{http://dx.doi.org/10.1016/j.nuclphysb.2009.09.018}{{\em Nucl.Phys.}
  {\bfseries B825} (2010) 119--150},
\href{http://arxiv.org/abs/0907.4682}{{\ttfamily arXiv:0907.4682 [hep-ph]}}.
\mciteBstWouldAddEndPunctfalse
\mciteSetBstMidEndSepPunct{\mcitedefaultmidpunct}
{}{\mcitedefaultseppunct}\relax
\EndOfBibitem
\bibitem{Goodsell:2014pla}
M.~D. Goodsell, K.~Nickel and F.~Staub, ``{Two-loop corrections to the Higgs
  masses in the NMSSM},''
  \href{http://dx.doi.org/10.1103/PhysRevD.91.035021}{{\em Phys.Rev.}
  {\bfseries D91} no.~3, (2015) 035021},
\href{http://arxiv.org/abs/1411.4665}{{\ttfamily arXiv:1411.4665 [hep-ph]}}.
\mciteBstWouldAddEndPunctfalse
\mciteSetBstMidEndSepPunct{\mcitedefaultmidpunct}
{}{\mcitedefaultseppunct}\relax
\EndOfBibitem
\bibitem{Ham:2001wt}
S.~Ham, S.~Oh and D.~Son, ``{Neutral Higgs sector of the next-to-minimal
  supersymmetric standard model with explicit CP violation},''
  \href{http://dx.doi.org/10.1103/PhysRevD.65.075004}{{\em Phys.Rev.}
  {\bfseries D65} (2002) 075004},
\href{http://arxiv.org/abs/hep-ph/0110052}{{\ttfamily arXiv:hep-ph/0110052}}
\mciteBstWouldAddEndPunctfalse
\mciteSetBstMidEndSepPunct{\mcitedefaultmidpunct}
{}{\mcitedefaultseppunct}\relax
\EndOfBibitem
\bibitem{Funakubo:2004ka}
K.~Funakubo and S.~Tao, ``{The Higgs sector in the next-to-MSSM},''
  \href{http://dx.doi.org/10.1143/PTP.113.821}{{\em Prog.Theor.Phys.}
  {\bfseries 113} (2005) 821--842},
\href{http://arxiv.org/abs/hep-ph/0409294}{{\ttfamily arXiv:hep-ph/0409294}}.
\mciteBstWouldAddEndPunctfalse
\mciteSetBstMidEndSepPunct{\mcitedefaultmidpunct}
{}{\mcitedefaultseppunct}\relax
\EndOfBibitem
\bibitem{Cheung:2010ba}
K.~Cheung, T.-J. Hou, J.~S. Lee and E.~Senaha, ``{The Higgs Boson Sector of
  the Next-to-MSSM with CP Violation},''
  \href{http://dx.doi.org/10.1103/PhysRevD.82.075007}{{\em Phys.Rev.}
  {\bfseries D82} (2010) 075007},
\href{http://arxiv.org/abs/1006.1458}{{\ttfamily arXiv:1006.1458 [hep-ph]}}.
\mciteBstWouldAddEndPunctfalse
\mciteSetBstMidEndSepPunct{\mcitedefaultmidpunct}
{}{\mcitedefaultseppunct}\relax
\EndOfBibitem
\bibitem{Cheung:2011wn}
K.~Cheung, T.-J. Hou, J.~S. Lee and E.~Senaha, ``{Higgs Mediated EDMs in the
  Next-to-MSSM: An Application to Electroweak Baryogenesis},''
  \href{http://dx.doi.org/10.1103/PhysRevD.84.015002}{{\em Phys.Rev.}
  {\bfseries D84} (2011) 015002},
\href{http://arxiv.org/abs/1102.5679}{{\ttfamily arXiv:1102.5679 [hep-ph]}}.
\mciteBstWouldAddEndPunctfalse
\mciteSetBstMidEndSepPunct{\mcitedefaultmidpunct}
{}{\mcitedefaultseppunct}\relax
\EndOfBibitem
\bibitem{Domingo:2015qaa}
F.~Domingo, ``{A New Tool for the study of the CP-violating NMSSM},''
  \href{http://dx.doi.org/10.1007/JHEP06(2015)052}{{\em JHEP} {\bfseries 06}
  (2015) 052},
\href{http://arxiv.org/abs/1503.07087}{{\ttfamily arXiv:1503.07087 [hep-ph]}}.
\mciteBstWouldAddEndPunctfalse
\mciteSetBstMidEndSepPunct{\mcitedefaultmidpunct}
{}{\mcitedefaultseppunct}\relax
\EndOfBibitem
\bibitem{Muhlleitner:2014vsa}
M.~Muhlleitner, D.~T. Nhung, H.~Rzehak and K.~Walz, ``{Two-loop contributions
  of the order $ \mathcal{O}\left({\alpha}\_t{\alpha}\_s\right) $ to the masses
  of the Higgs bosons in the CP-violating NMSSM},''
  \href{http://dx.doi.org/10.1007/JHEP05(2015)128}{{\em JHEP} {\bfseries 05}
  (2015) 128},
\href{http://arxiv.org/abs/1412.0918}{{\ttfamily arXiv:1412.0918 [hep-ph]}}.
\mciteBstWouldAddEndPunctfalse
\mciteSetBstMidEndSepPunct{\mcitedefaultmidpunct}
{}{\mcitedefaultseppunct}\relax
\EndOfBibitem
\bibitem{Spira:1995rr}
M.~Spira, A.~Djouadi, D.~Graudenz, and P.~Zerwas, ``{Higgs boson production at
  the LHC},'' \href{http://dx.doi.org/10.1016/0550-3213(95)00379-7}{{\em
  Nucl.Phys.} {\bfseries B453} (1995) 17--82},
\href{http://arxiv.org/abs/hep-ph/9504378}{{\ttfamily arXiv:hep-ph/9504378}}.
\mciteBstWouldAddEndPunctfalse
\mciteSetBstMidEndSepPunct{\mcitedefaultmidpunct}
{}{\mcitedefaultseppunct}\relax
\EndOfBibitem
\bibitem{Baglio:2013iia}
J.~Baglio, R.~Grober, M.~Muhlleitner, D.~Nhung, H.~Rzehak {\em et~al.},
  ``{NMSSMCALC: A Program Package for the Calculation of Loop-Corrected Higgs
  Boson Masses and Decay Widths in the (Complex) NMSSM},''
  \href{http://dx.doi.org/10.1016/j.cpc.2014.08.005}{{\em Comput.Phys.Commun.}
  (2014) },
\href{http://arxiv.org/abs/1312.4788}{{\ttfamily arXiv:1312.4788 [hep-ph]}}.
\mciteBstWouldAddEndPunctfalse
\mciteSetBstMidEndSepPunct{\mcitedefaultmidpunct}
{}{\mcitedefaultseppunct}\relax
\EndOfBibitem
\bibitem{Skands:2003cj}
P.~Z. Skands {\em et~al.}, ``{SUSY Les Houches accord: Interfacing SUSY
  spectrum calculators, decay packages, and event generators},''
  \href{http://dx.doi.org/10.1088/1126-6708/2004/07/036}{{\em JHEP} {\bfseries
  07} (2004) 036},
\href{http://arxiv.org/abs/hep-ph/0311123}{{\ttfamily arXiv:hep-ph/0311123}}
\mciteBstWouldAddEndPunctfalse
\mciteSetBstMidEndSepPunct{\mcitedefaultmidpunct}
{}{\mcitedefaultseppunct}\relax
\EndOfBibitem
\bibitem{Allanach:2008qq}
B.~C. Allanach {\em et~al.}, ``{SUSY Les Houches Accord 2},''
  \href{http://dx.doi.org/10.1016/j.cpc.2008.08.004}{{\em Comput. Phys.
  Commun.} {\bfseries 180} (2009) 8--25},
\href{http://arxiv.org/abs/0801.0045}{{\ttfamily arXiv:0801.0045 [hep-ph]}}.
\mciteBstWouldAddEndPunctfalse
\mciteSetBstMidEndSepPunct{\mcitedefaultmidpunct}
{}{\mcitedefaultseppunct}\relax
\EndOfBibitem
\bibitem{Bechtle:2008jh}
P.~Bechtle, O.~Brein, S.~Heinemeyer, G.~Weiglein and K.~E. Williams,
  ``{HiggsBounds: Confronting Arbitrary Higgs Sectors with Exclusion Bounds
  from LEP and the Tevatron},''
  \href{http://dx.doi.org/10.1016/j.cpc.2009.09.003}{{\em Comput.Phys.Commun.}
  {\bfseries 181} (2010) 138--167},
\href{http://arxiv.org/abs/0811.4169}{{\ttfamily arXiv:0811.4169 [hep-ph]}}
\mciteBstWouldAddEndPunctfalse
\mciteSetBstMidEndSepPunct{\mcitedefaultmidpunct}
{}{\mcitedefaultseppunct}\relax
\EndOfBibitem
\bibitem{Bechtle:2011sb}
P.~Bechtle, O.~Brein, S.~Heinemeyer, G.~Weiglein and K.~E. Williams,
  ``{HiggsBounds 2.0.0: Confronting Neutral and Charged Higgs Sector
  Predictions with Exclusion Bounds from LEP and the Tevatron},''
  \href{http://dx.doi.org/10.1016/j.cpc.2011.07.015}{{\em Comput.Phys.Commun.}
  {\bfseries 182} (2011) 2605--2631},
\href{http://arxiv.org/abs/1102.1898}{{\ttfamily arXiv:1102.1898 [hep-ph]}}
\mciteBstWouldAddEndPunctfalse
\mciteSetBstMidEndSepPunct{\mcitedefaultmidpunct}
{}{\mcitedefaultseppunct}\relax
\EndOfBibitem
\bibitem{Bechtle:2013gu}
P.~Bechtle, O.~Brein, S.~Heinemeyer, O.~St\aa{}l, T.~Stefaniak {\em et~al.},
  ``{Recent Developments in HiggsBounds and a Preview of HiggsSignals},'' {\em
  PoS} {\bfseries CHARGED2012} (2012) 024,
\href{http://arxiv.org/abs/1301.2345}{{\ttfamily arXiv:1301.2345 [hep-ph]}}
\mciteBstWouldAddEndPunctfalse
\mciteSetBstMidEndSepPunct{\mcitedefaultmidpunct}
{}{\mcitedefaultseppunct}\relax
\EndOfBibitem
\bibitem{Bechtle:2013wla}
P.~Bechtle, O.~Brein, S.~Heinemeyer, O.~Stål, T.~Stefaniak {\em et~al.},
  ``{$\mathsf{HiggsBounds}-4$: Improved Tests of Extended Higgs Sectors against
  Exclusion Bounds from LEP, the Tevatron and the LHC},''
  \href{http://dx.doi.org/10.1140/epjc/s10052-013-2693-2}{{\em Eur.Phys.J.}
  {\bfseries C74} (2014) 2693},
\href{http://arxiv.org/abs/1311.0055}{{\ttfamily arXiv:1311.0055 [hep-ph]}}.
\mciteBstWouldAddEndPunctfalse
\mciteSetBstMidEndSepPunct{\mcitedefaultmidpunct}
{}{\mcitedefaultseppunct}\relax
\EndOfBibitem
\bibitem{Aad:2015xua}
{\bfseries ATLAS Collaboration}, G.~Aad {\em et~al.}, ``{Constraints on the
  off-shell Higgs boson signal strength in the high-mass $ZZ$ and $WW$ final
  states with the ATLAS detector},''
  \href{http://dx.doi.org/10.1140/epjc/s10052-015-3542-2}{{\em Eur. Phys. J.}
  {\bfseries C75} no.~7, (2015) 335},
\href{http://arxiv.org/abs/1503.01060}{{\ttfamily arXiv:1503.01060 [hep-ex]}}.
\mciteBstWouldAddEndPunctfalse
\mciteSetBstMidEndSepPunct{\mcitedefaultmidpunct}
{}{\mcitedefaultseppunct}\relax
\EndOfBibitem
\bibitem{Khachatryan:2015mma}
{\bfseries CMS Collaboration}, V.~Khachatryan {\em et~al.}, ``{Limits on the
  Higgs boson lifetime and width from its decay to four charged leptons},''
  \href{http://dx.doi.org/10.1103/PhysRevD.92.072010}{{\em Phys. Rev.}
  {\bfseries D92} no.~7, (2015) 072010},
\href{http://arxiv.org/abs/1507.06656}{{\ttfamily arXiv:1507.06656 [hep-ex]}}
\mciteBstWouldAddEndPunctfalse
\mciteSetBstMidEndSepPunct{\mcitedefaultmidpunct}
{}{\mcitedefaultseppunct}\relax
\EndOfBibitem
\bibitem{Khachatryan:2016ctc}
{\bfseries CMS Collaboration}, V.~Khachatryan {\em et~al.}, ``{Search for
  Higgs boson off-shell production in proton-proton collisions at 7 and 8 TeV
  and derivation of constraints on its total decay width},''
  \href{http://dx.doi.org/10.1007/JHEP09(2016)051}{{\em JHEP} {\bfseries 09}
  (2016) 051},
\href{http://arxiv.org/abs/1605.02329}{{\ttfamily arXiv:1605.02329 [hep-ex]}}.
\mciteBstWouldAddEndPunctfalse
\mciteSetBstMidEndSepPunct{\mcitedefaultmidpunct}
{}{\mcitedefaultseppunct}\relax
\EndOfBibitem
\bibitem{Caola:2013yja}
F.~Caola and K.~Melnikov, ``{Constraining the Higgs boson width with ZZ
  production at the LHC},''
  \href{http://dx.doi.org/10.1103/PhysRevD.88.054024}{{\em Phys. Rev.}
  {\bfseries D88} (2013) 054024},
\href{http://arxiv.org/abs/1307.4935}{{\ttfamily arXiv:1307.4935 [hep-ph]}}.
\mciteBstWouldAddEndPunctfalse
\mciteSetBstMidEndSepPunct{\mcitedefaultmidpunct}
{}{\mcitedefaultseppunct}\relax
\EndOfBibitem
\bibitem{Englert:2014aca}
C.~Englert and M.~Spannowsky, ``{Limitations and Opportunities of Off-Shell
  Coupling Measurements},''
  \href{http://dx.doi.org/10.1103/PhysRevD.90.053003}{{\em Phys. Rev.}
  {\bfseries D90} (2014) 053003},
\href{http://arxiv.org/abs/1405.0285}{{\ttfamily arXiv:1405.0285 [hep-ph]}}.
\mciteBstWouldAddEndPunctfalse
\mciteSetBstMidEndSepPunct{\mcitedefaultmidpunct}
{}{\mcitedefaultseppunct}\relax
\EndOfBibitem
\bibitem{lapack}
{\url{https://www.netlib.org/lapack/}}\relax
\mciteBstWouldAddEndPunctfalse
\mciteSetBstMidEndSepPunct{\mcitedefaultmidpunct}
{}{\mcitedefaultseppunct}\relax
\EndOfBibitem
\bibitem{Lepage:1977sw}
G.~P. Lepage, ``{A New Algorithm for Adaptive Multidimensional Integration},''
\href{http://dx.doi.org/10.1016/0021-9991(78)90004-9}{{\em J. Comput. Phys.}
  {\bfseries 27} (1978) 192}.
\mciteBstWouldAddEndPunctfalse
\mciteSetBstMidEndSepPunct{\mcitedefaultmidpunct}
{}{\mcitedefaultseppunct}\relax
\EndOfBibitem
\bibitem{Harlander:2012pb}
R.~V. Harlander, S.~Liebler and H.~Mantler, ``{SusHi: A program for the
  calculation of Higgs production in gluon fusion and bottom-quark annihilation
  in the Standard Model and the MSSM},''
  \href{http://dx.doi.org/10.1016/j.cpc.2013.02.006}{{\em Computer Physics
  Communications} {\bfseries 184} (2013) 1605--1617},
\href{http://arxiv.org/abs/1212.3249}{{\ttfamily arXiv:1212.3249 [hep-ph]}}
\mciteBstWouldAddEndPunctfalse
\mciteSetBstMidEndSepPunct{\mcitedefaultmidpunct}
{}{\mcitedefaultseppunct}\relax
\EndOfBibitem
\bibitem{Liebler:2015bka}
S.~Liebler, ``{Neutral Higgs production at proton colliders in the
  CP-conserving NMSSM},''
  \href{http://dx.doi.org/10.1140/epjc/s10052-015-3432-7}{{\em Eur. Phys. J.}
  {\bfseries C75} no.~5, (2015) 210},
\href{http://arxiv.org/abs/1502.07972}{{\ttfamily arXiv:1502.07972 [hep-ph]}}
\mciteBstWouldAddEndPunctfalse
\mciteSetBstMidEndSepPunct{\mcitedefaultmidpunct}
{}{\mcitedefaultseppunct}\relax
\EndOfBibitem
\bibitem{Harlander:2016hcx}
R.~V. Harlander, S.~Liebler and H.~Mantler, ``{SusHi Bento: Beyond NNLO and
  the heavy-top limit},''
\href{http://arxiv.org/abs/1605.03190}{{\ttfamily arXiv:1605.03190 [hep-ph]}}.
\mciteBstWouldAddEndPunctfalse
\mciteSetBstMidEndSepPunct{\mcitedefaultmidpunct}
{}{\mcitedefaultseppunct}\relax
\EndOfBibitem
\bibitem{Heinemeyer:2013tqa}
{\bfseries LHC Higgs Cross Section Working Group}, J.~R. Andersen {\em
  et~al.}, ``{Handbook of LHC Higgs Cross Sections: 3. Higgs Properties},''
\href{http://arxiv.org/abs/1307.1347}{{\ttfamily arXiv:1307.1347 [hep-ph]}}.
\mciteBstWouldAddEndPunctfalse
\mciteSetBstMidEndSepPunct{\mcitedefaultmidpunct}
{}{\mcitedefaultseppunct}\relax
\EndOfBibitem
\bibitem{Lai:2010vv}
H.-L. Lai, M.~Guzzi, J.~Huston, Z.~Li, P.~M. Nadolsky, J.~Pumplin and C.~P.
  Yuan, ``{New parton distributions for collider physics},''
  \href{http://dx.doi.org/10.1103/PhysRevD.82.074024}{{\em Phys. Rev.}
  {\bfseries D82} (2010) 074024},
\href{http://arxiv.org/abs/1007.2241}{{\ttfamily arXiv:1007.2241 [hep-ph]}}.
\mciteBstWouldAddEndPunctfalse
\mciteSetBstMidEndSepPunct{\mcitedefaultmidpunct}
{}{\mcitedefaultseppunct}\relax
\EndOfBibitem
\bibitem{deFlorian:2016spz}
{\bfseries LHC Higgs Cross Section Working Group}, D.~de~Florian {\em et~al.},
  ``{Handbook of LHC Higgs Cross Sections: 4. Deciphering the Nature of the
  Higgs Sector},''
\href{http://arxiv.org/abs/1610.07922}{{\ttfamily arXiv:1610.07922 [hep-ph]}}.
\mciteBstWouldAddEndPunctfalse
\mciteSetBstMidEndSepPunct{\mcitedefaultmidpunct}
{}{\mcitedefaultseppunct}\relax
\EndOfBibitem
\bibitem{ATLAS:2016nke}
{\bfseries ATLAS Collaboration},
``{Measurement of fiducial, differential and production cross sections in the
  $H\to\gamma\gamma$ decay channel with 13.3 fb$^{-1}$ of 13 TeV proton-proton
  collision data with the ATLAS detector},'' ATLAS-CONF-2016-067.
\mciteBstWouldAddEndPunctfalse
\mciteSetBstMidEndSepPunct{\mcitedefaultmidpunct}
{}{\mcitedefaultseppunct}\relax
\EndOfBibitem
\bibitem{CMS:2016ixj}
{\bfseries CMS Collaboration},
``{Updated measurements of Higgs boson production in the diphoton decay channel
  at $\sqrt{s}=13~\textrm{TeV}$ in pp collisions at CMS.},'' CMS-PAS-HIG-16-020.
\mciteBstWouldAddEndPunctfalse
\mciteSetBstMidEndSepPunct{\mcitedefaultmidpunct}
{}{\mcitedefaultseppunct}\relax
\EndOfBibitem
\bibitem{Khachatryan:2016vau}
{\bfseries ATLAS and CMS Collaborations}, G.~Aad {\em et~al.}, ``{Measurements
  of the Higgs boson production and decay rates and constraints on its
  couplings from a combined ATLAS and CMS analysis of the LHC pp collision data
  at $ \sqrt{s}=7 $ and 8 TeV},''
  \href{http://dx.doi.org/10.1007/JHEP08(2016)045}{{\em JHEP} {\bfseries 08}
  (2016) 045},
\href{http://arxiv.org/abs/1606.02266}{{\ttfamily arXiv:1606.02266 [hep-ex]}}.
\mciteBstWouldAddEndPunctfalse
\mciteSetBstMidEndSepPunct{\mcitedefaultmidpunct}
{}{\mcitedefaultseppunct}\relax
\EndOfBibitem
\bibitem{mathematica}
{\url{http://reference.wolfram.com/language/ref/ListConvolve.html}}\relax
\mciteBstWouldAddEndPunctfalse
\mciteSetBstMidEndSepPunct{\mcitedefaultmidpunct}
{}{\mcitedefaultseppunct}\relax
\EndOfBibitem
\bibitem{Gianotti:2002xx}
F.~Gianotti {\em et~al.}, ``{Physics potential and experimental challenges of
  the LHC luminosity upgrade},''
  \href{http://dx.doi.org/10.1140/epjc/s2004-02061-6}{{\em Eur. Phys. J.}
  {\bfseries C39} (2005) 293--333},
\href{http://arxiv.org/abs/hep-ph/0204087}{{\ttfamily arXiv:hep-ph/0204087
  [hep-ph]}}.
\mciteBstWouldAddEndPunctfalse
\mciteSetBstMidEndSepPunct{\mcitedefaultmidpunct}
{}{\mcitedefaultseppunct}\relax
\EndOfBibitem
\bibitem{CMS:2016ilx}
{\bfseries CMS Collaboration},
``{Measurements of properties of the Higgs boson and search for an additional
  resonance in the four-lepton final state at sqrt(s) = 13 TeV},'' CMS-PAS-HIG-16-033.
\mciteBstWouldAddEndPunctfalse
\mciteSetBstMidEndSepPunct{\mcitedefaultmidpunct}
{}{\mcitedefaultseppunct}\relax
\EndOfBibitem
\bibitem{Cacciapaglia:2014rla}
G.~Cacciapaglia, A.~Deandrea, G.~Drieu La~Rochelle and J.-B. Flament, ``{Higgs
  couplings: disentangling New Physics with off-shell measurements},''
  \href{http://dx.doi.org/10.1103/PhysRevLett.113.201802}{{\em Phys. Rev.
  Lett.} {\bfseries 113} no.~20, (2014) 201802},
\href{http://arxiv.org/abs/1406.1757}{{\ttfamily arXiv:1406.1757 [hep-ph]}}.
\mciteBstWouldAddEndPunctfalse
\mciteSetBstMidEndSepPunct{\mcitedefaultmidpunct}
{}{\mcitedefaultseppunct}\relax
\EndOfBibitem
\bibitem{Logan:2014ppa}
H.~E. Logan, ``{Hiding a Higgs width enhancement from off-shell $gg(\to h^*)\to ZZ$
  measurements},'' \href{http://dx.doi.org/10.1103/PhysRevD.92.075038}{{\em
  Phys. Rev.} {\bfseries D92} no.~7, (2015) 075038},
\href{http://arxiv.org/abs/1412.7577}{{\ttfamily arXiv:1412.7577 [hep-ph]}}.
\mciteBstWouldAddEndPunctfalse
\mciteSetBstMidEndSepPunct{\mcitedefaultmidpunct}
{}{\mcitedefaultseppunct}\relax
\EndOfBibitem
\bibitem{Englert:2015zra}
C.~Englert, I.~Low and M.~Spannowsky, ``{On-shell interference effects in
  Higgs boson final states},''
  \href{http://dx.doi.org/10.1103/PhysRevD.91.074029}{{\em Phys. Rev.}
  {\bfseries D91} no.~7, (2015) 074029},
\href{http://arxiv.org/abs/1502.04678}{{\ttfamily arXiv:1502.04678 [hep-ph]}}
\mciteBstWouldAddEndPunctfalse
\mciteSetBstMidEndSepPunct{\mcitedefaultmidpunct}
{}{\mcitedefaultseppunct}\relax
\EndOfBibitem
\bibitem{Englert:2015bwa}
C.~Englert, M.~McCullough and M.~Spannowsky, ``{Combining LEP and LHC to bound
  the Higgs Width},''
  \href{http://dx.doi.org/10.1016/j.nuclphysb.2015.11.017}{{\em Nucl. Phys.}
  {\bfseries B902} (2016) 440--457},
\href{http://arxiv.org/abs/1504.02458}{{\ttfamily arXiv:1504.02458 [hep-ph]}}.
\mciteBstWouldAddEndPunctfalse
\mciteSetBstMidEndSepPunct{\mcitedefaultmidpunct}
{}{\mcitedefaultseppunct}\relax
\EndOfBibitem
\bibitem{Cornwall:1981zr}
J.~M. Cornwall, ``{Dynamical Mass Generation in Continuum QCD},''
\href{http://dx.doi.org/10.1103/PhysRevD.26.1453}{{\em Phys. Rev.} {\bfseries
  D26} (1982) 1453}.
\mciteBstWouldAddEndPunctfalse
\mciteSetBstMidEndSepPunct{\mcitedefaultmidpunct}
{}{\mcitedefaultseppunct}\relax
\EndOfBibitem
\bibitem{Cornwall:1989gv}
J.~M. Cornwall and J.~Papavassiliou, ``{Gauge Invariant Three Gluon Vertex in
  QCD},''
\href{http://dx.doi.org/10.1103/PhysRevD.40.3474}{{\em Phys. Rev.} {\bfseries
  D40} (1989) 3474}
\mciteBstWouldAddEndPunctfalse
\mciteSetBstMidEndSepPunct{\mcitedefaultmidpunct}
{}{\mcitedefaultseppunct}\relax
\EndOfBibitem
\bibitem{Papavassiliou:1989zd}
J.~Papavassiliou, ``{Gauge Invariant Proper Selfenergies and Vertices in Gauge
  Theories with Broken Symmetry},''
\href{http://dx.doi.org/10.1103/PhysRevD.41.3179}{{\em Phys. Rev.} {\bfseries
  D41} (1990) 3179}
\mciteBstWouldAddEndPunctfalse
\mciteSetBstMidEndSepPunct{\mcitedefaultmidpunct}
{}{\mcitedefaultseppunct}\relax
\EndOfBibitem
\bibitem{Papavassiliou:1994pr}
J.~Papavassiliou, ``{Gauge independent transverse and longitudinal self
  energies and vertices via the pinch technique},''
  \href{http://dx.doi.org/10.1103/PhysRevD.50.5958}{{\em Phys. Rev.} {\bfseries
  D50} (1994) 5958--5970},
\href{http://arxiv.org/abs/hep-ph/9406258}{{\ttfamily arXiv:hep-ph/9406258}}.
\mciteBstWouldAddEndPunctfalse
\mciteSetBstMidEndSepPunct{\mcitedefaultmidpunct}
{}{\mcitedefaultseppunct}\relax
\EndOfBibitem
\bibitem{Degrassi:1992ue}
G.~Degrassi and A.~Sirlin, ``{Gauge invariant selfenergies and vertex parts of
  the Standard Model in the pinch technique framework},''
\href{http://dx.doi.org/10.1103/PhysRevD.46.3104}{{\em Phys. Rev.} {\bfseries
  D46} (1992) 3104--3116}
\mciteBstWouldAddEndPunctfalse
\mciteSetBstMidEndSepPunct{\mcitedefaultmidpunct}
{}{\mcitedefaultseppunct}\relax
\EndOfBibitem
\bibitem{Hashimoto:1994ct}
S.~Hashimoto, J.~Kodaira, Y.~Yasui, and K.~Sasaki, ``{The Background field
  method: Alternative way of deriving the pinch technique's results},''
  \href{http://dx.doi.org/10.1103/PhysRevD.50.7066}{{\em Phys. Rev.} {\bfseries
  D50} (1994) 7066--7076},
\href{http://arxiv.org/abs/hep-ph/9406271}{{\ttfamily arXiv:hep-ph/9406271}}
\mciteBstWouldAddEndPunctfalse
\mciteSetBstMidEndSepPunct{\mcitedefaultmidpunct}
{}{\mcitedefaultseppunct}\relax
\EndOfBibitem
\bibitem{Watson:1994tn}
N.~J. Watson, ``{Universality of the pinch technique gauge boson
  selfenergies},'' \href{http://dx.doi.org/10.1016/0370-2693(95)00244-F}{{\em
  Phys. Lett.} {\bfseries B349} (1995) 155--164},
\href{http://arxiv.org/abs/hep-ph/9412319}{{\ttfamily arXiv:hep-ph/9412319}}
\mciteBstWouldAddEndPunctfalse
\mciteSetBstMidEndSepPunct{\mcitedefaultmidpunct}
{}{\mcitedefaultseppunct}\relax
\EndOfBibitem
\bibitem{Binosi:2009qm}
D.~Binosi and J.~Papavassiliou, ``{Pinch Technique: Theory and Applications},''
  \href{http://dx.doi.org/10.1016/j.physrep.2009.05.001}{{\em Phys. Rept.}
  {\bfseries 479} (2009) 1--152},
\href{http://arxiv.org/abs/0909.2536}{{\ttfamily arXiv:0909.2536 [hep-ph]}}.
\mciteBstWouldAddEndPunctfalse
\mciteSetBstMidEndSepPunct{\mcitedefaultmidpunct}
{}{\mcitedefaultseppunct}\relax
\EndOfBibitem
\end{mcitethebibliography}
\ifx\mcitethebibliography\mciteundefinedmacro
\PackageError{unsrtM.bst}{mciteplus.sty has not been loaded}
{This bibstyle requires the use of the mciteplus package.}\fi

\end{document}